\newif\iflandscape
\newif\ifportrait
\typein[\lorp]%
{Typein "l" (for landscape, twocolumn) or "p" (for portrait, onecolumn)}
\if l\lorp \landscapetrue \else \portraittrue \fi
\newlength{\extralineskip}
\ifportrait
  \documentstyle[]{article}
\fi
\iflandscape
  \documentstyle[twocolumn]{article}
\fi

\def\bea{\begin{eqnarray}}
\def\eea{\end{eqnarray}}
\def\nn{\nonumber}

\def\be{\begin{equation}\new\begin{array}{c}}
\def\ee{\end{array}\end{equation}}

\makeatletter
\newdimen\normalarrayskip              
\newdimen\minarrayskip                 
\normalarrayskip\baselineskip
\minarrayskip\jot
\newif\ifold             \oldtrue            \def\new{\oldfalse}
\def\arraymode{\ifold\relax\else\displaystyle\fi} 
\def\eqnumphantom{\phantom{(\theequation)}}     
\def\@arrayskip{\ifold\baselineskip\z@\lineskip\z@
     \else
     \baselineskip\minarrayskip\lineskip2\minarrayskip\fi}
\def\@arrayclassz{\ifcase \@lastchclass \@acolampacol \or
\@ampacol \or \or \or \@addamp \or
   \@acolampacol \or \@firstampfalse \@acol \fi
\edef\@preamble{\@preamble
  \ifcase \@chnum
     \hfil$\relax\arraymode\@sharp$\hfil
     \or $\relax\arraymode\@sharp$\hfil
     \or \hfil$\relax\arraymode\@sharp$\fi}}
\def\@array[#1]#2{\setbox\@arstrutbox=\hbox{\vrule
     height\arraystretch \ht\strutbox
     depth\arraystretch \dp\strutbox
     width\z@}\@mkpream{#2}\edef\@preamble{\halign \noexpand\@halignto
\bgroup \tabskip\z@ \@arstrut \@preamble \tabskip\z@ \cr}%
\let\@startpbox\@@startpbox \let\@endpbox\@@endpbox
  \if #1t\vtop \else \if#1b\vbox \else \vcenter \fi\fi
  \bgroup \let\par\relax
  \let\@sharp##\let\protect\relax
  \@arrayskip\@preamble}
\def\eqnarray{\stepcounter{equation}%
              \let\@currentlabel=\theequation
              \global\@eqnswtrue
              \global\@eqcnt\z@
              \tabskip\@centering
              \let\\=\@eqncr
              $$%
 \halign to \displaywidth\bgroup
    \eqnumphantom\@eqnsel\hskip\@centering
    $\displaystyle \tabskip\z@ {##}$%
    &\global\@eqcnt\@ne \hskip 2\arraycolsep
         $\displaystyle\arraymode{##}$\hfil
    &\global\@eqcnt\tw@ \hskip 2\arraycolsep
         $\displaystyle\tabskip\z@{##}$\hfil
         \tabskip\@centering
    &{##}\tabskip\z@\cr}
\def\theequation{\thesection.\arabic{equation}}

\newfont{\hr}{msbm10}
\newfont{\ams}{msam10}

\begin{document}
\begin{titlepage}
\setcounter{footnote}0
\begin{center}
\hfill ITEP-M2/95\\
\hfill {\it January 1995}          \\

\vspace{0.3in}

\centerline{A.Morozov}

\centerline{ITEP 117259 Moscow}

\bigskip

\centerline{\Large\bf MATRIX MODELS AS INTEGRABLE SYSTEMS}

\bigskip

\centerline{\it Presented at Banff Conference, August 1994}
\end{center}

\bigskip

\bigskip

\begin{quotation}
The theory of matrix models is reviewed from the point of view of
its relation to integrable hierarchies.
Determinantal formulas, relation to conformal field models and
the theory of Generalized Kontsevich model are discussed in some
detail. Attention is also paid to the group-theoretical interpretation
of $\tau$-functions which allows to go beyond the restricted set
of the (multicomponent) KP and Toda integrable hierarchies.
\end{quotation}
\end{titlepage}
\clearpage

\newpage

\tableofcontents

\newpage
\setcounter{page}{1}

\section{Introduction}

\setcounter{equation}{0}

The purpose of these notes is to review one of the branches of modern string
theory: the theory of matrix models. We put emphasize on their intrinsic
integrable structure and almost ignore direct physical applications which
are broadly discussed in the literature. Also most of technical details
and references are omited: they can be found in a recent review \cite{UFNmm}.

Both ``matrix models'' and ``integrability'' are somewhat misleading
names for the field to be discussed, they refer more to the history
of the subject than to its real content. In fact the problem which is
actually addressed is that of description of {\it non-perturbative
partition functions} in quantum theory. The term ``non-perturbative
partition function'' is now widely used to denote the generating
functional of {\it all} the {\it exact} correlation functions in
a given quantum model. Such quantity is given by a functional
integral where the weight in the sum over trajectories is defined
by effective action, which contains either all the possible
(local or non-local) counterterms or generic coupling to external
fields, so that any correlator can be obtained as derivative with respect
to appropriate coupling constants or background fields.
These exact generating functionals possess new peculiar properties,
resulting from the possibility to perform {\it arbitrary} change of
integration variables in the functional integral. Such properties are
never studied in the orthodox quantum field theory because there the freedom
to change integration variables is severely restricted by requirements of
locality and renormalizability, which at last lost their role as
fundamental principles of physics with creation of string theory.

Every change of integration variables can be alternatively described as
some change of parameters of non-perturbative partition function
(i.e. the coupling constants or background fields in effective action).
Thus invariance of the integral implies certain relations (Ward identities)
between partition functions at different values of parameters. Since all
the fields of the model are integrated over, the set of relations is
actually exhaustively large: more or less any two sets of parameters
are related. Exact formulation of this property is yet unknown.
The natural first step in these
investigations is to look at the finite-dimensional integrals and
then proceed to functional integrals by increasing the number
of integrations. In turn, the natural way to do it is to make use
of the well studied ``matrix models'', which deal with $N\times N$ matrix
integrals and their behaviour in the large-$N$ limit. It appears
that non-perturbative partition functions of matrix models, at least
when they can be handled with the presently available techniques, are
closely related to ``$\tau$-functions'', introduced
originally in the study of integrable hierarchies. It looks very
probable that some crucial characteristics of such partition functions
are in fact not so peculiar for these simple examples, but remain
true in the absolutely general setting. Extraction of such properties
and construction of the adequate notion of ``generalized $\tau$-function''
is the main task of further work in the theory of ``matrix models''
and ``integrable systems''.

One of the most straightforward and still
promissing approaches is based on interpretation of Ward identites for
non-perturbative partition functions as Hirota-like equations
(supplemented by a much smaller set of ``string equations''), while
generalized $\tau$-functions, which are solutions to these
equations, are interpreted as group-theoretical objects (generating
functionals of all the matrix elements of a group element in particular
representation). If successfull, this approach can provide description
of exact correlation functions in terms of some (originally hidden)
symmetry of the given class of theories, thus raising to the new height
the relation between physical theories and symmetries, which was the
guiding line for developement of theoretical physics during the last
decades.

Analysis of non-perturbative partition functions is very important for
one more reason. Construction of a generating functional is essentially
``exponentiation of perturbations'', i.e. it deformes original (bare)
action of the model. When perturbation parameters (extra coupling
constants or background fields) are non-infinitesimal, one in fact
obtains entire {\it set} of models instead of original one. Moreover,
original model is no longer distinguished within this class: non-perturbative
partition functions are associated with classes of models, not with
a single model. One can easily recognize here realization of the main
idea of the string programm (see for example \cite{UFN}). The study
of non-perturbative partition functions even for such a simple class
as ordinary matrix models can lead to much better understanding of
the general idea. As usual, this can help to figure out what the adequate
questions are and to develop effective technique to answer these
questions.

The purpose of these notes is to briefly illustrate the general ideas
with very simple examples. A lot of work is still required to obtain
applications to the really interesting problems. Still, simple
examples are enough to understand the ideas, and often conceptual level
is no less important than that of technical effectiveness.

We begin in the following sections from cosideration of the simplest
matrix models, to be refered to as discrete and Kontsevich (continuous)
models. Again, these names reflect more the history than the real
content of the subject (continuous models were originaly described
non-explicitly as specific (multiscaling) large-$N$ limits of the
discrete ones). In the context of non-perturbative partition
functions the difference is that discrete models possess effective
actions with all the possible counterterms added, while in Kontsevich
model an external field (source) is introduced.
Analysis of these theories includes their characterization
as {\it eigenvalue} models and derivation of related
determinant representations. We also consider description of
discrete models in the language of conformal field theory. It is
important for connecting these matrix models to the physically
relevant Liouville theory of $2d$ gravity and - more essential
for our notes - to the concept of KP and Toda $\tau$-functions.

Then we turn to the free-fermion (Grassmannian)
description of KP/Toda $\tau$-functions and list some results
from the theory of Generalized Kontsevich model \cite{Ko}-\cite{AvM}.
Some other interesting models, which look {\it a priori} very different,
are in fact just particular examples of Kontsevich model, which
actually describes a big family of theories. This example should
be very instructive for the future understanding of the interplay
between perturbative and non-perturbative information contained
in the non-perturbative partition function.

The last topic of these notes concerns group-theoretical
interpretation of $\tau$-functions. The natural object, which
arises in this way, while possessing the most important
properties of conventional $\tau$-functions, is in fact much
more general. In this framework the KP/Toda $\tau$-functions are associated
with fundamental representations of $SL(N)$ and the closely
related theory of the simply-laced Kac-Moody algebras of
level $k=1$. In fact $\tau$-function can be easily defined for {\it any}
representation of {\it any} group (including also quantum groups -
this can be important for the future construction of string
field theory, where the idea of ``third quantization'' requires
consideration of operator-valued $\tau$-functions; since it takes our
non-perturbative partition function as input - effective action
of the full string field theory).

\section{The basic example: discrete 1-matrix model}

The sample example of matrix model is that of 1-matrix integral
\be
Z_N\{t\} \equiv c_N\int_{N\times N}
dH e^{\sum_{k=0}^{\infty} t_k {\rm Tr} H^k},
\label{1mamo}
\ee
where the integral is over $N\times N$ matrix $H$ and
$dH = \prod_{i,j} dH_{ij}$.\footnote{This integral is often
refered to as {\it Hermitean}. In most of our considerations
we do not need to specify integration contours in matrix integrals,
in particular eigenvalues of $H_{ij}$ do not need to be real.
What is indeed important is the ``flatness'' of the measure
$dH = \prod_{i,j} dH_{ij}$. Below ``Hermitean'' (as opposed,
for example, to ``unitary'') will imply just this choice of the
measure and not any reality condition.}
This measure is invariant under the conjugation
$H \rightarrow UHU^\dagger$ with any unitary $N\times N$ matrix
$U$, and the ``action'' $\sum_{k=0}^{\infty} t_k {\rm Tr} H^k$
in (\ref{1mamo}) is the most general one consistent with this
invariance. Thus $Z_N\{t\}$ is indeed an example of the
non-perturbative partition function in the sence, described in
the Introduction. All observables in the theory are given by
algebraic combinations of ${\rm Tr}H^k$ and their correlation
functions can be obtained by action of $t_k$-derivatives on
$Z_N\{t\}$.  Our goal now is to find some more invariant description
of this quantity, not so specific as the matrix integral (\ref{1mamo}).

\subsection{Ward identities}

Such description is provided by Ward identities. The integral
is invariant under any change of integration matrix-variable
$H \rightarrow f(H)$. It is convenient to choose the special basis in
the space of such transformations:
\be
\delta H = \epsilon_nH^{n+1}.
\ee
Here $\epsilon_n$ is some infinitesimal matrix and, of course, $n \geq -1$.
Invariance of the integral implies the following identity:
\be
\int_{N\times N} dH e^{\sum_{k=0}^{\infty} t_k {\rm Tr}H^k}
 = \int d(H + \epsilon_nH^{n+1})e^{\sum_{k=0}^{\infty} t_k {\rm Tr}
(H + \epsilon_nH^{n+1})^k},
\nn
\ee
i.e.
\be
\int dH e^{\sum_{k=0}^{\infty} t_k {\rm Tr}H^k} \left(
\sum_{k=0}^{\infty} kt_k {\rm Tr}H^{k+n} + {\rm Tr}
\frac{\delta H^{n+1}}{\delta H} \right) \equiv 0.
\label{vird1}
\ee
In order to evaluate the Jacobian
${\rm Tr}\frac{\delta H^{n+1}}{\delta H}$ let us restore the matrix indices:
\be
(\delta H^{n+1})_{ij} =
\sum_{k=0}^n (H^k \delta H H^{n-k})_{ij} =
\sum_{k=0}^n (H^k)_{il}( \delta H)_{lm}( H^{n-k})_{mj},
\nn
\ee
and to obtain  ${\rm Tr}\frac{\delta H^{n+1}}{\delta H}$ put  $l=i$ and
$m=j$, so that
\be
{\rm Tr}\frac{\delta H^{n+1}}{\delta H} =
\sum_{k=0}^n {\rm Tr}H^k {\rm Tr} H^{n-k}.
\label{vird2}
\ee
Any correlation function can be obtained as variation of the coupling
constants:
\be
<{\rm Tr} H^{a_1} ... {\rm Tr} H^{a_n}> =
\int dH e^{\sum_{k=0}^{\infty} t_k {\rm Tr}H^k}
{\rm Tr} H^{a_1} ... {\rm Tr} H^{a_n}  = \nn \\
=   \frac{\partial^n}{\partial t_{a_1}...\partial t_{a_n}} Z_N\{t\}.
\label{vird3}
\ee
This relation together with (\ref{vird2}) can be used to rewrite
(\ref{vird1}) as:
\be
L_n Z_N\{t\} = 0, \ \ \ n\geq -1
\label{virdid}
\ee
with
\be
L_n \equiv \sum_{k=0}^{\infty} kt_k\frac{\partial}{\partial t_{k+n}} +
    \sum_{k=0}^n \frac{\partial^2}{\partial t_k\partial t_{n-k}}.
\label{virdop}
\ee
Note that according to the definition (\ref{1mamo})
\be
\frac{\partial}{\partial t_0}Z_N = NZ_N. \nn
\ee


\subsubsection{Details and comments}

Several remarks are now in order.

First of all, expression in brackets in (\ref{vird1}) represents just $all$
the equations of motion for the model (\ref{1mamo}), and (\ref{virdid}) is
nothing but another way to represent the same set of equations. This
example illustrates what ``exhaustively large'' set of Ward identities
is: it should be essentially the same as the set of all equations of motion.

Second, commutator of any two operators $L_n$ apearing in (\ref{virdid})
should also annihilate $Z_N\{t\}$. Another indication that we already
got a $complete$ set of constraints, is that $L_n$'s form a closed algebra:
\be
\phantom. [ L_n, L_m] = (n-m) L_{n+m}, \ \ \ n,m\geq -1.
\label{virdal}
\ee
Its particular representation (\ref{virdop}) is refered to as
``discrete Virasoro algebra'' (to emphasize the difference with
``continuous Virasoro'' constraints, see eq.(\ref{vircidder}) below).

Third, (\ref{virdid}) can be considered as invariant formulation of what is
$Z_N$: it is a solution of this set of compatible differential equations.
{}From this point of view eq.(\ref{1mamo}) is rather a particular
representation of $Z_N$  and it is sensible to look for
other representations as well (we shall later discuss two of them: one in
terms of CFT, another in terms of Kontsevich integrals).

Fourth, one can try to analyze the uniqueness of the solutions to
(\ref{virdid}). If there are not too many solutions, the set of constraints
can be considered complete. A natural approach to classification of solutions
to the algebra of constraints is in terms of the orbits of the corresponding
group \cite{GMMMMO}.
Let us consider an oversimplified example, which can still be usefull
to understand implications of the complete set of WI as well as to clarify
the meaning of classes of universality and of integrability.

Imagine, that instead of (\ref{virdid}) with $L_n$'s defined in (\ref{virdop})
we would obtain somewhat simpler equations:
\footnote{One can call them "classical" approximation to (\ref{virdid}), since
they would arise if the variation of measure (i.e. a
"quantum effect") was not taken into account in the derivation of
(\ref{virdid}). Though  this concept is often
used in physics it does not have much sense in the present context, when we
are analyzing $exact$ properties of functional (matrix) integrals. }
\be
l_n Z = 0, \ \ n\geq 0 \ \ {\rm with} \ \
l_n = \sum_{k=1}^{\infty} kt_k\frac{\partial}{\partial t_{k+n}}. \nn
\ee
Then operator $l_1$ can be interpreted as generating the shifts
\be
t_2 \longrightarrow t_2 + \epsilon_1 t_1, \nn \\
t_3 \longrightarrow t_3 + 2\epsilon_1 t_2, \nn \\
\ldots                    \nn
\ee
We can use it to shift $t_2$ to zero, and eq. $l_1Z = 0$
then implies that
\be
Z(t_1,t_2,t_3,...) = Z(t_1,0,\tilde t_3,...)
\nn
\ee
$(\tilde t_k = t_k - \frac{(k-1)t_2t_{k-1}}{t_1}, \ k\geq 3)$.

Next, operator $l_2$ generates the shifts
\be
t_3 \longrightarrow t_3 + \epsilon_2 t_1, \nn \\
t_4 \longrightarrow t_4 + 2\epsilon_2 t_2, \nn \\
.\ldots \nn
\ee
and does $not$ affect $t_2$. We can now use eq. $l_2Z = 0$ to
argue that
\be
Z(t_1,t_2,t_3,t_4,...) = Z(t_1,0,\tilde t_3,\tilde t_4,...) =
Z(t_1,0,0,\tilde{\tilde t}_4,...)  \nn
\ee
etc. Assuming that $Z$ is not very much dependent on $t_k$ with
$k \rightarrow \infty$,
\footnote{This, by the way,  is hardly correct in this particular example,
when the group has no compact orbits.}
we can conclude, that
\be
Z(t_1,t_2,t_3,...) = Z(t_1,0,0,...) =
Z(1,0,0,...) \nn
\ee
(at the last step we also used the equation $l_0Z = 0$ to rescale
$t_1$ to unity).

All this reasoning had sense provided $t_1 \neq 0$. Otherwise we would get
$Z(0,1,0,0,...)$, if $t_1 = 0,\ t_2\neq 0$, or
$Z(0,0,1,0,...)$, if $t_1 = t_2 = 0,\ t_3\neq 0$ etc.
In other words, we obtain classes of universality (such that the value of
partition function is just the same in the whole class), which in this
oversimplified example are labeled just by the first non-vanishing
time-variable. Analysis of the orbit structure for the actually important
realizations of groups, like that connected to eq.(\ref{virdop}) has never
been performed in the context of matrix model theory.

In this oversimplified case the constraints actually allow one to eliminate
all the dependence on the time-variables, i.e. to solve equations for
$Z$ exactly. In realistic examples one deals with less trivial
representations of the constraint algebra, like (\ref{virdal}). It appears
that in this general framework constraints somehow imply the integrability
structure of the model, what can thus be considered as a slightly more
complicated version of the same solvability phenomenon.

\subsection{CFT interpretation of 1-matrix model}

Given a complete set of the constraints on partition function of
infinitely many variables which form some closed algebra we can now ask an
inverse question: how these equations can be solved or what is the integral
representation of partition function. One approach to this problem
is analysis of orbits, briefly mentioned at the end of the previous
subsection. Now we turn to another technique \cite{comamo}, which makes use
of the knowledge from conformal field theory. This constructions can have
some meaning from the ``physical'' point of
view, which implies certain duality between the 2-dimensional world surfaces
and the spectral surfaces, associated to configuration space of the string
theory. However, our present goal is more formal than discussion of this
duality: we are going to use the methods of CFT for
solving the constraint equations.

This is especialy natural when the algebra of constraints is Virasoro
algebra, as is the case with the 1-matrix model, or some other algebra
known to arise as a chiral algebra in some simple conformal models.
In fact the approach to be discussed is rather general and can
be applied to construction of matrix models, associated with many different
algebraic structures: the only requirement is existence of the (massless)
free-field representation.

We begin from the set of "discrete Virasoro constraints" (\ref{virdid}).
The CFT formulation of interest should
provide the solution to these equations in the form of some correlation
function in some conformal field theory. Of course, it becomes natural if we
somehow identify the operators $L_n$ (\ref{virdop}) with the
harmonics of a stress-tensor $T_n$, which satisfy the same algebra, and
manage to relate the constraint that $L_n$ annihilate the correlator to the
statement that $T_n$ annihilate the vacuum state. Thus the procedure is
naturally split into two steps. First, we should find a $t$-dependent
operator ("Hamiltonian") $H(t)$, such that
\be
L_n(t) \langle e^{H(t)}| \  = \langle e^{H(t)}|T_n
\label{lcft1}
\ee
This will relate differential operators $L_n$ to $T_n$'s expressed through
the fields of conformal model. Second, we need to enumerate the states, that
are annihilated by the operators $T_n$ with $n \geq -1$, i.e. solve equation
\be
T_n \mid G \rangle = 0
\ee
for the ket-states, what is an internal problem of
conformal field theory. If both ingredients $H(t)$ and $\mid G \rangle$ are
found, solution to the problem is given by
\be
\langle e^{H(t)}\mid G\rangle.
\ee

To be more explicit, for the case of the discrete Virasoro constraints we can
just look for solutions in terms of the simplest possible conformal model:
that of a one holomorphic scalar field
\be
\phi (z) =  \hat q + \hat p \log z  + \sum _{k\neq 0} {J_{-k}\over k}
z^{k}\nn\\
\  [J_n,J_m] = n\delta _{n+m,0},  \ \ \     [\hat q,\hat p] = 1.
\ee
Then the procedure is as follows. Define vacuum states
\be
J_k|0\rangle   = 0, \ \ \  \langle N|J_{-k} = 0, \ \ \    k > 0\nn\\
\hat p|0\rangle   = 0, \ \ \   \langle N|\hat p = N\langle N|,
\ee
the stress-tensor
\be
T(z) = {1\over 2}[\partial \phi (z)]^2 = \sum    T_nz^{-n-2},\quad
T_n = \sum _{k>0}J_{-k}J_{k+n} +
{1\over 2}\sum _{{a+b=n}\atop{a,b\geq 0}}J_aJ_b,
\ee
and the Hamiltonian
\be
H(t)  = {1\over \sqrt{2}} \sum _{k>0}t_kJ_k =
{1\over \sqrt{2}}\oint_{C_0}U(z)J(z)\nn\\
U(z)  = \sum _{k>0}t_kz^k, \ \  \   J(z) = \partial \phi (z).
\ee
It is easy to check that
\be
L_n\langle N|e^{H(t)}\ = \langle N|e^{H(t)}T_n
\ee
and
\be
 T_n|0\rangle  = 0,  \ \ \    n \geq  -1 .
\ee
As an immediate consequence, any correlator of the
form
\be
Z_N\{t\mid G\} = \langle N|e^{H(t)}G|0\rangle
\label{confsol}
\ee
gives a solution to (\ref{virdid}) provided
\be
[T_n,G] = 0, \ \ \  n \geq  -1.
\label{crGop}
\ee
In fact operators $G$ that commute with the stress tensor are well known:
these are just any functions of the "screening charges"
\footnote{For notational simplicity we omit the normal ordering signs, in
fact  the relevant operators are  $:e^H:$ and $:e^{\pm \sqrt{2}\phi}:$}
\be
Q_\pm  = \oint J_\pm  = \oint
e^{\pm \sqrt{2}\phi }.
\ee
The correlator (\ref{confsol}) will be non-vanishing only if the matching
condition for zero-modes of $\phi$ is satisfied. If we demand the operator to
depend only on $Q_{+}$, this implies that only
one term of the expansion in powers of $Q_{+}$ will contribute to
(\ref{confsol}), so that the result is essentially independent on the choice
of the function $G(Q_+)$, we can for example take $G(Q_+) = e^{Q_+}$
and obtain:
\be
Z_N\{t\} \sim \frac{1}{N!}\langle N|e^{H(t)}(Q_+)^N|0\rangle .
\label{comamo1mm"}
\ee
This correlator is easy to evaluate using Wick theorem and the propagator
$\phi(z)\phi(z')\sim \log(z-z')$. Finally we get
\be
Z_N\{t\}  = \frac{1}{N!} \langle N \mid
:e^{{1\over \sqrt{2}}\oint_{C_0}U(z)\partial\phi(z)}:
\prod_{i=1}^N \oint_{C_i} dz_i :e^{\sqrt{2}\phi(z_i) }: \mid 0  \rangle  =
\nn \\
 = \frac{1}{N!}\prod_{i=1}^N\oint_{C_i} dz_i e^{U(z_i)}
\prod_{i<j}^N (z_i-z_j)^2
\label{comamo1mm}
\ee
in the form of a multiple integral.
This integral does not yet look like the matrix integral (\ref{1mamo}).
However, it is the same: (\ref{comamo1mm}) is an
``eigenvalue representation'' of matrix integral, see \cite{BIPZ} and
eq.(\ref{1mamoev}) in the next subsection \ref{ev1m}.

\subsubsection{Details and comments}

Thus in the simplest case we resolved the inverse problem: reconstructed an
integral representation  from the set of discrete Virasoro constraints.
However, the answer we got seems a little more general than (ref{1mamo}) and
(\ref{1mamoev}): the r.h.s. of eq.(\ref{comamo1mm}) still depends on the
contours of integration. Moreover, we can also recall that the operator $G$
above could depend not only on $Q_+$, but also on $Q_-$. The most general
formula is a little more complicated than (\ref{comamo1mm}):
\be
\new
\begin{array}{c}
Z_{N}\{t\mid C_i, C_r\} \sim \frac{1}{(N+M)!M!}\langle N|e^{H(t)}(Q_+)^{N+M}
(Q_-)^M|0\rangle =     \\
= \frac{1}{(N+M)!M!}\prod_{i=1}^{N+M}\oint_{C_i} dz_i e^{U(z_i)}
\prod_{r=1}^M\oint_{C'_r} dz'_r e^{U(z'_r)}
\cdot \\
\cdot
\frac{\prod_{i<j}^{N+M} (z_i-z_j)^2  \prod_{r<s}^N (z'_r-z'_s)^2 }
{\prod_{i}^{N+M}\prod_r^M (z_i-z_r)^2}.
\end{array}
\label{comamo1mm'}
\ee
We refer to the papers \cite{comamo} for discussion of the issue of
contour-dependence. In certain sense all these different integrals can be
considered as branches of the same analytical function $Z_N\{t\}$. Dependence
on $M$ is essentially eliminated by Cauchy integration around the poles in
denominator in (\ref{comamo1mm'}).

Above construction can be straightforwardly applied to any other
algebras of constraints, provided:

(i) The free-field representation of the algebra is known in the
CFT-framework, such that the generators are $polinomials$ in the
fields $\phi$ (only in such case it is straightforward to construct a
Hamiltonian $H$, which relates CFT-realization of the algebra to that
in terms of differential operators w.r.to the $t$-variables; in fact
under this  condition $H$  is usually linear in $t$'s and $\phi$'s).
There are examples (like Frenkel-Kac representation of level $k=1$
simply-laced Kac-Moody algebras \cite{FK} or generic reductions of the WZNW
model \cite{GMMOS},\cite{BO}-\cite{GereF})
when generators are $exponents$ of free fields, then this construction
should be slightly modified.

(ii) It is easy to find vacuum, annihilated by the relevant generators (here,
for example, is the problem with application of this approach to the case of
"continuous" Virasoro and $W$-constraints). The resolution to this problem
involves consideration of correlates on Riemann surfaces with non-trivial
topologies, often - of infinite genus.

(iii) The free-field representation of the "screening charges", i.e. operators
that commute with the generators of the group within the conformal model, is
explicitly known.

These conditions are fulfilled in many case in CFT, including conventional
{\bf W}-algebras \cite{Zam} and ${\cal N} = 1$
\footnote{In the case of ${\cal N} = 2$ supersymmetry a problem arises because
of the lack of reasonable screening charges. At the most naive level the
relevant operator to be integrated over superspace (over $dzd^{\cal N}\theta$)
in order to produce screening charge has dimension $1-\frac{1}{2}{\cal N}$,
which $vanishes$ when ${\cal N} = 2$.
}
supersymmetric models.

For illustrative purposes we present here several formulas from the last paper
of ref.\cite{comamo} for the case of the
${\bf W}_{r+1}$-constraints, associated with the simply-laced algebras
${\cal G}$ of rank $r$.

Partition function in such "conformal multimatrix model" is a function of
"time-variables" $t_k^{(\lambda)},\ k = 0...\infty,\ \lambda = 1...r={\rm
rank}{\cal G}$, and also depends on the integer-valued $r$-vector
${\vec N} = \{N_1...n_r\}$.
The ${\bf W}_{r+1}$-constraints imposed on partition function are:
\be
W_n^{(a)}(t)Z_{{\vec N}}^{\cal G}\{t\} = 0, \ \ n\geq 1-a, \ \ a= 2...r+1.
\ee
The form of the $W$-operators is somewhat complicated, for example, in the
case of $r+1=3$ (i.e. for ${\cal G} = SL(3)$)
\be
\new
\begin{array}{c}
W^{(2)}_n =  \sum ^\infty _{k=0}(kt_k\frac{\partial}{\partial t_{k+n}} +
k\bar t_k\frac{\partial}{\partial \bar t_{k+n}}) + \\
+ \sum _{a+b=n}(\frac{\partial ^2}{\partial t_a\partial t_b} +
\frac{\partial ^2}{\partial \bar t_a\partial \bar t_b})
\end{array}
\ee
\be
\new
\begin{array}{c}
W^{(3)}_n = \sum _{k,l>0}(kt_klt_l\frac{\partial}{\partial t_{k+n+l}} -
k\bar t_kl\bar t_l\frac{\partial }{\partial t_{k+n+l}}
-2kt_kl\bar t_l\frac{\partial }{\partial \bar t_{k+n+l}})+ \\
+ 2\sum  _{k>0}\left[ \sum _{a+b=n+k}(kt_k\frac{\partial ^2}{\partial t_a
\partial t_b} - kt_k\frac{\partial ^2}{\partial \bar t_a\partial \bar t_b} -
2k\bar t_k\frac{\partial ^2}{\partial t_a\partial \bar t_b)}\right]  + \\
+ {4\over 3}\sum _{a+b+c=n}(\frac{\partial ^3}
{\partial t_a\partial t_b\partial t_c} -
\frac{\partial ^3}{\partial t_a\partial \bar t_b\partial \bar t_c}),
\end{array}
\label{wopex}
\ee
and two types of time-variables, denoted through $t_k$ and  $\bar t_k$.
are associated with two $orthogonal$ directions
in the  Cartan plane of $A_2$:
${\bf e} = {{\vec\alpha} _1\over\sqrt{2}}$,
$\bar{\bf e} = {\sqrt{3}{\vec\nu} _2\over\sqrt{2}}$.
\footnote{Such orthogonal basis is especially convenient for discussion of
integrability properties of the model, these $t$ and $\bar t$ are linear
combinations of time-variables $t_k^{\lambda}$ appearing in eqs.
(\ref{hamAr}) and (\ref{comamoAr}).}

All other formulas, however, are very simple:
Conformal model is usually that of the $r$ free fields,
$S \sim \int\bar\partial{\vec\phi}\partial{\vec\phi} d^2z$,
which is used to describe representation
of the level one Kac-Moody algebra, associated with ${\cal G}$. Hamiltonian
\be
H(t^{(1)}\ldots t^{(r+1)}) =
\sum_{\lambda = 1}^{r+1}\sum _{k>0}t^{(\lambda )}_k
{\vec\mu}_\lambda {\vec J}_k,
\label{hamAr}
\ee
where $\{{\vec\mu}_{\lambda}\}$ are associated with "fundamental
weight" vectors ${\vec\nu}_{\lambda}$ in Cartan hyperplane and in the
simplest case of ${\cal G} = SL(r+1)$  satisfy
$$
{\vec\mu}_{\lambda}\cdot
{\vec\mu}_{\lambda'}=\delta_{\lambda\lambda'}-{1\over{r+1} }, \ \ \
	\sum_{\lambda=1}^{r+1} {\vec\mu}_{\lambda}=0,
$$
thus only $r$ of
the time variables $t^{(1)}\ldots t^{(r+1)}$ are linearly
independent.  Relation between differential operators $W_n^{(a)}(t)$
and operators ${\rm W}_n^{(a)}$ in the CFT is now defined by

\be
W^{(a)}_n\langle {\vec N}|e^{H(t)}\ = \langle {\vec N}|e^{H(t)}{\rm W}^{(a)}_i,
\nn \\   a=2,\ldots,p;  \ \ \ i\geq 1-a, \ee where \be {\rm
W}^{(a)}_n = \oint z^{a+n-1}{\rm W}^{(a)}(z)\nn\\ {\rm W}^{(a)}(z) =
\sum  _\lambda  [{\vec\mu} _\lambda \partial {\vec\phi} (z)]^a + \ldots
\ee
are spin-$a$ generators of the ${\bf W}^{\cal G}_{r+1}$ algebra.
The screening charges, that commute with all the ${\rm W}^{(a)}(z)$
are given by
\be
Q^{(\alpha)}  = \oint J^{(\alpha)}  = \oint
e^{{\vec\alpha} {\vec\phi} }
\ee
$\{{\vec\alpha} \}$ being roots of
finite-dimensional simply laced Lie algebra ${\cal G}$.

Thus partition function arises in the form:

\be
Z^{\cal G}_{{\vec N}}\{t\} = \langle {\vec N}|e^{H(t}G\{Q
^{(\alpha)} \}|0\rangle
\ee
where  $G$  is an exponential function of screening charges.
Evaluation of the free-feild correlator gives:

\be
\new
\begin{array}{c}
Z^{\cal G}_{{\vec N}}\{t\} \sim  \int   \prod  _\alpha
\left[ \prod ^{N_\alpha }_{i=1}dz^{(\alpha )}_i \exp \left(
\sum _{\lambda ;k>0}t^{(\lambda )}_k({\vec\mu} _\lambda {\vec\alpha} )
(z^{(\alpha)}_i)^k\right) \right] \times \\
\times \prod _{(\alpha ,\beta )}\prod ^{N_\alpha }_{i=1}
\prod ^{N_\beta }_{j=1}(z^{(\alpha )}_i- z^{(\beta )}_j)^{{\vec\alpha}
{\vec\beta} }
\end{array}
\label{comamoAr}
\ee
In fact this expression can be rewritten in terms of an $r$-matrix
integral -- a "conformal multimatrix model":

\be
Z^{\cal G}_{\vec N}\{t^{(\alpha)}\} =
c_N^{p-1}\int_{N\times N} dH^{(1)}...dH^{(p-1)}
\prod_{\alpha = 1}^{p-1} e^{\sum_{k=0}^{\infty}t_k^{(\alpha)}{\rm Tr}
H_{(\alpha)}^k}\cdot    \nn \\
\cdot\prod_{(\alpha ,\beta )}
{\rm Det} \left(H^{(\alpha)}\otimes I - I\otimes
H^{(\alpha+1)}\right)^{\vec\alpha\vec\beta }
\label{comamo"}
\ee
In the simplest case of ${\bf W}_3$ algebra
eq.(\ref{comamoAr}) with insertion of only two (of the six) screenings
$Q_{\alpha _1}$ and  $Q_{\alpha _2}$ turns into

\be
\new
\begin{array}{c}
Z^{SL(3)}_{N_1,N_2}(t,\bar t)  = {1\over N_1!N_2!}
\langle N_1,N_2|e^{H(t,\bar t)}(Q^{(\alpha _1)})^{N_1}
(Q^{(\alpha _2)})^{N_2}|0\rangle  =
 \\
= {1\over N_1!N_2!}  \prod_i \int   dx_i e^{U(x_i)}
\prod_j\int dy_j e^{\bar U(y_i)}  \Delta (x)\Delta (x,y) \Delta (y),
\end{array}
\label{comamo"3}
\ee
where $\Delta(x,y) \equiv \Delta(x)\Delta(y)\prod _{i,j}(x_i - y_j)$.
This model is associated with the algebra ${\cal G} = SL(3)$, while the
original 1-matrix model (\ref{comamo1mm"})-(\ref{comamo1mm'})
- with ${\cal G} = SL(2)$.

The whole series of models (\ref{comamoAr}-\ref{comamo"})
for ${\cal G} = SL(r+1)$ is distinguished by its relation to the level
$k=1$ simply-laced Kac-Moody algebras. In this particular situation the
underlying conformal model has integer central charge $ c = r = {\rm rank}
\ {\cal G}$ and can be "fermionized".\footnote{
This is possible only for very special Kac-Moody algebras, and such
formulation is important in order to deal with $conventional$ formulation of
integrability, which usually involves $commuting$ Hamiltonian flows (not just
a closed algebra of flows) and fermionic realization of the universal module
space (universal Grassmannian). In fact these restrictions are quite
arbitrary and can be removed (though this is not yet done in full details),
see \cite{UFNmm} and sections \ref{se4}, \ref{se5} below for more detailed
discussion.}
The main feature of this formulation is that the Kac-Moody currents (which
after integration turn into "screening charges" in the above construction)
are quadratic in fermionic fields, while they are represented by exponents in
the free-boson formulation.

In fact fermionic (spinor) model naturally possesses $GL(r+1)$ rather than
$SL(r+1)$ symmetry (other simply-laced algebras can be embedded into larger
$GL$-algebras and this provides fermionic descriprion for them in the case of
$k=1$). The model contains $r+1$ spin-1/2 fields $\psi_i$ and their conjugate
$\tilde\psi_i$ ($b,c$-systems);
\be
S = \sum_{j=1}^{r+1} \int \tilde\psi_j\bar\partial\psi_j d^2z,
\nn
\ee
central charge $c=r+1$, and operator algebra is
\be
\tilde\psi_j(z)\psi_k(z')  = \frac{\delta_{jk}}{z-z'}\ +
:\tilde\psi_j(z)\psi_k(z'): \nn \\
\psi_j(z)\psi_k(z')  =  (z-z')\delta_{jk}:\psi_j(z)\psi_k(z'):
   + \ (1-\delta_{jk}):\psi_j(z)\psi_k(z'): \nn\\
\tilde\psi_j(z)\tilde\psi_k(z')  =
    (z-z')\delta_{jk}:\tilde\psi_j(z)\tilde\psi_k(z'):
   + \ (1-\delta_{jk}):\tilde\psi_j(z)\tilde\psi_k(z'): \nn
\ee
The Kac-Moody currents of the level-one $\hat{GL(r+1)}_{k=1}$ are just
$J_{jk} = :\tilde\psi_j\psi_k:\ \ j,k = 1\ldots r+1$, and screening charges
are
$Q^{(\alpha)} = iE_{jk}^{(\alpha)}\oint :\tilde\psi_j\psi_k:$, where
$E_{jk}^{(\alpha)}$ are representatives of the roots ${\vec\alpha}$ in the
matrix
representation of $GL(r+1)$. Cartan subalgebra is represented by $J_{jj}$,
while positive and negative Borel subalgebras - by $J_{jk}$ with $j<k$ and
$j>k$ respectively. In eq.(\ref{comamo1mm'})
$Q_+ = i\oint\tilde\psi_1\psi_2,\ \ Q_- = i\oint\tilde\psi_2\psi_1\ $
while in eq.(\ref{comamo"3})
$Q^{(\alpha_1)} = i\oint\tilde\psi_1\psi_2,\ \ Q^{(\alpha_2)} =
i\oint\tilde\psi_1\psi_3\ $
(and $Q^{(\alpha_3)} = i\oint\tilde\psi_2\psi_3,\ \ Q^{(\alpha_4)} =
i\oint\tilde\psi_2\psi_1,\ \ Q^{(\alpha_5)} = i\oint\tilde\psi_3\psi_1,\ \
Q^{(\alpha_6)} = i\oint\tilde\psi_3\psi_2$). $Q^{(\alpha_6)}$ can be
substituted instead of $Q^{(\alpha_2)}$ in (\ref{comamo"3}) without changing
the answer. For generic $r$ the similar choice of "adjacent" (not simple!)
roots (such that their scalar products are $+1$ or $0$)  leads to
selection of the following $r$ screening operators
$Q^{(1)} = i\oint\tilde\psi_1\psi_2\,\ \
Q^{(2)} = -i\oint\psi_2\tilde\psi_3,\ \
Q^{(3)} = i\oint\tilde\psi_3\psi_4,\ldots$, i.e.
$Q^{(j)} = i\oint\tilde\psi_j\psi_{j+1}$ for odd $j$ and
$Q^{(j)} = -i\oint\psi_j\tilde\psi_{j+1}$ for even $j$.


\subsection{1-matrix model in eigenvalue representation \label{ev1m}}

In the last section we found solution of Virasoro constraints
(\ref{virdid}) in the form of the multiple integral (\ref{comamo1mm}).
Now we shall see that this expression is in fact equal to original
matrix integral (\ref{1mamo}) and arises after ``auxiliary''
angular variables are explicitly integrated out. These angular
variables are in fact the ones associated with physical vector
bosons and the possibility to solve this sector of the model
explicitly is peculiar feature of the simplest class of matrix
models, naturaly named {\it eigenvalue models}. The theory of
eigenvalue models is in a sense equivalent to the theory of conventional
integrable hierarchies and thus is rather straightforward to work
on. Inclusion of non-trivial angular integration in the general scheme
is still a sophisticated task, with no universal solution
found so far. Also unknown is solution of inverse problem:
how can be arbitrary
eigenvalue model - with integration over eigenvalues of some
matrix - ``lifted to'' the full matrix model where integration goes
over entire matrix\footnote{see also ref.\cite{AMMS}}; in other words what
is the way to couple  vector bosons to the ``topological'' eigenvalue
sector so that the two sectors are interacting only in the
``solvable'' fashion and angular integrations can be easily performed.

Let us now turn to particular example of the 1-matrix integral (\ref{1mamo}).
First of all, this model possesses gauge
symmetry, associated with the unitary (angular) rotation of matrices,
$H_{\alpha} \longrightarrow U_{\alpha}^{\dagger}H_{\alpha}U_{\alpha}$.
This illustrates the general phenomenon:
matrix models are usually $gauge$ theories. In the case of eigenvalue
models this symmetry is realized without "gauge fields" $V_{\alpha\beta}$,
which would depend on pairs of indices $\alpha$, $\beta$ and transform like
$V_{\alpha\beta} \longrightarrow
U_{\alpha}^{\dagger}V_{\alpha\beta}U_{\beta}$.
In other words, eigenvalue models are gauge theories without gauge fields,
i.e. are pure topological.
The case of the 1-matrix model (\ref{1mamo})
\be
Z_N\{t\} \equiv c_N\int_{N\times N} dH e^{\sum_{k=0}^{\infty} t_k {\rm Tr}
H^k},
\label{1mamo"}
\ee
is especially simple, because
separation of eigenvalue and angular variables does not involve any
information about unitary-matrix integrals. Take
\be
H = U^{\dagger}DU,
\label{diag}
\ee
where $U$ is a unitary matrix and
diagonal matrix $D = {\rm diag}(h_1...h_N)$ has eigenvalues of $H$ as its
entries. Then integration measure\footnote{
In order to derive eq.(\ref{Dyson}) one can consider the norm of
infinitesimal variation
\be
\mid\mid \delta H\mid\mid^2  \equiv \sum_{i,j=1}^N \mid \delta H_{ij}\mid^2 =
 \sum_{i,j=1}^N \delta H_{ij}\delta H_{ji} = {\rm Tr} (\delta H)^2 = \nn \\
 = {\rm Tr}\left(-U^{\dagger}\delta UU^{\dagger}DU + U^{\dagger}D\delta U
+ U^{\dagger}\delta DU \right)^2 = \nn \\
 = {\rm Tr} (\delta D)^2 + 2i{\rm Tr}{\delta u}[\delta D,D] +
2{\rm Tr}\left(-{\delta u} D {\delta u} D + (\delta u)^2D^2 \right), \nn
\ee
where $\delta u \equiv \frac{1}{i}{\delta U}U^{\dagger} = \delta u^{\dagger}$
and $\delta D = {\rm diag}(\delta h_1\ldots \delta h_N)$.
The second term at the r.h.s. vanishes because both $D$ and $\delta D$ are
diagonal and commute. Therefore
\be
\mid\mid \delta H\mid\mid^2 = \sum_{i=1}^N (\delta h_i)^2 +
\sum_{i,j=1}^N (\delta u)_{ij}(\delta u)_{ji}(h_i-h_j)^2. \nn
\ee
Now it remains to recall the basic relation between the infinitesimal norm and
the measure:
${\rm if}\ \ \mid\mid \delta l \mid\mid^2 = G_{ab}\delta l^a\delta l^b \ \
{\rm then}\ \ [dl] = \sqrt{{\rm det}_{ab}G_{ab}} \prod_a dl^a, $
to obtain eq.(\ref{Dyson}) with Haar measure
$[dU] = \prod_{ij}^N du_{ij}$ being associated with the infinitesimal norm
\be
\mid\mid \delta u\mid\mid^2 = {\rm Tr}(\delta u)^2 = \sum_{i,j=1}^N \delta
u_{ij}\delta u_{ji} = \sum_{i,j=1}^N \mid \delta u_{ij}\mid^2 \nn
\ee
and $[dU_{Cartan}] \equiv \prod_{i=1}^N du_{ii}$.
}
\be
dH = \prod_{i,j=1}^N dH_{ij} = \frac{[dU]}{[dU_{Cartan}]} \prod_{i-1}^N dh_i
\Delta^2(h),
\label{Dyson}
\ee
where "Van-der-Monde determinant" $\Delta (h) \equiv det_{(ij)} h_i^{j-1} =
\prod_{i>j}^N (h_i - h_j)$
and $[dU]$ is Haar measure of integration over unitary matrices.

It remains  to note that the "action"
${\rm Tr}\ U(H) \equiv \sum_{k=0}^{\infty} t_k {\rm Tr}H^k$
with $H$ substituted in the form (\ref{diag}) is independent of $U$:
\be
{\rm Tr}\ U(H) = \sum_{i=1}^N U(h_i). \nn
\ee
Thus
\be
Z_N\{t\} = \frac{1}{N!} \prod_{i=1}^N \int dh_i e^{U(h_i)} \prod_{i>j}^N
(h_i-h_j)^2 = \nn \\
= \frac{1}{N!} \prod_{i=1}^N \int dh_i e^{U(h_i)} \Delta^2(h),
\label{1mamoev}
\ee
provided $c_N$ in (\ref{1mamo}) and (\ref{1mamo"}) is chosen to be
\be
c_N^{-1} = N!\frac{{\rm Vol}_{U(N)}}{({\rm Vol}_{U(1)})^N},
\label{cN}
\ee
where the volume of unitary group in Haar measure is equal to
\be
{\rm Vol}_{U(N)} = \frac{(2\pi)^{N(N+1)/2}}{\prod_{k=1}^N k!}.
\label{volun}
\ee

\subsection{Kontsevich-like representation of 1-matrix model}

Matrix-integral representation (\ref{1mamo}) is, however, not the
only possible one for the given eigenvalue model (\ref{comamo1mm}).
Expression (\ref{1mamo}) involves the ``most general action'',
consistent with the symmetry $H \longrightarrow UHU^\dagger$.
As was already mentioned in the introduction, alternative representation
of the partition function should instead involve the general
coupling to background (source) field. In the theory of matrix
models such representations are known under the name of Kontsevich
models. They will be the subject of detailed discussion in the next
sections of this paper. What we need now is one simple identity,
relating original (\ref{1mamo}) and Kontsevich-like representations
of the 1-matrix theory:

\be
\frac{Z_N\{t_0 = 0; t_k = -\frac{1}{k}{\rm tr}\Lambda ^{-k} +
\frac{1}{2}\delta_{k,2}\}}
{Z_N\{t_k = \frac{1}{2}\delta_{k,2}\}} =
\frac{\int_{N\times N} dH e^{\sum_{k=0}^{\infty} t_k {\rm Tr}H^k} }
{\int_{N\times N} dH e^{\frac{1}{2}H^2}} = \nn \\
= \frac{e^{-{\rm tr}\frac{\Lambda ^2}{2}}}
{(2\pi)^{\frac{n^2}{2}}({\rm det}\Lambda )^N}
\int_{n\times n}  dX ({\rm det}X)^N e^{-{\rm tr}\frac{X^2}{2} +
{\rm tr}\Lambda X} = {\cal Z}_{\frac{X^2}{2}}\{N,t\},
\label{1mamoidgako}
\ee
where $Z_N\{t_k = \frac{1}{2}\delta_{k,2}\} = (-2\pi)^{\frac{N^2}{2}}c_N$.
This relation follows from another identity:

\be
\frac{\int_{N\times N}dH e^{\frac{1}{2}{\rm Tr}H^2} Det(\Lambda \otimes I-
I\otimes H)}{ \int_{N\times N}dH e^{\frac{1}{2}{\rm Tr}H^2}} =   \nn \\
= \frac{\int_{n\times n}dX e^{-\frac{1}{2}{\rm tr}X^2} {\rm
det}^N(X+\Lambda )}{\int_{n\times n}dX e^{-\frac{1}{2}{\rm tr}X^2}},
\ee
which is valid for any $\Lambda$ and can be proved by different
methods:  see \cite{UFNmm} and references therein.  Note that
integrals on the right and left hand sides are of differents sizes:
$N\times N$ at the l.h.s. and $n\times n$ at the r.h.s.  While
$N$-dependence is explicit at both sides of the equation, the
$n$-dependence at the l.h.s. enters only implicitly: through the
allowed domain of variation of variables $t_k = -\frac{1}{k}{\rm
tr}\Lambda ^{-k} + \frac{1}{2}\delta_{k,2}$.  This is the usual
feature of Kontsevich integrals: explicit $n$-depenence disappears
once the integral is expressed through the $t$-like variables.

Eq.(\ref{1mamoidgako}) can be used to perform analytical continuation
in $N$ and define what is $Z_N$ for $N$, which are not positive
integers. Since $c_N = 0$ for all {\it negative} integers (see
\cite{UFNmm}), the same is true for $Z_N$. This property
($\tau_N = 0$ for all negative integers $N$) is in fact
characteristic for $\tau$-functions of {\it forced} hierarchies of
which the partition function (\ref{1mamoev}) is an example.

\section{Generalized Kontsevich Model (GKM)}
\setcounter{equation}0

Let us now proceed to investigation of Kontsevich models of a rather
general type. Further generalizations, leading directly to theories
with physical vector bosons (see for example \cite{GKM2}),
are beyond the scope of the present notes. The basic mathematical
fact, responsible for solvability (integrability) of Kontsevich models
is Duistermaat-Heckmann theorem, which allows to evaluate explicitly
the celebrated non-linear Harish-Chandra-Itzykson-Zuber integral
over unitary matrices and thus transform matrix integral into
an eigenvalue model.

\subsection{Kontsevich integral. The first step}

Kontsevich integral is defined as
\be
{\cal F}_{V,n}\{ L\} =
\int_{n\times n} dX e^{-{\rm tr} V(X) + N{\rm tr}\log X + {\rm tr}LX}.
\label{KI'}
\ee
In fact it depends only on {\it eigenvalues} of the matrix $L$. Indeed,
substitute $X = U_X^{\dagger}D_XU_X;  \ \  L = U_{ L}^{\dagger}
D_{ L}U_{ L}$ in (\ref{KI'}) and denote $U \equiv U_XU_{ L}^{\dagger}$.
Then
\be
 {\cal F}_{V,n}\{ L\} = \nn \\
 = \prod_{a=1}^n \int dx_a x_a^N e^{-V(x_a)}
\Delta^2(x)
\int_{n\times n} \frac{[dU]}{[dU_{Cartan}]}
\exp\left({\sum_{a,b =1}^n x_a l_b\mid
U_{a\delta}\mid^2} \right).
\label{KI'2}
\ee
In order to proceed further one needs to evaluate the integral over unitary
matrices, which appeared at the r.h.s.

This integral can actually be represented in two different forms:
\be
I_n\{X, L\} \equiv  \int_{n\times n} \frac{[dU]}{[dU_{Cartan}]}
e^{{\rm tr} XU L U^{\dagger}} =
\label{IZa} \\
=  \int_{n\times n} \frac{[dU]}{[dU_{Cartan}]}
e^{\sum_{a,b =1}^n x_a l_b\mid U_{ab}\mid^2}
\label{IZb}
\ee
(the U's in the two integrals are related by  transformation
$U \longrightarrow U_XUU_{ L}^{\dagger}$ and Haar measure is both left
and right invariant).
Formula (\ref{IZa}) implies that $I_n\{X, L\}$ satisfies a set of simple
equations \cite{Migeq}:
\be
\left( {\rm tr} \left(\frac{\partial}{\partial X_{tr}}\right)^k -
       {\rm tr}  L^k \right) I_n\{X, L\} = 0, \ \ k\geq 0, \nn\\
\left( {\rm tr} \left(\frac{\partial}{\partial  L_{tr}}\right)^k -
       {\rm tr} X^k \right) I_n\{X, L\} = 0, \ \ k\geq 0,
\label{Migeq}
\ee
which by themselves are not very restrictive. However, another formula,
(\ref{IZb}), implies that $I_n\{X, L\}$ in fact depends only on the
eigenvalues of $X$ and $ L$, and for $such$ $I_n\{X, L\} =
\hat I\{x_a, l_b\}$ eqs.(\ref{Migeq}) become very restrictive
\footnote{When acting on $\hat I$, which depends only on eigenvalues,  matrix
derivatives turn into:
\be
{\rm tr} \frac{\partial}{\partial X_{tr}} \hat I  =
   \sum_a \frac{\partial}{\partial x_a} \hat I; \nn \\
{\rm tr} \frac{\partial^2}{\partial X_{tr}^2} \hat I  =
   \sum_a \frac{\partial^2}{\partial x_a^2} \hat I  +
\sum_{a\neq b} \frac{1}{x_a-x_b}
\left(\frac{\partial}{\partial x_a} -
\frac{\partial}{\partial x_b}\right) \hat I;
\nn
\ee
etc.
} and allow to determine $\hat I\{x_a, l_b\}$ unambigously as a
formal power seria in positive powers of $x_a$ and $ l_b$.
The final answer is
\be
I_n\{X, L\} = \frac{(2\pi)^{\frac{n(n-1)}{2}}}{n!}
\frac{{\rm det}_{ab}
   e^{x_a l_b}}{\Delta(x)\Delta( l)}.
\label{IZ}
\ee
Normalization constant can be defined by taking $ L = 0$, when
\be
I_n\{X,L=0\} =
\frac{{\rm Vol}_{U(n)}}{({\rm Vol}_{U(1)})^n} =
\frac{(2\pi)^{\frac{n(n-1)}{2}}}{\prod_{k=1}^n k!},
\nn
\ee
and using the fact that
\be
\left.\frac{{\rm det}_{ab} f_a(l_b)}{\Delta(l)}\right|
_{\{l_b = 0\}} =
\left(\prod_{k=0}^{n-1}\frac{1}{k!} \right) {\rm det}_{ab}
\partial^{b-1}f_a(0).
\nn
\ee

\subsection{Itzykson-Zuber integral and Duistermaat-Heckmann theorem}

Eq.(\ref{IZ}) is usually refered to as the Itzykson-Zuber formula
\cite{IZ}. In mathematical literature it was earlier derived by
Kharish-Chandra \cite{KhCh}, and in fact the integral (\ref{IZa})
is the basic example of the coadjoint orbit integrals \cite{STS}-\cite{AS},
which can be exactly evaluated with the help of the Duistermaat-Heckmann
theorem  \cite{DH}-\cite{NT}.
We now iterrupt our discussion of Kontsevich model for a brief illustration
of this important phenomenon. Duistermaat-Heckmann theorem claims that under
certain restrictive
conditions (that dynamical flow is consistent with the action of
a compact group) some integrals can be expressed in a simple way through
extrema of the integrand. This almost sounds like a statement that
quasiclassical approximation can be exact, with two correction
that {\it all} the extrema - not only the deepest minimum of the action -
should be taken into account, and that the quantum measure should be
adjusted appropriately. When applicable, the theorem states that
\be
\int [d\phi] e^{-S(\phi)} \sim
\sum_{\phi:\ \ \partial S/\partial\phi = 0}
\left(\frac{\partial^2S}{\partial\phi^2}\right)^{-1/2}e^{-S(\phi)}
\label{DH}
\ee
The simplest example is given by the integral
$\int_0^\pi [\sin\theta d\theta] e^{\mu\cos\theta} =
\frac{e^{\mu} - e^{-\mu}}{\mu}$.
We shall not dwell upon the reasons why DH theorem is true in the case
of the Itzykson-Zuber integral (the basic requirement: existence of
the compact group action - that of unitary group - is obviously
fulfilled in this case).
Instead we just evaluate the r.h.s. of (\ref{DH}) provided
the l.h.s. is given by $\int [dU] \exp\left({\rm tr}XULU^\dagger\right)$.
Then equation of motion for $U$ looks like
\be
\left[X,ULU^{\dagger}\right] = 0
\label{eqmU}
\ee
We assume that $X$ and $L$ are already diagonal matrices.
Then (\ref{eqmU}) has an obvious solution $U = I$ (identity matrix),
but it is not unique. For generic diagonal $X$, $L$ the most general
solution is givem by arbitrary permutation matrix $P$: $U = P$.
The ``classical action'' on such solution is equal to
${\rm tr}XULU^\dagger = \sum_a x_al_{P(a)}$,
while the preexponential
factors provide Van-der-Monde determinants $\Delta(x)\Delta(l)$
in denominator and the sign factor $(-)^P$. Since
$$
\sum_P (-)^P \exp\left(\sum_a x_a l_{P(a)}\right) =
\det_{ab} e^{x_a l_b}
$$
we immediately obtain the IZ formula (\ref{IZ}).  Unfortunately the
Duistermaat-Heckmann theory is not yet well developed and even if
the vacuum average is exactly calculable, it does not
provide immediate prescription for evaluation of correlators
(or, what is essentially the same, corrections to DH formula -
when it is not exactly true - are not yet described in a universal
way).   The very important {\it general}
technique of $exact$ evaluation of $non-Gaussian$ unitary-matrix
integrals is now doing its first steps (see \cite{KMSW1}-\cite{Mat}).

\subsection{Kontsevich integral. The second step}

Now we turn back to the eigenvalue representation of Kontsevich integral
(\ref{KI'}). Substitution of (\ref{IZ}) into (\ref{KI'2}) gives:
\be
{\cal F}_{V,n}\{ L\} = \frac{(2\pi)^{\frac{n(n-1)}{2}}}{\Delta( l)}
\prod_{b=1}^n \int dx_b e^{-V(x_b)} \Delta(x)
\frac{1}{n!} {\rm det}_{ab} e^{x_a l_b} = \nn \\
= \frac{(2\pi)^{\frac{n(n-1)}{2}}}{\Delta( l)}
\prod_{b =1}^n \int dx_b x_b^N
e^{-V(x_b) + x_b l_b}
\Delta(x),
\label{KIev}
\ee
where we used antisymmetry of $\Delta(x)$ under permutations of
$x_a$'s in order to change $\ \frac{1}{n!} {\rm det}_{ab}
e^{x_a l_b}\ $ for $\ e^{\sum_b x_b l_b}\ $
under the sign of the $x_b$ integration.

We can now use the fact that $\Delta(x) = {\rm
det}_{ab}x_b^{a -1}$ in order to rewrite the r.h.s. of (\ref{KIev}):
\be
{\cal F}_{V,n}\{ L\} =
(2\pi)^{\frac{n(n-1)}{2}}\frac{{\rm det}_{ab}
\hat\varphi_{a+N}(l_b)}{\Delta( l)},
\label{KIev'}
\ee
where
\be
\hat\varphi_a(l) \equiv \int dx x^{a-1}e^{-V(x)+lx},
\ \ \ a\geq 1.
\label{hatvarphi}
\ee
These functions $\hat\varphi(l)$ satisfy a simple recurrent relation:
\be
\hat\varphi_a = \frac{\partial\hat\varphi_{a-1}}{\partial  l} =
\left(\frac{\partial}{\partial  l}\right)^{a-1}\hat\Phi
\label{preKasch}
\ee
with
\be
\hat\Phi( l)\equiv \hat\varphi_1(l) =
\int dx e^{-V(x)+ lx}.
\label{preKasch'}
\ee

This completes the transformation of Kontsevich integral to the form
of eigenvalue model.

\subsection{``Phases'' of Kontsevich integral.
GKM as the ``quantum piece'' of ${\cal F}_V\{L\}$ in Kontsevich phase}

One of the natural things to do in the study of functions,
defined in the integral form, is to investigate various
asimptotics when the arguments tend to various distinguished
limits. In the case of Kontsevich integral the arguments are
just eigenvalues $l_a$ of the matrix $L$, while their distinguished
values are either zero or positions of singularities of
potential $V(x)$. For simplicity we assume that $V(x)$ has only
a pole of the order $p+1$ at infinity, i.e. V(x) is a polinomial
of degree $p+1$. Then it remains to consider separately only two
different asymptotics: small $l_a$ and large $l_a$. Of course,
since there are many different $l_a$ one actualy has a vast variety
of possibilities: some of $l_a$'s are small and the rest are large.
The two extreme cases are when all the $l_a$'s are either small
or large, and they are refered to as the ``character phase'' and
``Kontsevich phase'' respectively. The word ``phase'' is used instead
of the more exact ``asymptotics'' in order to emphasize
the relation of these two cases to the strong and weak coupling
phases in lattice models of Yang-Mills theories.
We refer to \cite{MMS} for some more discussion of these phases and
their properties, here only some basic things will be mentioned.

In the character phase the main observation is that the r.h.s.
of the Itzykson-Zuber formula (\ref{IZ}) is essentially the
character of the group element $g = e^L$ of $SL(n)$:
\be
\chi_R(e^L) = \frac{\det e^{l_ax_b}}{\Delta(e^l)}{\Delta(x)},
\ee
where the set $\{x_a\}$ specifies representation $R$ of the $SL$
group\footnote{
Correction factor $\Delta(l)/\Delta(e^l)$ can be restored if one
considers appropriate generalization of Kontsevich integral (which
describes not just a puncture but a hole on a surface - in terms of
the naive string theory), see, for example, ref.\cite{GKM2}.
Actually one needs to substitute
$\int [dU] \exp\left({\rm tr} XULU^\dagger\right)$
by the loop integral
$$
\int [dU(s)] \exp\left(\int ds\ {\rm tr} X \left(U(s)\partial_s U^\dagger(s)
+ U(s)LU^\dagger(s)\right)\right) = \frac{\chi_R(e^L)}{d_R}
$$
When this integral is evaluated for integer $x_a$'s one essentially
substitutes every
item in the product $\Delta(l) = \prod_{a<b}(l_a - l_b)$ by
a new infinite product over harmonics,
$\ \ l_a - l_b \ \rightarrow \ \prod_k (l_a - l_b + 2\pi ik) \sim
\sinh\left(\frac{l_a-l_b}{2}\right) \sim e^{l_a}-e^{l_b}$
}.
Accordingly the integral (\ref{KI'}) can be represented as a
linear combination of characters with various $R$, with coefficients
depending on the choice of potential $V(x)$ and $N$. This explains
the reason why this limit (when everything is assumed to be
expandable in positive powers of $l_a$ or $e^{l_a}$) is refered to
as the {\it character} phase. One should keep in mind, of course,
that the most natural, from this point of view, is the situation
when integral over $x_a$'s is changed for a discrete sum over
integer $x_a$'s - this is one of the possible directions of the search
for ``quantum deformations'' of GKM.

Let us now turn to the limit of large $l_a$. Then the natural
expansion would be in negative powers of the arguments, but the
integral (\ref{KI'}) does not have such expansion, as it is.
One should first extract a ``quasiclassical factor'' and it will
be the remaining ``quantum part'' that will possess such an
expansion. This quantum piece is the most interesting
one, and it is for it that the name GKM is usualy used.
So, in Kontsevich limit, integral (\ref{KI'}) can be evaluated with
the help of the steepest descent method, with the classical
solution defined from $V'(X_0) = L$. (Note that when doing so we include
the logarithmic piece in the measure, not in the action.)
Let us call solution of this equation $X_0 = \Lambda$. One could
use $\Lambda$ from he very beginning instead of $L$ to parametrize
Kontsevich integral,
writing ${\rm tr}V'(\Lambda)X$ instead of ${\rm tr}LX$ in the
exponent in (\ref{KI'}). This is a natural parameter in Kontsevich
phase, while $L = V'(\Lambda)$ plays this role in the character
phase. The quasiclassical contribution to (\ref{KI'}), i.e.
exponent of the classical action divided by determinant of quadratic
fluctuations, is equal to:
\be
{\cal C}_V\{\Lambda|N\} =
(2\pi)^{n^2/2}\frac{\exp [{\rm tr}(\Lambda V'(\Lambda) - V(\Lambda))]}
{\sqrt{\det V''(\Lambda)}}(\det \Lambda)^N
\label{357}
\ee
and partition function of GKM is by definition
\be
{\cal Z}_{V,n}\{\Lambda|N\} \equiv \frac{{\cal F}_V\{\Lambda |N\}}
{{\cal C}_V\{\Lambda|N\}}
\ee
This function can be expanded in negative integer powers of
$\lambda_a$ (i.e. in {\it fractional} negative powers of original $l_a$),
at least as a formal series. Moreover, there is a symmetry between all
the eigenvalues $\lambda_a$, thus $Z^{GKM}$ is in fact a function
(formal series) of the ``time-variables'' (the name is historic trace
from the theory of integrable hierarchies)
\be
T_k \equiv \frac{1}{k}{\rm tr}\Lambda^{-k}
\label{Tvar}
\ee
Remarkably, if considered as a function of these $T_k$'s
${\cal Z}$ becomes independent of $n$! We refer to \cite{GKM} and
\cite{UFNmm} for more details (including exact definition of $V''(\Lambda)$
in (\ref{357})).

What remains to be considered here is the
eigenvalue representation of ${\cal Z}$.
If (\ref{KIev'}) is divided by the  quasiclassical factor
${\cal C}_V\{\Lambda|N \}$, we get:
\be
{\cal Z}_V\{N,T\} = \frac{1}{({\rm det} \Lambda )^N}\cdot\frac{{\rm
det}_{ab} \varphi_{a +N}(\lambda _b)}{\Delta(\lambda )}.
\label{KIev"N}
\ee
Extraction of the quasiclassical factor converts $\hat\varphi(l)$ into
the properly normalized expansions in negative integer powers of $\lambda$:
\be
\varphi_a(\lambda )  =
\frac{e^{-\lambda V'(\lambda )+V(\lambda )}\sqrt{V''(\lambda )}}{\sqrt{2\pi}}
\hat\varphi_a(V'(\lambda )) =
\lambda ^{a -1}(1 + {\cal O}(\lambda ^{-1})),
\label{varphiko}
\ee
It also changes $\Delta( l) = \Delta(V'(\lambda ))$ in the denominator of
(\ref{KIev'}) for  $\Delta(\lambda )$ in (\ref{KIev"N}).

Note that, as a corolary of normalization condition (\ref{varphiko}),
whenever one puts $\lambda_n = \infty$ the $n\times n$ determinant
in (\ref{KIev"N}) just turns into the $(n-1)\times (n-1)$ determinant
of the same form. One can easily understand that this implies the
$n$-independence of ${\cal Z}$ as a function of $T$-variables
(\ref{Tvar}).

Instead of the simple recurrent relations (\ref{preKasch}) for $\hat\varphi$
the normalized functions $\varphi$ satisfy:
\be
\varphi_a(\lambda ) = {\cal A}\varphi_{a -1}(\lambda ) =
{\cal A}^{a -1}\Phi(\lambda ),
\label{Kasch}
\ee
where $\Phi(\lambda ) = \varphi_1(\lambda )$ and operator
\be
{\cal A} = \frac{1}{V''(\lambda )}\cdot\frac{\partial}{\partial \lambda }
-\frac{1}{2}\frac{V'''(\lambda )}{(V''(\lambda ))^2} + \lambda
\label{Kaschop}
\ee
now depends on the shape of potential $V(x)$.

\subsection{Relation between time- and potential-dependencies
\label{Krich}}

Consider the vector space ${\cal H}$ of all formal Laurent series in
some variable $\lambda$. The set of functions
$$
\left\{\lambda^{-a},\ \lambda^{a-1}\ | \ a=1,2,\ldots\right\}
$$
is one of the many possible basises in this vector space. Let us
consider a ``half-space'' ${\cal H}_+$ consisting of all
the series in non-negative powers of $\lambda$. Then
$\left\{\lambda^{a-1}\ | \ a=1,2,\ldots\right\}$
is a possible basis in ${\cal H}_+$. Every rotation of the
linear subspace ${\cal H}_+$ within entire ${\cal H}$ can be
represented by projection of some basis in rotated subspace onto
original one. In other words, any semi-infinite set of functions
$\{ \phi_a(\lambda)\}$, such that
\be
\phi_a(\lambda) =   \lambda ^{a -1}(1 + {\cal O}(\lambda ^{-1}))
= \lambda^{a-1} (1 + \sum_{b>0} S_{ab}\lambda^{-b})
\label{phiko}
\ee
can be considered as describing some particular rotation of
${\cal H}_+$ in ${\cal H}$. Of course, the same rotation can be
represented by different matrices $S_{ab}$, and in fact
rotations  are in one-to-one correspondence with the factor
of the set $\{\phi_a\}$ modulo triangular transformations,
which has the natural name of the Universal Grassmannian ${\cal GR}$.
Eq.(\ref{KIev"N}) is obviously invariant under
such triangular transformations, thus ${\cal Z}$, as a function of
its argument $V(x)$ can be considered as a function on ${\cal GR}$
(if the shape of potential is changed, the set $\{\phi_a\}$ is
also changed).

Normalization condition (\ref{phiko}) is, however, invariant not
only under tringular transformations in $\{\phi_a\}$, but also
under the changes $\lambda \longrightarrow \lambda(1 + {\cal O}
(\lambda^{-1}))$. Such transformations change the point of
Grassmannian and they also induce a triangular linear transformation
of time-variables: $T_k \longrightarrow T_k + {\rm lin}(T_{k+1},
T_{k+2},\ldots)$. In other words, ${\cal Z}$ depends on the choice
of variable $\lambda$ on the ``spectral curve'' and on the point of
${\cal GR}$, i.e. is essentially a function on the tensor
product ${\cal GR}\times {\cal GR}$ of two different Grassmannians.
One of them is a space of various models (related to the choice of
potential in GKM), another - specifies the basis in the space of
observables in a given model (related to the choice of time-variables).
As one expects from the physical arguments and as we just saw on
a more formal level these two dependencies are in fact interrelated.
When consideration is restricted to the set of GKM's (from that of all the
models of string theory) a much more definite statement can be made
\cite{KriTimes}. Being a priori a function of two distinct types of
variables: the times
$\left\{T_k = \frac{1}{k}{\rm tr}\Lambda^{-k}\right\}$
and potential $V(x)$, the GKM partition function in fact depends only
on the {\it type} of singularities of $V(x)$ (which is a kind of
``discrete'' information) and on peculiar combination
of these variables. If $V(x)$ is a polinomial of degree $p+1$
(i.e. has only finite order pole at infinity), then
\be
{\cal Z}_V\{T_k\} \sim
\tau_p\left(\frac{1}{k}{\rm tr} W(\Lambda)^{-k/p} +
\frac{p}{k(p-k)}res\ W(x)^{1-k/p}dx\right),
\label{T-Vdep}
\ee
where $W(x) \equiv V'(x)$, and the shape of the function $\tau_p$
depends only on the value of $p$. (In order to substitute ``$\sim$''
by ``$=$'' in (\ref{T-Vdep}) one should slightly redefine the
quasiclassical factor and thus ${\cal Z}$: one should in fact work
with the variable $L^{1/p} = W(\Lambda)^{1/p}$ instead of $\Lambda$.
As often happens, different variables are nice for different purposes.)

If all the arguments $\hat T_k$ with $k>p$ are put equal to zero,
we get a reduced $\tau$-function
$$
\tilde \tau_p(\hat T_1,\ldots,\hat T_p) \equiv
\tau_p(\hat T_k)|_{T_k = 0\ {\rm for}\ k>p}
$$
It appears to be a solution to ``quasiclassical KP-hierarchy'',
which arises from pure algebraic construction and can be also
identified as partition function of topological Landau-Ginsburg
model with the superpotential $W(x)$.

\subsection{Kac-Schwarz problem}

The function $\tau_p(T_k)$ is of course very far from arbitrary.
First of all it possesses peculiar determinant representation
of the type (\ref{KIev"N}), which in turn implies some restrictive
bilinear (Hirota) equations and as result $\tau_p$ appears to be
a KP $\tau$-function. Second, $\tau_p$ is further distinguished
even among $\tau$-functions by peculiar features of the functions
$\varphi_a(\lambda)$.     In particular case of GKM one of the
ways to represent these properties is to use the recursive relations
(\ref{Kasch}), $\varphi_a = {\cal A}^{a-1}\Phi$,
suplemented by another obvious property
$W(\lambda)\Phi(\lambda) = \varphi_{p+1}(\lambda)$ (it follows from
invariance of the integral $\Phi(\lambda)$ under the change of
integration $x$-variable). These two relations together give rise
to an equation on $\varphi_1(\lambda) = \Phi(\lambda)$:
$({\cal A}^p - W(\lambda))\Phi(\lambda) = 0$, which is just a
$p$-th order differential equation.

In this way one specifies non-perturbative partition function
of some string model (in the case of GKM this is in fact a
$(p,1)$-minimal model coupled to $2d$-gravity) in terms of
invariant points of certain operators acting on the Universal
Module space ${\cal GR}$. This reformulation, though far more
abstract than the original one, can be very useful for non-trivial
generalizations - and be a natural step in the search for generic
configuration space of the string theory. It was first introduced
by V.Kac and A.Schwarz \cite{Kasch}, and is not yet studied as deep
as it deserves, even if one deals only with the space of
KP $\tau$-functions, determinant formulas and ordinary Universal
Grassmannian. In this (actually somewhat narrow) context, the
general problem is to describe common invariant points in ${\cal GR}$
of two operators, acting on formal Laurent series in $\lambda$:
$\forall a$
\be
{\cal A}\phi_a \in {\rm Span}\{\phi_b\},
\label{Aop}
\ee
\be
{\cal K}\phi_a \in {\rm Span}\{\phi_b\}
\label{Bop}
\ee
In the case of GKM ${\cal A}$ and ${\cal K}$ are differential
operators of the 1-st and 0-th order respectively. Moreover,
${\cal A}$ is a ``gap-one'' operator (the gap is equal to $g$ if
${\cal A}\phi_a = \sum_{b =1}^{a+g} {\cal A}_{ab}\phi_b$).
See \cite{AMMS} for some more comments on the gap-one case.
When both gaps are different from unity, the system (\ref{Aop})-(\ref{Bop})
usualy describes a multi-parametric set of invariant points,
the simplest example being associated with
$(p,q)$-minimal models (where $p$ and $q$ are in fact the values
of gaps for ${\cal K}$ and ${\cal A}$ respectively).
In this situation Kontsevich integral desribes only
duality transformation of $(p,q)$-model into the $(q,p)$-one
\cite{KhaMar} (note that these models do not coincide after
they are coupled to $2d$ gravity).

\subsection{Ward identities for GKM}

Advantage of the Kac-Schwarz reformulation of GKM is that it is
very easy to deform, since there are no real restriction imposed
on the choice of operators ${\cal A}$ and ${\cal K}$. However,
instead, this formulation does not introduce any reach
structure and does not immediately provide any valuable information
about the form and properties of solutions to (\ref{Aop})-(\ref{Bop}), i.e.
does not explicitly reveal any nice properties which could be common for
all the string models. Thus it could serve as a starting, but not the
final point of analysis of non-perturbative partition functions.
Alternative approach, making use of
the Ward identites, provides a better description of
GKM, but can appear too restrictive to allow for any interesting
deformations. From the point of view of Grassmannian, the
Ward identities specify some subset of points in ${\cal GR}$,
which are invariant under certain subalgebras of ${\cal U}GL(\infty)$
(the symmetry group of entire Grassmannian). The corresponding
``homogeneus spaces'' are normally not discrete and contain
a vast variety of points. By themselves the Ward identites are not
enough to specify particular points in ${\cal GR}$ uniquely. They
should still be supplemented by some extra conditions, like
``reduction constraints'' (in fact one should keep one of the
Kac-Schwarz constraints, (\ref{Bop}), and only another one can be usualy
substituted by the symmetry-like relation, coming from the Ward
identities).

\subsubsection{Gross-Newmann equation}

We refer to \cite{UFNmm} for a very detailed review of the Ward
identities in GKM. They are all corollaries of the simple Gross-Newmann
(GN) equation \cite{GN},
imposed on Kontsevich integral (\ref{KI'}) as result of
its invariance under arbitrary change of the integration variable
$X$:
$$
\int dX e^{-{\rm tr}V(x) + N{\rm tr}\log X + {\rm tr}LX}
\left(-V'(X) + NX^{-1} + L \right) = 0.
$$
or
\be
\left( V'\left(\partial/\partial L_{tr}\right) - L -
N\left(\partial/\partial L_{tr}\right)^{-1}\right){\cal F}_V = 0.
\label{GNe}\label{GNeq}
\ee
Being just equation of motion for ${\cal F}_V\{L\}$ eq.(\ref{GNe})
provides complete information about this function.
However, this statement needs to be formulated more carefully.
One of the reasons is that (\ref{GNe}) does
not account explicitly for a very important property of ${\cal
F}_V\{ L\}$: that it actualy depends only on eigenvalues of
$ L$. This information should still be taken into account explicitly.
If this is kept in mind, it becomes
a tedious but straightforward work to substitute
${\cal F}_V = {\cal C}_V{\cal Z}_V$ and express $L$-derivatives through
$T_k$-derivatives in order to derive Virasoro and $W$-constraints in
conventional form \cite{UFNmm}. We shall briefly sketch some pieces
of this derivation and related problems in the remaining part of
this section.

\subsubsection{$\tilde W$-operators in Kontsevich models}

First of all, the Gross-Newmann equation  (\ref{GNeq})
for Kontsevich models can be easily expressed in terms of the
so called $\tilde W$-operators. Namely, we shall prove the
following identity
\cite{tildeW}:
\be
\left(\frac{\partial}{\partial
\Lambda_{tr}}\right)^{m+1} {\cal Z}\{ T_k\} =
(\pm)^{m+1}
\sum_{l\geq 0} \Lambda^{-l-1} \tilde W_{l-m}^{(m+1)}( T)
{\cal Z}\{ T_k\},
\label{tiwko}
\ee
valid for $any$ function ${\cal Z}$ which depends on
$ T_k = \mp\frac{1}{k}{\rm tr}\Lambda^{-k},\ \ k\geq 1\ $ and
$ T_0 = \pm{\rm tr}\log \Lambda$.

The $\tilde W$-operators  are defined \cite{tildeW} by the
following construction. Consider the action of
${\rm Tr} \frac{\partial^m}{\partial  L_{tr}^m}  L^n$
on  $e^{{\rm Tr}U( L)} = e^{\sum_k t_k {\rm Tr} L^k}$. It
gives some linear combination of terms like
\be
{\rm tr} L^{a_1} ...{\rm tr} L^{a_l} e^{{\rm tr} U( L)} =
\frac{\partial^l}{\partial t_{a_1}...\partial t_{a_l}} e^{-{\rm tr} U( L)}
\nn
\ee
i.e. we obtain a combination of differential operators with $t$-derivatives,
to be denoted $\tilde W(t)$:
\be
\tilde W_{n-m}^{(m+1)}(t)
e^{{\rm tr} U( L)} \equiv
{\rm Tr} \frac{\partial^m}{\partial  L_{tr}^m}  L^n
e^{{\rm tr} U( L)}, \ \ \ m,n \geq 0.
\label{twop}
\ee
For example,
\be
\new
\begin{array}{c}
\tilde W_n^{(1)} = \frac{\partial}{\partial t_n}, \ \ n\geq 0;  \\
\tilde W_n^{(2)} = \sum_{k=0}^{\infty} kt_k\frac{\partial}{\partial t_{k+n}} +
    \sum_{k=0}^n \frac{\partial^2}{\partial t_k\partial t_{n-k}},
\ \ n\geq -1;   \\
\tilde W_n^{(3)} = \sum_{k,l = 1}^{\infty} kt_klt_l
\frac{\partial}{\partial t_{k+l+n}}
+ \sum_{k=1}^{\infty} kt_k
  \sum_{a+b=k+n}\frac{\partial^2}{\partial t_a\partial t_b}  + \\
+ \sum_{k=1}^{\infty} kt_k
  \sum_{a+b=n+1}\frac{\partial^2}{\partial t_a\partial t_{b+k-1}}
+ \sum_{a+b+c=n} \frac{\partial^3}{\partial t_a\partial t_b\partial t_c}
+ \frac{(n+1)(n+2)}{2}\frac{\partial}{\partial t_n};
 \\
\ldots
\end{array}
\label {twopex}
\ee
Note, that while $\tilde W_n^{(1)}$ and $\tilde W_n^{(2)}$
are just the ordinary $(U(1)$-Kac Moody and Virasoro
operators respectively, the higher $\tilde
W^{(m)}$-operators do $not$ coincide with the generators of the
{\bf W}-algebras: already
\be
\tilde W_n^{(3)} \neq W_n^{(3)} =
\sum_{k,l = 1}^{\infty} kt_klt_l
\frac{\partial}{\partial t_{k+l+n}}
+ 2\sum_{k=1}^{\infty} kt_k
  \sum_{a+b=k+n}\frac{\partial^2}{\partial t_a\partial t_b} + \nn \\
+ \frac{4}{3}
  \sum_{a+b+c=n} \frac{\partial^3}{\partial t_a\partial t_b\partial t_c}.
\nn
\ee
$\tilde W$-operators (in variance with ordinary $W$-operators) satisfy
recurrent relation:
\be
\tilde W_n^{(m+1)} = \sum_{k=1}^{\infty} kt_k\tilde W_{n+k}^{(m)} +
          \sum_{k=0}^{m+n-1} \frac{\partial}{\partial t_k}
                      \cdot \tilde W_{n-k}^{(m)}, \ \ \ n\geq -m.
\ee
Actually not too much is already known about the ${\tilde W}$ operators and
the structure of $\tilde{\bf W}$-algebras (in particular it remains unclear
whether the negative harmonics $\tilde W_n^{(m+1)}$ with $n < -m$ can be
introduced in any reasonable way), see \cite{tildeW} for some preliminary
results.

Now we can come back to the identity (\ref{tiwko}).
Its most straightforward application is to the
Gaussian Kontsevich model with potential  $V(x) = \frac{x^2}{2}$,
see the next subsection.
In other cases calculations with the use of identity
(\ref{tiwko}), accounting for the quasiclassical factor ${\cal C}_V\{ L\}$
and the difference between $ L = V'(\Lambda)$ and
$\Lambda$ become somewhat more involved, though still seem
sufficiently straightforward.  Also for particular potentials $V(X)$
partition function ${\cal Z}_V\{T\}$  is actually independent
of certain (combinations of) time-variables (for example, if
$V(X) = \frac{X^{p+1}}{p+1}$, it is independent of all the
$T_{pk},\ k\in Z_+$),
and this is important for appearence of the constraints in the standard
form, i.e. for certain {\it reduction} of $\tilde W$-constraints to the
ordinary $W$-constraints. This relation between $\tilde W$- and
$W$-operators deserves further investigation.

The proof of eq.(\ref{tiwko}) is provided by the following trick. Let us make
a sort of Fourier transformation:
\be
{\cal Z}\{ T\} =
\int dH\ {\cal G}\{H\} e^{\sum_{k=0}^{\infty} T_k{\rm Tr}H^k},
\label{mafoutr}
\ee
where integral is over $N\times N$ Hermitean matrix $H$.\footnote{
Here it is for the first time that we encounter an important idea: matrix
models - the ordinary 1-matrix model (\ref{1mamo})
in this case - can be considered as defining integral transformations. This
view on matrix models can to large extent define their role in the future
developement of string theory.
}
Then it is clear that once the identity (\ref{tiwko}) is established for
${\cal Z}\{ T\}$ substituted by $ e^{{\rm Tr} U(H)},\ \
 U(H) = \sum_{k=0}^{\infty} T_k{\rm Tr}H^k$, with any matrix $H$,
it is valid for $any$ function ${\cal Z}\{ T\}$. The advantage of such
substitution is that we can now make use of the definition (\ref{twop})
of the $\tilde W$ operators in order to rewrite (\ref{tiwko}) in a very
explicit form:
\be
\left(\frac{\partial}{\partial \Lambda_{tr}}\right)^{m+1}
e^{{\rm Tr} U(H)} =
(\pm)^{m+1}\sum_{l \geq 0}^{\infty} \Lambda^{-l-1}
 \tilde W_{l-m}^{(m+1)}( T) e^{{\rm Tr} U(H)} = \nn \\
= (\pm)^{m+1}
\sum_{l \geq 0}^{\infty} \Lambda^{-l-1} {\rm Tr}
\left(\frac{\partial}{\partial H_{tr}}\right)^m H^l
e^{{\rm Tr} U(H)} = \nn \\
= (\pm)^{m+1}
{\rm Tr}\left(\frac{\partial}{\partial H_{tr}}\right)^m
\frac{1}{\Lambda\otimes I - I\otimes H} e^{{\rm Tr} U(H)}.
\label{tiwkoder1}
\ee
Now expression for $ T$'s in terms of $\Lambda$ should be used. Then
\be
e^{{\rm Tr} U(H)} =  Det^{\pm 1} (\Lambda\otimes I - I\otimes H)
\nn
\ee
and substituting this into (\ref{tiwkoder1}) we see that (\ref{tiwko}) is
equivalent to
\be
\left( \left(\frac{\partial}{\partial \Lambda_{tr}}\right)^{m+1} -
(\pm)^{m+1} I\cdot{\rm Tr} \left(\frac{\partial}{\partial H_{tr}}\right)^m
\cdot \frac{1}{\Lambda\otimes I - I\otimes H}
\right)\cdot \nn \\
\cdot  Det^{\pm 1} (\Lambda\otimes I - I\otimes H) = 0
\label{tiwkoder2}
\nn
\ee
Here ``${\rm Tr}$'' stands for the trace in the $H$-space only, while
$Det = {\rm Det}\otimes {\rm det}$ - for determinant in both $H$ and
$\Lambda$ spaces.
After one $\Lambda$-derivative is taken explicitly, we get:
\be
\left(I\otimes{\rm Tr}\right) \left( \left(\frac{\partial}{\partial
\Lambda_{tr}}\right)^m\otimes I -  I\otimes \left(\pm\frac{\partial}{\partial
H_{tr}}\right)^m
\right)
\cdot \nn \\ \cdot
\frac{Det^{\pm 1} (\Lambda\otimes I -  I\otimes H)}
{\Lambda\otimes I - I\otimes H} = 0.
\label{tiwkoder3}
\ee
This is already a matrix identity,
valid for any $\Lambda$ and $H$ of the sizes
$n\times n$ and $N\times N$ respectively. For example, if $m=0$ ($\tilde
W^{(1)}$-case), it is obviously satisfied. If both $n=N=1$, it is also
trivially true, though for different reasons for different choice of signs:
for the upper signs, the ratio at the l.h.s. is just unity and all derivatives
vansih; for the lower signs we have:
\be
\left(\frac{\partial}{\partial \lambda }\right)^m  -
\left(-\frac{\partial}{\partial h}\right)^m =
\left(\sum_{\stackrel{a+b=m-1}{a,b\geq 0}}
\left(\frac{\partial}{\partial \lambda }\right)^a
\left(-\frac{\partial}{\partial h}\right)^b \right)
\left(\frac{\partial}{\partial \lambda }  + \frac{\partial}{\partial h}\right),
\nn
\ee
and this obviously vanishes since $(\frac{\partial}{\partial \lambda } +
\frac{\partial}{\partial h})f(\lambda -h) \equiv 0$ for any $f(x)$.

If $m>0$ and $\Lambda,\ H$ are indeed {\it matrices},
direct evaluation becomes much more
sophisticated. We present the first two nontrivial examples: $m=1$ and $m=2$.
The following relations will be usefull. Let
$Q \equiv  \frac{1}{\Lambda\otimes I - I\otimes H}$. Then
\be
Det^{\pm 1}Q \frac{\partial}{\partial \Lambda_{tr}}  Det^{\mp 1} Q =
\pm\left[(I\otimes {\rm Tr})Q\right]; \nn \\
Det^{\pm 1}Q \frac{\partial}{\partial H_{tr}}  Det^{\mp 1} Q =
\mp\left[({\rm tr}\otimes I)Q\right]; \nn \\
\left(\frac{\partial}{\partial \Lambda_{tr}}\otimes I\right) Q =
-\left[({\rm tr}\otimes I)Q\right]Q;
\nn \\
\left(I\otimes\frac{\partial}{\partial H_{tr}}\right) Q =
\left[(I\otimes {\rm Tr})Q\right]Q.
\label{tiwkoder5}
\ee

This is already enough for the proof in the case of $m=1$. Indeed:
\be
Det^{\pm 1}Q \left(\frac{\partial}{\partial \Lambda_{tr}}\otimes I \mp
    I\otimes \frac{\partial}{\partial H_{tr}}\right)Q Det^{\mp 1} Q = \nn \\
= \{-\left[({\rm tr}\otimes I)Q\right]Q
  \pm \left[(I\otimes {\rm Tr})Q\right]Q \}\mp \nn \\
\mp\{ \left[(I\otimes {\rm Tr})Q\right]Q \mp
 \left[({\rm tr}\otimes I)Q\right]Q \} = 0. \nn
\ee
The first two terms at the r.h.s. come from $\Lambda$-, while the
last two -- from $H$-derivatives.

In the case of $m=2$ one should take derivatives once again. This is a little
more tricky, and the same compact notation are not sufficient. In addition to
(\ref{tiwkoder5}) we now need:
\be
\left(\frac{\partial}{\partial \Lambda_{tr}}\otimes I\right)
\left[({\rm tr}\otimes I)Q\right]Q =
- \left[({\rm tr}\otimes I)Q\right]^2 Q - {\cal B}.
\label{tiwkoder7}
\ee
Here
\be
\left[({\rm tr}\otimes I)Q\right]^2 =
\left[({\rm tr}\otimes I)\left[({\rm tr}\otimes I)Q\right]Q\right],
\label{tiwkoder8}
\ee
while in order to write ${\cal B}$ explicitly we need to restore matrix
indices (Greek for the $\Lambda$-sector and Latin - for the $H$ one). The
$(\alpha i,\gamma k)$-component of (\ref{tiwkoder7}) looks like:
\be
\left(\frac{\partial}{\partial \Lambda_{\beta\alpha}}\delta^{im}\right)
Q_{\delta\delta}^{mj}Q_{\beta\gamma}^{jk} =
- Q_{\delta\delta}^{ij}Q_{\beta\beta}^{jl}Q_{\alpha\gamma}^{lk} -
  Q_{\delta\beta}^{il}Q_{\alpha\delta}^{lj}Q_{\beta\gamma}^{jk}
\ee
and appearence of the second term at the r.h.s. implies, that
${\cal B}_{\alpha\gamma}^{ik} =
Q_{\delta\beta}^{il}Q_{\alpha\delta}^{lj}Q_{\beta\gamma}^{jk}$.
Further,
\be
 \left(\frac{\partial}{\partial \Lambda_{tr}}\otimes I\right)
\left[(I\otimes {\rm Tr})Q\right]Q =  \nn \\
 -\left[(I\otimes {\rm Tr})\left[({\rm tr}\otimes I)Q\right]Q\right]Q
-\left[(I\otimes {\rm Tr})\left[(I\otimes {\rm Tr})Q\right]Q\right]Q; \nn \\
 \left( I\otimes \frac{\partial}{\partial H_{tr}}\right)
\left[({\rm tr}\otimes I)Q\right]Q = \nn \\
 + \left[({\rm tr}\otimes I)\left[(I\otimes {\rm Tr})Q\right]Q\right]Q +
  \left[(I\otimes {\rm Tr})\left[({\rm tr}\otimes I)Q\right]Q\right]Q; \nn \\
 \left( I\otimes \frac{\partial}{\partial H_{tr}}\right)
\left[(I\otimes {\rm Tr})Q\right]Q =
+ \left[(I\otimes {\rm Tr})\left[(I\otimes {\rm Tr})Q\right]Q\right]Q
+{\cal B}.
\ee
It is important that ${\cal B}$ that appears in the last relation
in the form of
${\cal B}_{\alpha\gamma}^{ik} =
Q_{\alpha\delta}^{lj}Q_{\delta\beta}^{il}Q_{\beta\gamma}^{jk}$
is exactly the same ${\cal B}$ as in eq.(\ref{tiwkoder7}).

Now we can prove (\ref{tiwkoder3}) for $m=2$:
\be
 Det^{\pm 1}Q \left(\left(\frac{\partial}{\partial
\Lambda_{tr}}\right)^2\otimes I -
I\otimes \left(\frac{\partial}{\partial H_{tr}}\right)^2\right)
Q Det^{\mp 1} Q =  \nn \\
 = \left\{
\pm\left[(I\otimes{\rm Tr})Q\right]\left(
- \left[({\rm tr}\otimes I)Q\right]Q \pm \left[(I\otimes{\rm Tr})Q\right]Q
\right) - \right. \nn \\  \left. - \left(
- \left[({\rm tr}\otimes I)\left[({\rm tr}\otimes I)Q\right]Q\right]Q - {\cal
B}
\right) \pm \right. \nn \\  \left. \pm \left(
- \left[(I\otimes{\rm Tr})\left[({\rm tr}\otimes I)Q\right]Q\right]Q -
  \left[({\rm tr}\otimes I)\left[(I\otimes{\rm Tr})Q\right]Q\right]Q
\right)\right\} - \nn \\
 - \left\{
\mp\left[({\rm tr}\otimes I)Q\right]\left(
\left[(I\otimes{\rm Tr})Q\right]Q \mp \left[({\rm tr}\otimes I)Q\right]Q
\right) + \right. \nn \\  \left. + \left(
\left[(I\otimes{\rm Tr})\left[(I\otimes{\rm Tr})Q\right]Q\right]Q + {\cal B}
\right) \mp \right. \nn \\  \left. \mp \left(
\left[({\rm tr}\otimes I)\left[(I\otimes{\rm Tr})Q\right]Q\right]Q +
\left[(I\otimes{\rm Tr})\left[({\rm tr}\otimes I)Q\right]Q\right]Q
\right)\right\}
\ee
where the terms 1,2,3,4,5,6 in the first braces cancel the terms 1,3,2,4,6,5
in the second braces and identity (\ref{tiwkoder8}) and its counterpart with
$({\rm tr}\otimes I) \rightarrow  (I\otimes {\rm Tr})$ is used.

Explicit proof of eq.(\ref{tiwkoder3}) for generic $m$ is unknown.

\subsubsection{Discrete Virasoro constraints for the
Gaussian Kontsevich model}

As a simplest illustration
we derive now the constraints for the Gaussian Kontsevich
model \cite{ChMa} with potential $V(X) = \frac{1}{2}X^2$:
\be
{\cal Z}_{\frac{X^2}{2}}\{N,T\} =
\frac{e^{-{\rm tr}\frac{ L^2}{2}}}{({\rm det} L)^N}
\int dX ({\rm det}X)^N e^{-{\rm tr}\frac{X^2}{2} +  L X}.
\label{gako}
\ee
In this case $ L = V'(\Lambda) = \Lambda$, and the time-variables are just
\be
T_k = \frac{1}{k}{\rm tr} \Lambda^{-k} = \frac{1}{k}{\rm tr} L^{-k}.
\label{gakoT}
\ee
The model is non-trivial becuase of the presence of "zero-time" variable
$N$ \cite{Toda}. The Gross-Newmann equation (\ref{GNeq}) looks like
\be
\new
\begin{array}{c}
\frac{e^{-{\rm tr}\frac{ L^2}{2}}}{({\rm det} L)^N}
\left(\frac{\partial}{\partial L_{tr}}\right)^{n+1}\cdot
\left( \frac{\partial}{\partial L_{tr}} -
N\left(\frac{\partial}{\partial L_{tr}}\right)^{-1} -  L
\right) \cdot \\
\cdot
({\rm det} L)^Ne^{+{\rm tr}\frac{ L^2}{2}}
{\cal Z}_{\frac{X^2}{2}}\{N,T\} = 0.
\end{array}
\label{gako2}
\ee
In order to get rid of the integral operator
$(\frac{\partial}{\partial L})^{-1}$
one should take here $n \geq 0$ rather than $n \geq -1$. In fact all the
equations with $n > 0$ follow from the one with $n=0$, and we restrict
our consideration to the last one. For $n=0$ we obtain from
(\ref{gako2}):
\be
\left(\left( \frac{\partial}{\partial L_{tr}} + \frac{N}{ L} +
{ L}\right)^2 - 2N -  L
\left( \frac{\partial}{\partial L_{tr}} + \frac{N}{ L} +
{ L}\right) \right) {\cal Z} = 0 \nn
\ee
or
\be
\left(\left( \frac{\partial}{\partial L_{tr}}\right)^2 +
     \left( L +
\frac{2N}{ L}\right)\frac{\partial}{\partial L_{tr}}
+ \frac{N^2}{ L^2} - \frac{N}{ L}{\rm tr}\frac{1}{ L}
\right) {\cal Z} = 0,
\label{gako3}
\ee
One can now use eq.(\ref{tiwko}) to obtain:
\be
\sum_{m=-1}^{\infty}  \frac{1}{ L^{m+2}}
\left( \sum_{k=1+\delta_{m,-1}}^{\infty}
\left({\rm tr}\frac{1}{ L^{k}}\right)
\frac{\partial}{\partial T_{k+m}} +
  \sum_{k=1}^{m-1} \frac{\partial^2}{\partial T_k\partial T_{m+k}} -
\right. \nn \\  \left.
-  \frac{\partial}{\partial T_{m+2}} - 2N\frac{\partial}{\partial T_{m}} +
N^2\delta_{m,0} - N \left({\rm tr}\frac{1}{ L}\right) \delta_{m,-1}
\right) {\cal Z} = \nn
\\
 = \sum_{m=-1}^{\infty} \frac{1}{ L^{m+2}}
e^{NT_0}  L_m(T+r) e^{-NT_0} {\cal Z}
= 0.
\label{gako4}
\ee
Here $L_m(t) = \tilde W^{(2)}_m(t)$ are just the generators
(\ref{virdop}) of ``discrete'' Virasoro algebra (\ref{virdid}):
\be
e^{Nt_0}  L_m(t) e^{-Nt_0} =
e^{Nt_0}  \left( \sum_{k=1}^{\infty} kt_k\frac{\partial}{\partial t_{k+m}} +
\sum_{k=0}^m \frac{\partial^2}{\partial t_k\partial t_{m-k}}
\right) e^{-Nt_0}.
\ee
and at the r.h.s. of (\ref{gako4}) $r_k =
-\frac{1}{2}\delta_{k,2}$.\footnote{
This small correction is manifestation of a very general
phenomenon which was already mentioned in s.\ref{Krich} above:
from the point of view of symmetries (Ward identities)
it is more natural to consider $Z_V$ not as a function of $T$-variables, but
of some more complicated combination $\hat T_k + r_k$, depending on the shape
of potential $V$. If $V$ is a polinomial of degree $p+1$, $\hat T_k =
\frac{1}{k}{\rm tr} (V'(\lambda))^{-k/p},$ while $r_k = \frac{p}{k(p-k)}{\rm
Res}\left(V'(\mu)\right)^{1-\frac{k}{p}}d\mu$. For monomial potentials these
expressions become very simple: $\hat T_k = T_k$ and $r_k =
-\frac{p}{p+1}\delta_{k,p+1}$. See \cite{KriTimes} for
more details. In most places in these notes we prefer to use invariant
potential-independent times $T_k$, instead of $\hat T_k$, but then Ward
identites acquire some extra terms with $r_k$.
}

Thus we found that the Ward identities for the Gaussian Kontsevich model
(\ref{gako}) coincide with those for the ordinary 1-matrix model
(\ref{1mamo}), moreover the size of the
matrix $N$ in the latter model is associated with the "zero-time" in the
former one. This result \cite{ChMa} of course implies, that the two models
are identical:
\be
e^{-NT_0}{\cal Z}_{\frac{X^2}{2}}\{N,T_1,T_2,\ldots\} \sim
Z_N\{T_0,T_1,T_2,\ldots\}.
\ee
See \cite{Toda} and \cite{UFNmm} for explicit proof of this identity.

\subsubsection{Continuous Virasoro constraints for the $V = \frac{X^3}{3}$
Kontsevich model}

This example is a little more complicated than that in the previous
subsection, and we do not present calculations in full details (see
\cite{MMM} and \cite{GKM}). Our goal is to demonstrate that the
constraints which arise in this model,
though still form  (Borel subalgebra of) some Virasoro
algebra, are $different$  from (\ref{virdid}). From the point of view of the
CFT-formulation the relevant model is that of the $twisted$ (in this
particular case - antiperiodic) free fields. These so called "continuous
Virasoro constraints" give the simplest illustration of the difference
between discrete and continuous matrix models: this is essentially the
difference between "homogeneous" (Kac-Frenkel) and "principal"
(soliton vertex operator) representation of the level $k=1$ Kac-Moody
algebra. From the point of view of integrable hierarchies this is the
difference between Toda-chain-like and KP-like hierarchies.

Another (historicaly first) aspect of the same relation also deserves
mentioning, since it also illustrates the interrelation between
different models.  The discrete 1-matrix model arises naturally in
description of quantum $2d$ gravity as sum over 2-geometries in the
formalism of random equilateral triangulations. The model, however,
decribes only lattice approximation to $2d$ gravity and
(double-scaling) coninuum limit should be taken in order to obtain
the real (continuous) theory of $2d$ gravity. This limit was
originally formulated \cite{FKN} in terms of the contraint algebra
(equations of motion or "loop" or "Schwinger-Dyson" equations -
terminology is taste-dependent), leaving open the problem of what is
the form of partition function ${\cal Z}^{cont}\{T\}$ of continuous
theory. Since the relevant algebra appeared to be just the set of
Ward identities for  Kontsevich model (with $V(X) = \frac{X^3}{3}$),
this proves that the latter one is exactly the continuous theory of
pure $2d$ gravity.  At the same time, Kontsevich model itself can be
naturally introduced as a theory of $topological$ gravity (in fact
this is how the model was originally discovered in \cite{Ko}). From
this point of view the constraint algebra, to be discussed below in
this subsection, plays the central role in the proof of equivalence
between pure $2d$ quantum gravity and pure topological gravity  (in
both cases ``pure'' means that ``matter'' fields are not included).

After these introductory remarks we proceed to  calculations.
Actually they just repeat those from the previous subsection for the
Gaussian model, but formulas get somewhat more complicated.
This time we do not include zero-time $N$ and use eq.(\ref{GNe})
with $V(X) = \frac{X^3}{3}$.  Also, this time it is much
more tricky (though possible) to work in  matrix notations (because
fractional powers of $ L$ will be involved) and we rewrite everything
in terms of the eigenvalues of $ L$.

Substitute
\be
\new
\begin{array}{c}
{\cal C}_{\frac{X^3}{3}} = \frac{\prod_b
e^{\frac{2}{3}\lambda_b ^{3/2}}}{\sqrt{\prod_{a ,b }
              (\sqrt{\lambda_b } + \sqrt{\lambda_a })}},  \\
\left( \frac{\partial^2}{\partial L_{tr}^2}\right)_{a a } =
 \frac{\partial^2}{\partial\lambda_a ^2} +
     \sum_{b \neq a }\frac{1}{\lambda_a -\lambda_b }
     \left(\frac{\partial}{\partial\lambda_a }
                 - \frac{\partial}{\partial\lambda_b }\right)
\end{array}
\nn
\ee
and introduce a special notation for
\be
\frac{{\cal D}}{{\cal D}\lambda_a } \equiv
{\cal C}_{\frac{X^3}{3}}^{-1} \frac{\partial}{\partial \lambda_a }
         {\cal C}_{\frac{X^3}{3}} =
 \frac{\partial}{\partial \lambda_a } + \sqrt{\lambda_a } -
\frac{1}{4\lambda_a }
- \frac{1}{2} \sum_{b \neq a }\frac{1}{\sqrt{\lambda_a }
(\sqrt{\lambda_b } +
           \sqrt{\lambda_a })}. \nn
\ee
Then (\ref{GNe}) turns into
\be
\left( \left(\frac{{\cal D}}{{\cal D}\lambda_a }\right)^2 +
\sum_{b \neq a }\frac{1}{\lambda_a -\lambda_b }
\left(\frac{{\cal D}}{{\cal D}\lambda_a } - \frac{{\cal D}}{{\cal
D}\lambda_b }\right)
\right) {\cal Z}_{\frac{X^3}{3}}\{T\} = 0.
\label{GNX3}
\ee
Now we need explicit expression for $T$:
\be
T_k = \frac{1}{k} L^{-k},
\label{KMX3T}
\ee
and  - as we already know from the previous subsection - we also
need
\be
r_k = - \frac{2}{3}\delta_{k,3}.
\ee
${\cal Z}_{\frac{X^3}{3}}\{T\}$ is in fact independent of all
the time-variables with {\it even} numbers (subscripts), see
\cite{GKM}, \cite{UFNmm} for the explanation.
Therefore we can take only $k=2l+1$ in (\ref{KMX3T}),
\be
T_{2l+1}  =
\frac{1}{2l+1} \sum_b  \lambda_b ^{-l-\frac{1}{2}}, \nn \\ r_{2l+1}
 = - \frac{2}{3}\delta_{l,1}
\ee
and
\be
\frac{\partial}{\partial
\lambda_a } {\cal Z}_{\frac{X^3}{3}}\{T\}  = \sum_{l=0}^{\infty}
\frac{\partial T_{2l+1}}{\partial \lambda_a }
\frac{\partial{\cal Z}}{\partial T_{2l+1}} =
-\frac{1}{2} \sum_{a=0}^{\infty}\lambda_a ^{-l-\frac{3}{2}}
\frac{\partial{\cal Z}}{\partial T_{2l+1}}; \nn \\
\frac{\partial^2}{\partial \lambda_a ^2} {\cal Z}_{\frac{X^3}{3}}\{T\}  =
\frac{1}{4}\sum_{l,m=0}^{\infty} \lambda_a ^{-l-m-3}
\frac{\partial{\cal Z}}{\partial T_{2l+1}\partial T_{2m+1}} +
\frac{1}{2} \sum_{l=0}^{\infty} (l + \frac{3}{2})
\lambda_a ^{-l-\frac{5}{2}}\frac{\partial{\cal Z}}{\partial T_{2l+1}}. \nn
\ee
These expressions should be now substituted into (\ref{GNX3}) and we obtain:
\be
\new
\begin{array}{c}
\frac{1}{4}\sum_{l,m=0}^{\infty} \lambda_a ^{-l-m-3}
   \frac{\partial{\cal Z}}{\partial T_{2l+1}\partial T_{2m+1}} +\\
+ \sum_{l=0}^{\infty} \left[\frac{1}{2} \sum_{a=0}^{\infty}
(l + \frac{3}{2})\lambda_a ^{-l-\frac{5}{2}} -
\frac{1}{2} \sum_{b \neq a }\frac{1}{\lambda_a -\lambda_b }
\left(\lambda_a ^{-l-\frac{3}{2}} - \lambda_b ^{-l-\frac{3}{2}}
\right) -
\right.
 \\
\left.
- \sum_{a=0}^{\infty}\left( \sqrt{\lambda_a } - \frac{1}{4\lambda_a }
- \frac{1}{2} \sum_{b \neq a }
\frac{1}{\sqrt{\lambda_a }(\sqrt{\lambda_b } + \sqrt{\lambda_a })}
\right)\lambda_a ^{-l-\frac{3}{2}}
\right]\frac{\partial{\cal Z}}{\partial T_{2l+1}} +  \\
+ \left[\ldots\right] {\cal Z}  \ \ \
= \ \ \ \sum_{n=-1}^{\infty} \frac{1}{\lambda_a ^{n+2}}{\cal L}_n{\cal Z}
\end{array}
\label{vircder}
\ee
with
\be
\new
\begin{array}{c}
{\cal L}_{2n} = \sum_{l=0}^{\infty}
    \left(l+\frac{1}{2}\right)\left(T_{2l+1}+r_{2l+1}\right)
\frac{\partial}{\partial T_{2l+2n+1}} +  \\ +
\frac{1}{4} \sum_{\stackrel{l+m=n-1}{l,m\geq 0}}
\frac{\partial^2}{\partial T_{2l+1}\partial T_{2m+1}}
+  \frac{1}{16}\delta_{n,0} + \frac{1}{4} T_1^2\delta_{n,-1} = \\
= \frac{1}{2}\sum_{{\rm odd}\ k=1}^{\infty}
       k(T_k+r_k)\frac{\partial}{\partial T_{k+2n}} +
 \frac{1}{4} \sum_{{\rm odd}\ k = 1}^{2n-1}
\frac{\partial^2}{\partial T_k\partial T_{2n-k}} +
  \frac{1}{16}\delta_{n,0} + \frac{1}{4} T_1^2\delta_{n,-1}.
\end{array}
\label{vircidder}
\ee
Factor $\frac{1}{2}$ in front of the first term at the r.h.s. in
(\ref{vircidder}) is important for ${\cal L}_{2n}$ to satisfy the properly
normalized Virasoro algebra:\footnote{
Therefore it could be reasonable to use a different notation: ${\cal L}_n$
instead of ${\cal L}_{2n}$. We prefer ${\cal L}_{2n}$, because it emphasises
the property of the model to be 2-reduction of KP hierarchy (to KdV).
}
\be
\phantom. [{\cal L}_{2n}, {\cal L}_{2m}] = (n-m){\cal L}_{2n+2m}.
\nn
\ee
Coefficient $\frac{1}{4}$ in front of the second term can be eliminated by
rescaling of time-variables: $T \rightarrow \frac{1}{2}T$, then the last term
turns into $\frac{1}{16}T_1^2\delta_{n,-1}$.

We shall not actually discuss evaluation of the coefficient in front of
${\cal Z}$ (with no derivatives), which is denoted by $[\ldots]$ in
(\ref{vircder}) (see \cite{MMM} and \cite{GKM}). In fact almost all the terms
in original complicated expression cancel, giving finally
\be
\left[ \ldots \right] =
\frac{1}{16\lambda_a ^2} + \frac{T_1^2}{4\lambda_a },
\nn
\ee
and this is represented by the terms with $\delta_{n,0}$ and $\delta_{n,-1}$
in expressions (\ref{vircidder}) for the Virasoro generators ${\cal L}_{2n}$.

The term with the double $T$-derivative in (\ref{vircder}) is already
of the necessary form. Of intermidiate complexity is evaluation of
the coefficient in front of $\frac{\partial{\cal Z}}{\partial T_{2l+1}}$
in (\ref{vircder}), which we shall briefly describe now.
First of all, rewrite this coefficient, reordering the items:
\be
\frac{1}{2}\left[ (l + \frac{3}{2})\lambda_a ^{-l-\frac{5}{2}} -
 \sum_{b \neq a }\frac{1}{\lambda_a -\lambda_b }
\left(\lambda_a ^{-l-\frac{3}{2}} -
\lambda_b ^{-l-\frac{3}{2}} \right)\right] + \nn \\
+ \left[ \frac{1}{4}\lambda_a ^{-l-\frac{5}{2}} +
       \frac{1}{2} \sum_{b \neq a }
\frac{\lambda_a ^{-l-2}}{\sqrt{\lambda_b } + \sqrt{\lambda_a }} \right]
- \lambda_a ^{-l-1}.
\label{vircder2}
\ee
The first two terms together are equal to the sum over {\it all} $b$
(including $b=a$):
\be
- \frac{1}{2} \sum_b  \frac{1}{\lambda_a -\lambda_b }
\left(\lambda_a ^{-l-\frac{3}{2}} - \lambda_b ^{-l-\frac{3}{2}} \right)
 =  \frac{1}{2} \sum_b  \frac{\lambda_a ^{l+\frac{3}{2}} -
             \lambda_b ^{l+\frac{3}{2}}}{\lambda_a -\lambda_b }
  \cdot\frac{1}{\lambda_a ^{l+\frac{3}{2}}\lambda_b ^{l+\frac{3}{2}}} =
\nn \\
= \frac{1}{2\lambda_a ^{l+2}} \sum_b
\frac{\lambda_a ^{l+2} - \lambda_a ^{\frac{1}{2}}
\lambda_b ^{l+\frac{3}{2}}}
{\lambda_a -\lambda_b }\cdot\frac{1}{\lambda_b ^{l+\frac{3}{2}}}.
\nn
\ee
Similarly, the next two terms can be rewritten as
\be
\frac{1}{2} \sum_b  \frac{\lambda_a ^{-l-2}}{\sqrt{\lambda_a } +
\sqrt{\lambda_b }} = \frac{1}{2\lambda_a ^{l+2}}
\sum_b  \frac{\sqrt{\lambda_a } - \sqrt{\lambda_b }}
{\lambda_a -\lambda_b } = \nn \\
= \frac{1}{2\lambda_a ^{l+2}} \sum_b
\frac{\lambda_a ^{\frac{1}{2}} \lambda_b ^{l+\frac{3}{2}} -
\lambda_b ^{l+2}}
{\lambda_a -\lambda_b }\cdot\frac{1}{\lambda_b ^{l+\frac{3}{2}}}.
\nn
\ee
The sum of these two expressions is equal to
\be
\frac{1}{2\lambda_a ^{l+2}} \sum_b
\frac{\lambda_a ^{l+2} - \lambda_b ^{l+2}}
{\lambda_a -\lambda_b }
\cdot\frac{1}{\lambda_b ^{l+\frac{3}{2}}}.
\nn
\ee
Note that powers $l+2$ are already integer and the remaining ratio can be
represented as a sum of $l+2$ terms. Adding also the last term from the l.h.s.
of (\ref{vircder2}), we finally obtain:
\be
-\frac{1}{\lambda_a ^{l+1}} + \frac{1}{2} \sum_{n=-1}^a
\frac{1}{\lambda_a ^{n+2}}\sum_b
\frac{1}{\lambda_b ^{l-n+\frac{1}{2}}} =   \nn \\ =
\frac{1}{2}\sum_{n=-1}^a \frac{1}{\lambda_a ^{n+2}}
(2a-2n+1)(T+r)_{2l-2n+1}
\nn
\ee
in accordance with (\ref{vircder}) and (\ref{vircidder}).

	Calculations can be repeated for every particular monomial potential
	$V(x) = \frac{x^{p+1}}{p+1}$, but they become far more tedious and no
	general derivation of $W^{(p)}$-constraints \cite{FKN} is yet found
	on these lines. See \cite{Mikh} for detailed examination of the
	$W^{(3)}$-constraints in the $\frac{X^4}{4}$-Kontsevich model.

\section{KP/Toda $\tau$-function in terms of free fermions \label{se4}}
\setcounter{equation}0

There are several different definitions of $\tau$-functions, but all of them
are particular realizations of the following idea:
$\tau$-function is a generating functional of all the matrix elements of
some group element in particular representation. Since methods of
geometrical quantization allow to express all the group theoretical objects
in terms of quantum theory of free  fields, generic $\tau$-functions
can be also considered as non-perturbative partition functions of such
models. The basic property of $\tau$-function, which can be
practically derived  in such a general context, is that it always
satisfy certain bilinear equations, of which
Hirota equation for conventional KP $\tau$-function is the simplest
example.

KP/Toda $\tau$-functions are associated with the free particles of a
peculiar type: free fermions in $1+1$ dimensions \cite{DJMS}.
Existence of fermionization is a very rare property of free field theory
(in varience with bosonization which is always available).  If existing
it leads to dramatic simplification of the formalism and to especially
simple determinant formlulas (instead of sophisticated and often somewhat
abstract objects like chiral determinants $\det \bar\partial$ in generic
case). In the case of Kac-Moody algebras the corresponding $\tau$-function
is nothing but non-perturbative partition function of the corresponding
Wess-Zumino-Novikov-Witten model.  Among simply-laced algebras only
$\hat{GL(N)}_{k=1}$ is straightforwardly
fermionized, and the formalism is much simpler in this case than for
generic Wess-Zumino-Witten model with arbitrary level $k$.
For $N=1$ we obtain KP/Toda $\tau$-functions, while $N\neq 1$ are
related to the ``$N$-component KP/Toda systems''.
Level-one Kac-Moody algebras $\hat{SL(N)}_{k=1}$ are distinguished
because their universal envelopping are essentially the same as those of
their Cartan subalgebras. This allows to define generation functions
with the help of sets of mutualy commuting generators and makes
evolution, described by commuting Hamiltonian flows, complete
(acting transitively on the orbits of the group). This is why such
systems are distinguished from the point of view of Hamiltonian
integrability -- and why they are the usual personages in the theory
of integrable hierarchies.  In general case ($k\neq 1$) one naturally
deals with the set of flows that form closed but non-Abelian algebra.
In the language of matrix models restriction to $k=1$ and free
{\it fermions} is essentially equivalent to restriction to {\it eigenvalue}
models. Serious consideration of non-eigenvalue models, aimed at revealing
their integrable (solvable) structure will certainly involve the theory of
generic $\tau$-functions.

\subsection{Explicit definition}

Let us introduce two fields (a spin-$1/2$ $b,c$-system)
$\tilde\psi(z)$ and $\psi(z)$ satisfying canonical commutation relation:
\be
\phantom. [\tilde\psi(\tilde z),\psi( z)]_+ =
\delta(\tilde z -  z) d\tilde z^{1/2}d z^{1/2}.
\ee
Then
\be
\tau\{A\} \sim
\langle 0 \mid \exp\left({\oint_{d\tilde z}\oint_{d z}
A( z,\tilde z)\psi( z)\tilde\psi(\tilde z)}\right)
\mid 0 \rangle.
\label{hrtf}
\ee
Now it is usual to expand in Laurent series around $ z = 0$:
\be
 \psi( z) = \sum_{n\in Z}\psi_n z^n d z^{1/2}; \ \ \
\tilde\psi( z) = \sum_{n\in Z}\tilde\psi_n z^{-n-1}d z^{1/2};
\nn \\
 \phantom. [\tilde\psi_m,\psi_n]_+ = \delta_{m,n}; \nn  \\
 \psi_m\mid 0 \rangle = 0 \ \ {\rm for}\ m<0; \ \ \
\tilde\psi_m \mid 0 \rangle = 0 \ \ {\rm for}\ m\geq 0; \nn \\
 A( z,\tilde z) = \sum_{m,n\in Z}  z^{-m-1}\tilde z^n
A_{mn}d z^{1/2}d\tilde z^{1/2}; \nn
\ee
so that
\be
\oint_{d\tilde z}\oint_{d z}
A( z,\tilde z)\psi( z)\tilde\psi(\tilde z) =
\sum_{m,n\in Z}  A_{mn}\psi_m\tilde\psi_n.
\nn
\ee
In fact this expansion could be around {\it any} pair of points
$z_0,\ z_\infty$ and on a
2-surface of any topology: topological effects can be easily included as
specific shifts of the functional $A( z,\tilde z)$ - by combinations
of the "hadle-gluing operators". Analogous shifts can imitate the change of
basic functions $ z^n$ for $ z^{n+\alpha}$ and more complicated
expressions (holomorphic 1/2-differentials with various boundary conditions
on surfaces of various topologies).

One can now wonder, whether $local$ functionals $A( z,\tilde z) =
U( z)\delta(\tilde z- z)d z^{1/2}d\tilde z^{1/2}$
play any special role. The corresponding contribution to the Hamiltonian looks
like
\be
H_{Cartan} =
\oint_{d z}U( z)\psi( z)\tilde\psi( z) =
\oint_{d z} U( z)J( z),
\ee
where
\be
J( z) = \psi( z)\tilde\psi( z) =
\sum_{n\in Z}J_n z^{-n-1}d z
\ee
is the $U(1)_{k=1}$ Kac-Moody current;
\be
J_n = \sum_{m\in Z}\psi_m\tilde\psi_{m+n}; \ \ \ [J_m,J_n] = m\delta_{m+n,0}.
\ee
If scalar function (potential) $U( z)$ is expanded as
$U( z) = \sum_{k\in Z} t_k z^k$, then
\be
H_{Cartan} =
\sum_{n\in Z} t_kJ_k.
\ee
This contribution to the whole Hamiltonian can be considered distinguished for
the following reason. Let us return to original expression (\ref{hrtf}) and
try to consider it as a generating functional for all the correlation
functions of $\tilde\psi$ and $\psi$.
Naively, variation w.r.to $A( z,\tilde z)$ should produce bilinear
combination $\psi( z)\tilde\psi(\tilde z)$ and this would solve the
problem. However, things are not just so trivial, because operators involved
do not commute (and in particular, the exponential operator in (\ref{hrtf})
should still be defined less symbolically, see next subsection). Things would
be much simpler, if we can consider $commuting$ set of operators: this is
where abelian $\hat{U(1)}_{k=1}$ subgroup of the entire $GL(\infty)_{k=1}$
(and even its purely commuting Borel subalgebra) enters the game. Remarkably,
it is sufficient to deal with this abelian subgroup in order to reproduce all
the correlation functions.\footnote{
We once again emphasize that this trick is specific for the free fermions and
for the level $k=1$ Kac-Moody algebras, which can be expressed entirely in
terms of free fields, associated with Cartan generators (modulo some
unpleasant details, related to "cocycle factors" in the Frenkel-Kac
representations \cite{FK},
which are in fact reminiscents of free fields associated with the non-Cartan
generators (parafermions) \cite{Turb}, - but can, however, be put under the
carpet or/and taken into account "by hands" as "unpleasant but
non-essential(?) sophistications).
}
The crucial point is the identity for free fermions (generalizable to any $b,
c$-systems):
\be
:\psi( \lambda)\tilde\psi(\tilde \lambda):\ = \ :\exp\left({
\int_{ \lambda}^{\tilde \lambda} J}\right):
\label{bosid}
\ee
which is widely known in the form of bosonization formulas:\footnote{
Formulas in brackets are indeed correct, they are preceded by the usual
symbolic relations. Using these formulas we get:
\be
:\psi(\lambda)\tilde\psi(\tilde \lambda):\  = \  :e^{
\phi(\tilde \lambda)-\phi( \lambda)}:\ = \ :e^{
\int_{ \lambda}^{\tilde \lambda} \partial\phi}: \ = \ :e^{
\int_{ \lambda}^{\tilde \lambda} J}:
\nn
\ee
This identity can be of course obtained within fermionic theory, one should
only take into account that $\psi$-operators are nilpotent, so that exponent
of a single $\psi$-operator would be just a sum of two terms (polinomial)
and carefully follow the normal oredering prescription.
}
if $J( z) = \partial\phi( z)$,
\be
\tilde\psi(\tilde \lambda) \sim \ :e^{
\phi(\tilde \lambda)}:
\ \ \ \left(\ :\psi(\infty)\tilde\psi(\tilde \lambda):\ =\ :e^{
(\phi(\tilde \lambda) - \phi(\infty))}:\ \right); \nn \\
\psi( \lambda) \sim \ :e^{-
\phi( \lambda)}:
\ \ \ \left(\ :\psi( \lambda)\tilde\psi(\infty):\ =\ :e^{
(\phi(\infty) - \phi( \lambda))}:\ \right).
\nn
\ee
This identity implies that one can generate any bilinear combinations of
$\psi$-operators by variation of potential $U( z)$ only, moreover this
variation should be of specific form:
\be
\Delta\oint UJ = \Delta\left( \sum_{k\in Z} t_kJ_k \right) =
\int_{ z}^{\tilde z} J = \sum_{k \in Z}
\int_{ z}^{\tilde z}   z^{-k-1} d z = \nn \\
= \sum_{k \in Z} \frac{1}{k}J_k \left(\frac{1}{ z^k} -
\frac{1}{\tilde z^k}\right),
\nn
\ee
i.e.
\be
\Delta t_k = \frac{1}{k}\left(\frac{1}{ z^k} -
\frac{1}{\tilde z^k}\right)
\ee
Note that this is {\it not} an infinitesimal variation and that it
has exactly the form, con\-sis\-tent with Miwa para\-metri\-zation.

Since any bilinear combination can be generated in this way from $U(
z)$, it is clear that the entire Hamiltonian $\sum
A_{mn}\tilde\psi_m\psi_n$ can be also considered as resulting from
some transformation of $V$ (i.e. of "time-variables" $t_k$). In other
words, \be \tau\{A\} = {\cal O}_A[t] \tau\{A = U\}.  \nn \ee These
operators ${\cal O}_A$ are naturally interpreted as elements of the
group $GL(\infty)$, acting on the Universal Grassmannian {\it GR}
\cite{SeWi}-\cite{Orlov}, parametrized by the matrices $A_{mn}$
modulo changes of coordinates $ z \rightarrow f( z)$. This
representation for $\tau\{A\}$ is, however, not very convenient, and
usually one considers {\it infinitesimal} version of the
transformation, which just shifts $A$ \be \tau\{t\mid A+\delta A\} =
\hat{\cal O}_{\delta A}[t] \tau\{t\mid A\}, \label{trtf} \ee note
that this transformation clearly distinguishes between the
dependencies of $\tau$ on $t$ and on all other components of $A$. The
possibility of such representation with the privileged role of Cartan
generators is the origin of all simplifications, arising in the case
of free-fermion $\tau$-functions. Relation (\ref{trtf}) is the basis
of the orbit interpretation of $\tau$-functions \cite{Kac}.

\subsection{Basic determinant formula for the free-fermion correlator}

Let us consider the following matrix element:
\be
\tau_N\{t,\bar t\mid G\} =
\langle N \mid e^H \ G \ e^{\bar H} \mid N \rangle
\label{pretaf}
\ee
where
\be
\new
\begin{array}{c}
\psi( z) = \sum_{n\in Z}\psi_n z^n d z^{1/2}; \ \ \
\tilde\psi( z) = \sum_{n\in Z}\tilde\psi_n z^{-n-1}d z^{1/2};\\
G = \exp\left( \sum_{m,n\in Z}{A}_{mn}\psi_m\tilde\psi_n \right);  \\
H = \sum_{k>0} t_kJ_k, \ \ \ \bar H = \sum_{k>0} \bar t_kJ_{-k} \\
J( z) = \psi( z)\tilde\psi( z) =
\sum_{n\in Z}J_n z^{-n-1}d z; \ \ \ J_n =
\sum_k\psi_k\tilde\psi_{k+n};   \\
\phantom. [\tilde\psi_m,\psi_n]_+ = \delta_{m,n}; \ \ \
[J_m,J_n] = m\delta_{m+n,0};  \\
\psi_m\mid N \rangle = 0, \ \ m<N; \ \ \ \
\langle N \mid \psi_m = 0, \ \ m\geq N;  \\
\tilde\psi_m\mid N \rangle = 0, \ \ m\geq N; \ \ \ \
\langle N \mid \tilde\psi_m = 0, \ \ m< N;  \\
J_m\mid N \rangle = 0,\ \ m>0; \ \ \ \
\langle N \mid J_m = 0, \ \ m<0.
\end{array}
\label{notfermcor}
\ee
The "$N$-th vacuum" $\mid N \rangle$ is defined as the Dirac sea,
filled up to the level $N$:
\be
\mid N \rangle =
\prod_{i=N}^\infty \tilde\psi_i\mid\infty\rangle =
\prod_{i=-\infty}^{N-1} \psi_i \mid-\infty\rangle \ ; \nn \\
\langle N \mid = \langle \infty\mid \prod_{i=N}^\infty \psi_i =
\langle -\infty\mid \prod_{i=-\infty}^{N-1} \tilde\psi_i,
\ee
where the "empty" (bare) and "completely filled" vacua  are defined so that:
\be
 \tilde\psi_m \mid -\infty \rangle = 0, \ \ \ \langle -\infty \mid \psi_m = 0,
\nn \\
 \psi_m \mid \infty \rangle = 0, \ \ \ \langle \infty \mid \tilde\psi_m = 0
\ee
for $any$ $m\in Z$. For the only reason that operators $J$, $H$, $\bar H$ and
$G$ are defined so that they always have $\tilde\psi$ at the very right and
$\psi$ at the very left, we get also:
\be
 J_m \mid -\infty \rangle = 0, \ \ \ \langle -\infty\mid  J_m = 0,
\nn \\
 G^{\pm 1} \mid-\infty \rangle = \mid-\infty \rangle; \ \ \
\langle -\infty \mid G^{\pm 1} = \langle -\infty \mid; \nn \\
 e^{\pm \bar H}  \mid-\infty \rangle = \mid-\infty \rangle; \ \ \
\langle -\infty \mid e^{\pm H} = \langle -\infty \mid.
\ee
Now we can use all these formulas to rewrite our original correlator
(\ref{pretaf}) as:
\be
\new
\begin{array}{c}
\langle N \mid e^H \ G\ e^{\bar H}\mid N \rangle =  \\
=  \langle  -\infty\mid \left(\prod_{i=-\infty}^{N-1} \tilde\psi_i\right)
                  e^H \ G\ e^{\bar H}
\left(\prod_{i=-\infty}^{N-1} \psi_i\right) \mid-\infty\rangle =  \\
= \langle  -\infty\mid e^{-H} \left(\prod_{i=-\infty}^{N-1}\tilde\psi_i\right)
                  e^H \ G\ e^{\bar H}
\left(\prod_{i=-\infty}^{N-1} \psi_i\right) e^{-\bar H}\mid-\infty\rangle =
 \\
= \langle  -\infty\mid \prod_{i=-\infty}^{N-1} \tilde\Psi_i[t]
\prod_{j=-\infty}^{N-1} \Psi_j^G[\bar t] \mid-\infty\rangle =
 \\
= {\rm Det}_{-\infty < i,j < N}
\langle  -\infty\mid \tilde\Psi_i[t] \Psi_j^G[\bar t] \mid-\infty\rangle =
 \\
= {\rm Det}_{i,j<0} {\cal H}_{i+N,j+N}.
\end{array}
\label{pretafdet}
\ee
The last two steps here were introduction of "$GL(\infty)$-rotated" fermions,
\be
\tilde\Psi_i[t] \equiv e^{-H}\psi_i e^H; \ \ \
\Psi_j[\bar t] \equiv e^{\bar H} \psi_j e^{-\bar H};\ \ \
\Psi_j^G[\bar t] \equiv G\Psi_j[\bar t] G^{-1},
\label{bigpsi}
\ee
and application of the Wick theorem to express multifermion correlation
function through pair correlators
\be
\new
\begin{array}{c}
{\cal H}_{ij}(t,\bar t) \equiv
\langle  -\infty\mid \tilde\Psi_i[t] \Psi_j^G[\bar t] \mid-\infty\rangle =
 \\ =
\langle  -\infty\mid \tilde\Psi_i[t]\ G \ \Psi_j[\bar t] \mid-\infty\rangle ,
\end{array}
\label{Hmatrcor}
\ee
(once again the fact that $G^{-1}\mid-\infty\rangle = \mid-\infty\rangle$
was used). The only non-trivial dynamical information entered through
applicability of the Wick theorem, and for that  it was crucial that all the
operators $e^H,\ e^{\bar H},\ G$ are $quadratic$ exponents, i.e. can only
modify the shape of the propagator, but do not destroy the quadratic form of
the action (fields remain $free$).
This is exactly equivalent to the statement that "Heisenberg" operators
$\Psi[t]$ are just "rotations"
of $\psi$, i.e. that transformations (\ref{bigpsi}) are $linear$.

We shall now describe these transformations in  a little more explicit form.
Namely, their entire time-dependence can be encoded in terms of the
ordinary Shur polinomials $P_n(t)$.
These are defined to have a very simple generating
function (which we already encountered many times in the theory of matrix
models):
\be
\sum_{n\geq 0} P_n(t)z^n = \exp\left({\sum_{k=1}^{\infty} t_kz^k}\right)
\ee
(i.e. $P_0 = 1,\ \ P_1 = t_1,\ \ P_2 = \frac{t_1^2}{2} + t_2$ etc.), and
satisfy the relation
\be
\frac{\partial P_n}{\partial t_k} = P_{n-k}.
\ee
Since
\be
\exp\left({\sum_{k=1}^{\infty} t_kz^k}\right) =
\prod_{k>0} \left(\sum_{n_k\geq 0}\frac{1}{n_k!}\ t_k^{n_k}z^{kn_k}\right),
\nn
\ee
Shur polinomials can be also represented as
\be
P_n(t) = \sum_{\stackrel{\{n_k\}}{\sum_{k>0} kn_k = n}}
\left(\prod_{k> 0}\frac{1}{n_k!}\ t_k^{n_k}\right).
\label{Shupoex}
\ee
Now, since
\be
e^{-B}Ae^B = A + [A,B] + \frac{1}{2!}[[A,B],B] +
\frac{1}{3!}[[[A,B],B],B] + \ldots
\nn
\ee
and
\be
\phantom. [\tilde\psi_i,J_k] = \tilde\psi_{i+k}, \ \
[ [\tilde\psi_i,J_{k_1}], J_{k_2}] = \tilde\psi_{i+k_1+k_2}, \ \ldots,
\nn
\ee
we have for every fixed $k$:
\be
e^{-t_kJ_k} \tilde\psi_i e^{t_kJ_k} =
\sum_{n_k\geq 0} \frac{t_k^{n_k}}{n_k!}\ \tilde\psi_{i+kn_k}.
\nn
\ee
It remains to note that all the harmonics of $J$ in $H = \sum_{k>0} t_kJ_k$
commute with each other, to obtain:
\be
\new
\begin{array}{c}
\tilde\Psi_i(t) = e^{-H} \tilde\psi_i e^H =
\left( \prod_{k>0} e^{-t_kJ_k}\right) \tilde\psi_i
\left( \prod_{k>0}e^{t_kJ_k}\right) =  \\
= \sum_{n\geq 0}\tilde\psi_{i+n} \left(
 \sum_{\stackrel{\{n_k\}}{\sum_{k>0} kn_k = n}}
\left(\prod_{k> 0}\frac{1}{n_k!}\  t_k^{n_k}\right)\right)
\stackrel{(\ref{Shupoex})}{=} \\
= \sum_{n\geq 0}\tilde\psi_{i+n} P_n(t) = \sum_{l\geq i}\tilde\psi_{l}
P_{l-i}(t).
\end{array}
\ee
Similarly, relation $[J_k,\psi_j] = \psi_{k+j}$ implies, that
\be
\Psi_j(\bar t) = e^{\bar H}\psi_j e^{-\bar H} =
\sum_{n\geq 0}\psi_{j+n} P_n(\bar t) = \sum_{m\geq j}\psi_{m}P_{m-j}(\bar t)
\label{bigpsishur}
\ee
and finally
\be
 {\cal H}_{ij} = \sum_{\stackrel{l\geq i}{m\geq j}}
\langle -\infty \mid \tilde\psi_l\ G\ \psi_m \mid-\infty \rangle\
P_{l-i}(t)P_{m-j}(\bar t) = \nn \\           =
\sum_{\stackrel{l\geq i}{m\geq j}} T_{lm} P_{l-i}(t)P_{m-j}(\bar t),
\label{HversTlm}
\ee
which implies also that
\be
\frac{\partial{\cal H}_{ij}}{\partial t_k} = {\cal H}_{i+k,j}; \nn \\
\frac{\partial{\cal H}_{ij}}{\partial\bar t_k} = {\cal H}_{i,j+k} .
\label{todaeqforH}
\ee
Matrix
\be
T_{lm} \equiv \langle -\infty \mid \tilde\psi_l\ G\ \psi_m \mid-\infty \rangle
\ee
is the one which defines fermion rotations under the action of
$GL(\infty)$-group element $G$:
\be
 G\psi_mG^{-1} = \sum_{l\in Z} \psi_lT_{lm}; \nn \\
 G^{-1}\tilde\psi_l G = \sum_{m\in Z} T_{lm}\tilde\psi_m,\ {\rm or}\
G\tilde\psi_l G^{-1} = \sum_{m\in Z} (T^{-1})_{lm}\tilde\psi_m.
\ee
If $G=1$, $T_{lm} = \delta_{lm}$. If all $t_k = \bar t_k = 0$,
${\cal H}_{ij} = T_{ij}$.

\subsection{KP hierarchy and other reductions}

In the previous subsection a formula
\be
\tau_N\{t,\bar t\mid G\} =
\left.{\rm Det}\right._{i,j<0} {\cal H}_{i+N,j+N}
\label{todatau}
\ee
was derived for the basic correlator, which defines
"Toda-lattice $\tau$-function". For obvious reasons the variables
$\bar t$ are often refered to as ``negative-times''. $\tau$-function
can be normalized by division over the same quantity with all the
time-variables vanishing, but this is not always convenient.
Eq.(\ref{todatau}) has generalizations - when similar
matrix elements in a multifermion system is considered - this leads to
"multicomponent Toda" (or AKNS) $\tau$-functions. Generalizations to
arbitrary conformal models should be considered as well.
It has also particular
"reductions", of which the most important are: KP (Kadomtsev-Petviashvili),
forced (semi-infinite) and Toda-chain $\tau$-functions.
This is the subject to be discussed in this subsection.

Idea of linear reduction is that the form of operator $G$, or, what is
the same, of
the matrix $T_{lm} $ in eq.(\ref{HversTlm}), can be adjusted in such a way,
that $\tau_N\{t,\bar t\mid G\}$ becomes independent of some variables, i.e.
equation(s)
\be
\left(\sum_k \alpha\frac{\partial}{\partial t_k} +
 \sum_k \bar\alpha\frac{\partial}{\partial \bar t_k} +
\sum_k \beta_k D_N(k) + \gamma \right) \tau_N\{t,\bar t\mid G\} = 0
\label{redtotau1}
\ee
can be solved as equations for $G$ for all the values of $t,\bar t$ and $N$
at once. (In (\ref{redtotau1}) $D_N(k)f_N \equiv f_{N+k} - f_N$.)
In this case the system of integrable equations (hierarchy), arising from
Hirota equation for $\tau$, gets reduced and one usually speaks about
"reduced hierarchy". Usually equation (\ref{redtotau1}) is imposed directly
on matrix ${\cal H}_{ij}$, of course than (\ref{redtotau1}) is just a
corollary.

We shall refer to the situation when (\ref{redtotau1}) is fulfilled for {\it
any} $t, \bar t, N$  as to "strong reduction". It is often reasonable to
consider also "weak reductions", when (\ref{redtotau1}) is satisfied on
particular infinite-dimensional hyperplanes in the space of time-variables.
Weak reduction is usually a property of entire $\tau$-function as well, but
not expressible in the from of a local linear equation, satisfied identicall
for {\it all} values of $t, \bar t, N$.
Now we proceed to concrete examples:

{\it Toda-chain hierarchy}. This is a {\it strong} reduction. The
corresponding constraint (\ref{redtotau1}) is just
\be
\frac{\partial {\cal H}_{ij}}{\partial t_k} = \frac{\partial {\cal
H}_{ij}}{\partial \bar t_k},
\ee
or, because of (\ref{todaeqforH}), ${\cal H}_{i+k,j} = {\cal H}_{i,j+k}$.
It has an obvious solution:
\be
{\cal H}_{i,j} = \hat{\cal H}_{i+j},
\ee
i.e. ${\cal H}_{ij}$ is expressed in terms of a one-index quantity $\hat{\cal
H}_i$. It is, however, not enough to say, what are restrictions on ${\cal
H}_{ij}$ - they should be fulfiled for all $t$ and $ \bar t$ at once, i.e.
should be resolvable as equations for $T_{lm}$. In the case under
consideration this is simple: $T_{lm}$ should be such that
\be
T_{lm} = \hat T_{l+m}.
\ee
Indeed, then
\be
 {\cal H}_{ij} = \sum_{l,m} T_{lm} P_{l-i}(t) P_{m-j}(\bar t) =
\sum_{l,m} \hat T_{l+m} P_{l-i}(t)P_{m-j}(\bar t) = \nn \\
 = \sum_{n\geq 0} \hat T_{n+i+j} \left(\sum_{k=0}^n P_k(t)P_{n-k}(\bar
t)\right),
\nn
\ee
and
\be
\hat{\cal H}_i = \sum_{n\geq 0} \hat T_{n+i} \left(\sum_{k=0}^n
P_k(t)P_{n-k}(\bar t)\right).
\label{todachainH}
\ee

{\it Volterra hierarchy}. Toda-chain $\tau$-function can be further {\it
weakly}
reduced to satisfy the identity
\be
\left.\frac{\partial \tau_{2N}}{\partial t_{2k+1}}\right|
_{\{t_{2l+1}=0\}}  = 0, \ \ \ {\rm for\ all}\ k,
\label{volterrared}
\ee
i.e. $\tau_{2N}$ is requested to be even function of all odd-times $t_{2l+1}$
(this is an example of "global characterization" of the weak reduction).
Note that (\ref{volterrared}) is imposed only on {\it Toda-chain}
$\tau$-function with {\it even} values of zero-time. Then (\ref{volterrared})
will hold whenever $\hat{\cal H}_i$ in (\ref{todachainH}) are even (odd)
functions of $t_{\rm odd}$ for even (odd) values of $i$. Since Shur
polinomials $P_k(t)$ are even (odd) functions of odd-times for even (odd) $k$,
it is enough that the sum in (\ref{todachainH}) goes over even (odd) $n$
when $i$ is even (odd). In other words, the restriction on $T_{lm}$ is that
\be
T_{lm} = \hat T_{l+m}, \ \ \ {\rm and} \ \ \ \hat T_{2k+1} = 0 \ \ {\rm for\
all}\ k.
\ee

{\it Forced hierarchies}. This is another important example of strong
reduction. It also provides an example of {\it singular}
$\tau$-functions,
arising when $G = \exp \left(\sum A_{mn}\psi_m\tilde\psi_n\right)$
blows up and normal ordered operators should be used to define
regularized
$\tau$-functions. Forced hierarchy appears when $G$ can be
represented in the form \cite{KMMOZ}  $G = G_0P_+$, where
projection operator $P_+$ is such that
\be
P_+ \mid N \rangle = \mid N \rangle \ \ {\rm for}\ N\geq N_0,
\nn \\
P_+ \mid N \rangle = 0 \ \ {\rm for}\ N < N_0.
\label{Pplus}
\ee
Explicit expression for this operator is\footnote{
Normal ordering sign $\ : \ \ :\ $ means that all operators
$\tilde\psi$ stand to the {\it left} of all operators $\psi$.
The product at the r.h.s. obviously implies both the property
(\ref{Pplus}) and projection property $P_+^2 = P_+$.
}
\be
P_+ = \ :\exp \left( - \sum_{l<N_0} \tilde\psi_l\psi_l\right): \ =
\prod_{l<N_0} (1 - \tilde\psi_l\psi_l) =
\prod_{l < N_0} \psi_l\tilde\psi_l.
\nn
\ee
Because of (\ref{Pplus}), $P_+\mid -\infty \rangle = )$, and
the identity $G\mid -\infty \rangle = \mid -\infty \rangle$,
which was essentially used in the derivation in (4.27), can be
satisfied only if $G_0$ is singular and $T_{lm} = \infty$.
In order to avoid this problem one usually introduces in the
vicinity of such
singular points in the universal module space a sort of normalized
(forced) $\tau$-function $\tau_N^f \equiv \frac{\tau_N}{\tau_{N_0}}$.
One can check that now $T^f_{lm} = \infty$ for all $l,m < N_0$, and
$\tau^f$ can be represented as determinant of a final-dimensional
matrix \cite{UT},\cite{KMMOZ}:
\be
 \tau_N^f = {\rm Det}_{N_0\leq i,j < N} {\cal H}^f_{ij}\ \ \ {\rm for}\ \
N>N_0;
\nn \\
 \tau_{N_0}^f = 1; \nn \\
 \tau_{N}^f = 0\ \ \ {\rm for }\ \ N < N_0.
\ee
For $N>N_0$ we have now determinant of a {\it finite}-dimensional
$(N-N_0)\times (N-N_0)$ matrix.
The choice of $N_0$ is not really essential, therefore it is better to put
$N_0 = 0$ in order to
simplify formulas, phraising  and relation with the discrete matrix models
($N_0$ is easily restored if everywhere $N$ is substituted by $N-N_0$).
For forced hierarchies one can also represent $\hat\tau$ as
\be
\tau_N^f = {\rm Det}_{0\leq i,j < N}
\partial_1^i\bar\partial_1^j {\cal H}^f,
\ee
where ${\cal H}^f = {\cal H}^f_{00}$ and $\partial_1 = \frac{\partial}{\partial
t_1}$, $\bar \partial_1 = \frac{\partial}{\partial \bar t_1}$. For
{\it forced Toda-chain} hierarchy this turns into even simpler expression:
\be
\tau_N^f = {\rm Det}_{0\leq i,j < N} \partial_1^{i+j}\hat{\cal H}^f,
\ee
while for the {\it forced Volterra} case we get a product of two Toda-chain
$\tau$-functions with twice as small value of $N$ \cite{Bowick}:
\be
\tau_{2N}^f  = \left({\rm Det}_{0\leq i,j < N} \partial_2^{i+j}\hat{\cal
H}^f\right) \cdot
       \left( {\rm Det}_{0\leq i,j < N} \partial_2^{i+j} (\partial_2
\hat{\cal H}^f)\right) = \nn \\
 = \tau_N^f[\hat{\cal H}^f]\cdot \tau_N^f[\partial_2\hat{\cal H}^f].
\ee

Forced $\tau^f_N$ can be {\it always} represented in the form of a
scalar-product matrix model. Indeed,
\be
{\cal H}_{ij} = \sum T_{lm}P_{l-i}(t)P_{m-j}(\bar t) =
\oint\oint e^{U(h)+\bar U(\bar h)} h^i\bar h^j T(h,\bar h) dhd\bar h,
\ee
where $T(h,\bar h) \equiv \sum_{lm} T_{lm} h^{-l-1}\bar h^{-m-1}$, and
$e^{U(h)} = e^{\sum_{k>0}t_kh^k} = \sum_{l\geq 0} h^lP_l(t)$.
Then, since ${\rm Det}_{0\leq i,j < N}h^i = \Delta_N(h)$ - this is where it
is essential that the hierarchy is forced -
\be
{\rm Det}_{0\leq i,j < N} {\cal H}_{ij} = \prod_i
\oint\oint e^{U(h_i)+\bar U(\bar h_i)}
T(h_i,\bar h_i)dh_id\bar h_i
\cdot \Delta_N(h)\Delta_N(\bar h),
\ee
i.e. we obtain a scalar-product model with
\be
d\mu_{h,\bar h} = e^{U(h)+\bar U(\bar h)}T(h,\bar h) dhd\bar h.
\ee
Inverse is also true: partition function of every scalar-product model is
a forced Toda-lattice $\tau$-function.

{\it KP hierarchy}.  In this case we just ignore the  dependence of
$\tau$-function on times $\bar t$. Every Toda-lattice $\tau$-function can be
considered also as KP $\tau$-function: just operator $G^{KP} \equiv Ge^{\bar
H}$ (a point of Grassmannian) becomes $\bar t$-dependent. Usually
$N$-dependence is also eliminated - this can be considered as a little more
sophisticated change of $G$. When $N$ is fixed, extra changes of
field-variables are allowed, including transformation from Ramond to
Neveu-Schwarz sector etc. Often KP hierarchy is from the very beginning
formulated in terms of Neveu-Schwarz (antiperiodic) fermionic fields
(associated with  principal
representations of Kac-Moody algebras), i.e. expansions in the first line of
(\ref{notfermcor}) are in semi-integer powers of $z$:
$\psi_{NS}( z) = \sum_{n\in Z}\psi_n z^{n-\frac{1}{2}} d z^{1/2}$.

Given a KP $\tau$-function one can usually construct a Toda-lattice one with
the {\it same} G, by introducing in appropriate way dependencies on $\bar t$
and $N$. For this purpose $\tau^{KP}$ should be represented in the form of
(\ref{todatau}):
\be
\tau^{KP}\{t\mid G\} = {\rm Det}_{i,j<0} {\cal H}_{ij}^{KP},
\label{addkptau1}
\ee
where ${\cal H}_{ij}^{KP} = \sum_l T_{lj} P_{l-i}(t)$. Since $T_{lm} $ is a
function of $G$ only, it does not change when we built up a Toda-lattice
$\tau$-function:
\be
 \tau_N\{t,\bar t\mid G\} = {\rm Det}_{i,j<0} {\cal H}_{i+N,j+N}; \nn \\
 {\cal H}_{ij} = \sum_{l,m}T_{lm} P_{l-i}(t)P_{m-j}(\bar t) =
\sum_m {\cal H}_{im}^{KP} P_{m-j}(\bar t).
\ee
Then
\be
\tau^{KP}\{ t\mid G\} = \tau_0\{t,0\mid G\}.
\ee
If we go in the opposite direction, when Toda-lattice $\tau$-function is
considered as KP $\tau$-function,
\be
\tau_0\{t,\bar t\mid G\} = \tau^{KP}\{t \mid \tilde G(\bar t)\}; \nn \\
\tilde {\cal H}^{KP}_{ij} = \sum_m {\cal H}_{im}P_{m-j}(\bar t)\ \
{\rm and} \nn \\
\tilde T_{lj}\{\tilde G(\bar t)\} = \sum_m T_{lm}\{G\}P_{m-j}(\bar t).
\label{addkptau2}
\ee

KP reduction in its turn has many further weak reductions
(KdV and  Boussinesq being the simplest examples).

\subsection{Fermion correlator in Miwa coordinates}

Let us now return to original correlator (\ref{pretaf}) and
discuss in a little more details the implications of bosonization
identity (\ref{bosid}). In order not to write down integrals of $J$, we
introduce scalar field:\footnote{
One can consider $\phi $ as introduced for simplicity of notation,
but it should be kept in mind that the scalar-field
representation is in fact more fundamental for {\it generic} $\tau$-functions,
not related to the level $k=1$ Kac-Moody algebras (this phenomenon is well
known in conformal filed theory, see \cite{GMMOS} for more details).
}
\be
\phi(z) = \sum_{\stackrel{k\neq 0}{k\in Z-0}}\frac{J_{-k}}{k}z^k
+ \phi_0 + J_0\log z,
\ee
such that $\partial\phi(z) = J(z)$. Then (\ref{bosid}) states that:
\be
:\psi( \lambda)\tilde\psi(\tilde \lambda):\ = \
:e^{\phi(\tilde \lambda)-\phi( \lambda)}:
\label{bosid'}
\ee
"Normal ordering" here means nothing more but the requirement to neglect all
mutual contractions (or correlators) of operators in between $:\ \ :$ when
Wick theorem is applied to evaluate corrletion functions.
One can also get rid of the normal ordering sign at the l.h.s. of
(\ref{bosid'}), then
\be
\psi( \lambda)\tilde\psi(\tilde \lambda) = \
:e^{\phi(\tilde \lambda)}:\  :e^{-\phi( \lambda)}:
\label{bosid"'}
\ee
In distinguished coordinates on a sphere, when the free field propagator is
just $\log( z-\tilde z)$, one also has:
\be
\psi( z)\tilde\psi(\tilde z) =
\frac{1}{ z-\tilde z}\ :\psi( z)\tilde\psi(\tilde z):
\nn
\ee

Our task now is to express operators $e^H$ and $e^{\bar H}$ through the field
$\phi$. This is simple:
\be
H = \oint_0 U( z)J( z) = \oint_0 U( z)\partial\phi( z) =
- \oint_0 \phi( z)\partial U( z).
\ee
Here as usual $U( z) = \sum_{k>0} t_k z^k$ and integral is around
$ z = 0$. This is very similar to generic linear functional of
$\phi_-(\lambda) \equiv -\sum_{k>0}\frac{1}{k}J_k\lambda^{-k}$,
\be
H = \int \phi_-(\lambda)f(\lambda) d\lambda,
\label{hamneg}
\ee
one should only require that\footnote{
The factor  $2\pi i$ is included into
the definition of contour integral $\oint$.
}
\be
\partial U( z) = \oint \frac{f(\lambda)}{ z - \lambda}d\lambda,
\nn
\ee
i.e.
\be
U( z) = \oint\log\left(1-\frac{ z}{\lambda}\right) f(\lambda)d\lambda.
\ee
In terms of time-variables this means that
\be
t_k = -\frac{1}{k}\int \lambda^{-k}f(\lambda)d\lambda.
\label{miwatransint}
\ee
Here we required that $U( z = 0) = 0$, sometimes it can be more natural
to introduce also
\be
t_0 = \int \log \lambda\ f(\lambda)d\lambda.
\ee
This change from the time-variables to "time density" $f(\lambda)$ is known as
Miwa
transformation. In order to establish relation with fermionic representation
and also with matrix models we shall need it in "discretized" form:
\be
t_k  = \frac{\xi}{k}\left(\sum_{a } \lambda_a ^{-k} -
\sum_a  \tilde \lambda_a ^{-k} \right), \nn \\
t_0  = -\xi\left(\sum_a  \log \lambda_a   - \sum_a \log \tilde
\lambda_a \right).
\label{miwatrans}
\ee
We changed integral over $\lambda$ for a discrete sum
(i.e. the density function $f(\lambda)$ is
a combination of $\delta$-functions, picked at some points
$\lambda_a ,\ \tilde \lambda_a $. This is
of course just another basis in the space of the linear functionals, but the
change from one basis to another one is highly non-trivial. The thing is, that
we selected the basis where amplitudes of different $\delta$-functions are the
{\it same}: parameter $\xi$ in (\ref{miwatrans}) is {\it independent} of
$a $. Thus the real parameters are just positions of the points
$\lambda_a , \ \tilde \lambda_a $,
while the amplitude is defined by the density
of these points in the integration (summation) domain. This domain does
not need to be {\it a priori} specified:
it can be real line, any other contour or - better -
some  Riemann surface.)
Parameter $\xi$ is also unnecessary to introduce, because basises with
different $\xi$ are essentially equivalent. We shall soon put it equal to {\it
one}, but not before Miwa transformation will be discussed in a little more
detail.

Our next steps will be as follows.
Substitution of (\ref{hamneg}) into (\ref{miwatrans}), gives:
\be
H = -\xi\sum_a  \phi_-(\lambda_a ) +
\xi\sum_a \phi_-(\tilde\lambda_a ).
\ee
In fact, what we need is not the operator $H$ itself, but the state which is
created by $e^H$ from the vacuum state $\langle N\mid$. Then, since
$\langle N\mid J_m = 0$ for $m<0$, $\langle N\mid e^{-\xi\phi_-(\lambda)}$ is
essentially equivalent to $\langle N\mid e^{-\xi\phi(\lambda)}$ with
$\phi_-(\lambda)$ substituted by entire $\phi(\lambda)$. If $\xi = 1 $,
$e^{-\phi(\lambda)}$ can be further changed for $\psi(\lambda)$ and we
obtain an expression for the correlator (\ref{pretaf}) where
$e^H$ is substituted by a product of operators $\psi(\lambda_a )$. The
same is of course true for $e^{\bar H}$. Then Wick theorem can be applied and a
new type of determinant formulas arises like, for example,
\be
\tau \sim \frac{\Delta(\lambda,
\tilde\lambda)}{\Delta^2(\lambda)\Delta^2(\tilde\lambda)}
{\rm det}_{a  b} \langle N\mid \psi(\lambda_a )
\tilde\psi(\tilde\lambda_ b)\ G \ \mid N\rangle
\ee
It can be also obtained directly from (\ref{pretafdet}), (\ref{Hmatrcor}) and
(\ref{HversTlm}) by Miwa transformation.
The rest of this subsection describes this derivation in somewhat more
details.

The first task is to substitute $\phi_-$ by $\phi$. For this purpose we
introduce operator
\be
\sum_{k=-\infty}^{\infty} t_kJ_k = H_+ + H_-,
\ee
where $H_+ = \sum_{k>0} t_kJ_k$ is just our old $H$, $H_- = \sum_{k\geq 0}
t_{-k}J_k$, and "negative times" $t_{-k}$ are defined by "analytical
continuation" of the same formulas (\ref{miwatransint}) and (\ref{miwatrans}):
\be
t_{-k} = \frac{1}{k} \int \lambda^k f(\lambda)d\lambda =
-\frac{\xi}{k} \left(\sum_a  \lambda_a ^k - \sum_a
\tilde\lambda^k_a  \right).
\ee
Then
\be
\sum_{k=-\infty}^{\infty} t_kJ_k = H_+ + H_- =
-\xi\left(\sum_a  \phi(\lambda_a ) - \sum_a
\phi(\tilde\lambda_a ) \right).
\label{miwadetder1}
\ee
Further,
\be
e^{H_+ + H_-} = e^{-\frac{1}{2}s(t)} e^{H_+}e^{H_-} =
      e^{\frac{1}{2}s(t)} e^{H_-}e^{H_+},
\ee
where
\be
\new
\begin{array}{c}
s(t) \equiv \sum_{k>0} kt_kt_{-k} =
-\xi^2 \sum_{k>0}\frac{1}{k} \left(\sum_a
\left(\lambda_a ^{-k} - \tilde\lambda_a ^{-k}\right)
\sum_ b \left(\lambda_ b^k - \tilde\lambda_ b^k \right)\right) = \\
= \xi^2 \log \left(\left. \prod_{a , b}\right.^\prime\
\frac{(1-\frac{\lambda_ b}{\lambda_a })
(1-\frac{\tilde\lambda_ b}{\tilde\lambda_a })}
{(1-\frac{\tilde\lambda_ b}{\lambda_a })
(1-\frac{\lambda_ b}{\tilde\lambda_a })}\right) \ + \ {\rm const},
\end{array}
\ee
where prime means that the terms with $a  =  b$ are excluded from the
product in the numerator and accounted for in the infinite "constant", added
at the r.h.s. In other words,
\be
e^{\frac{1}{2}s(t)}  = {\rm const}\cdot
\left(
\frac{\prod_{a  >  b} (\lambda_a  - \lambda_ b)
(\tilde\lambda_a  - \tilde\lambda_ b)}
{\prod_a  \prod_ b (\lambda_a  - \tilde\lambda_ b)}
\right)^{\xi^2} =  \nn \\  =
{\rm const}\cdot
\left(
\frac{\Delta^2(\lambda) \Delta^2(\tilde\lambda)}{\Delta(\lambda,
\tilde\lambda)}
\right)^{\xi^2}.
\label{miwadetder2}
\ee
Since $\ \langle N\mid J_m = 0\ $  for all $m<0$, we have
$\langle N\mid e^{H_-} = \langle N\mid$, and therefore
\be
\langle N\mid e^H \equiv \langle N\mid e^{H_+} =
\langle N\mid e^{H_-}e^{H_+} =
e^{-\frac{1}{2}s(t)}\langle N\mid e^{H_+ + H_-}.
\ee
{}From eq.(\ref{miwadetder1}),

\be
e^{H_+ + H_-} = {\rm const}\cdot \prod_a  \ :e^{-\xi\phi(\lambda_a )}:
\ :e^{\xi\phi(\tilde\lambda_a )}:
\ee
where "const" is exactly the same as in (\ref{miwadetder2}).
If $\xi = 1$, eq.(\ref{bosid"'}) can be used to write:

\be
\langle N\mid e^H \ = \
\frac{\Delta(\lambda,\tilde\lambda)}
{\Delta^2(\lambda) \Delta^2(\tilde\lambda)}
\langle N \mid \prod_a  \psi(\lambda_a )
\prod_a  \tilde\psi(\tilde\lambda_a )
\ee
Similarly,

\be
e^{\bar H} \mid N \rangle = \prod_ b \psi(\bar\lambda_ b)
\prod_ b \tilde\psi(\tilde{\bar\lambda}_ b) \mid N \rangle
\frac{\Delta(\bar\lambda,\tilde{\bar\lambda})}
{\Delta^2(\bar\lambda) \Delta^2(\tilde{\bar\lambda})},
\ee

where

\be
\bar t_k = - \frac{1}{k} \sum_ b \left(\bar\lambda_ b^k -
                  \tilde{\bar\lambda}_ b^k\right)
\ee
and we used the fact that $J_m \mid N \rangle = 0 $ for all $m>0$.
Finaly,
\be
\new
\begin{array}{c}
\tau_N\{t, \bar t \mid G\} =
\langle N \mid e^H \ G \ e^{\bar H} \mid N \rangle =
\frac{\Delta(\lambda,\tilde\lambda)}
{\Delta^2(\lambda) \Delta^2(\tilde\lambda)}
\frac{\Delta(\bar\lambda,\tilde{\bar\lambda})}
{\Delta^2(\bar\lambda) \Delta^2(\tilde{\bar\lambda})}
\cdot   \\ \cdot
\langle N \mid  \prod_a  \psi(\lambda_a )
\prod_a  \tilde\psi(\tilde\lambda_a )\ G\
\prod_ b \psi(\bar\lambda_ b)
\prod_ b \tilde\psi(\tilde{\bar\lambda}_ b)
\mid N \rangle .
\end{array}
\label{tautodamiwdet}
\ee
Singularities at the coinciding points are completely eliminated from this
expression, since poles and zeroes of the correlator are canceled by those
coming from the Van-der-Monde determinants.

Let us now put $N=0$ and define normalized $\tau$-function
\be
\hat \tau_0\{t,\bar t \mid G\} \equiv
\frac{\tau_0\{t,\bar t\mid G\}}{\tau_0\{0,0\mid G\}},
\ee
i.e. divide r.h.s. of (\ref{tautodamiwdet}) by $\langle 0 \mid  G  \mid
0 \rangle$.
Wick theorem now allows to rewrite the correlator at the r.h.s.
as a determinant of the block matrix:
\be
\new
{\rm det}\left(\begin{array}{cc}
\frac{\langle 0 \mid \psi(\lambda_a )
\tilde\psi(\tilde\lambda_ b)\ G \
\mid 0 \rangle}{\langle 0 \mid \ G \ \mid 0 \rangle}
    \frac{\langle 0 \mid \psi(\lambda_a )\ G \
\tilde\psi(\tilde{\bar\lambda}_ b) \mid 0 \rangle}{\langle 0 \mid \ G \
\mid 0 \rangle} \\
-\frac{\langle 0 \mid \tilde\psi(\tilde\lambda_ b)\ G \
\psi(\bar\lambda_a ) \mid 0 \rangle}{\langle 0 \mid \ G \ \mid 0 \rangle}
    \frac{\langle 0 \mid \ G\ \psi(\bar\lambda_a )
\tilde\psi(\tilde{\bar\lambda}_ b) \mid 0 \rangle}{\langle 0 \mid \ G \
\mid 0 \rangle}
\end{array} \right)
\label{blockform}
\ee

Special choices of points $\lambda_a , \ldots,
\tilde{\bar\lambda}_ b$ can lead to simpler formulas.
If $\tilde{\bar\lambda}_a  \rightarrow {\bar\lambda}_a $, so that
$\bar t_k \rightarrow 0$, the matrix elements at the right lower block in
(\ref{blockform}) blow up, so that the off-diagonal blocks can be neglected.
Then
\be
\new
\begin{array}{c}
\tau_0\{t,\bar t \mid G\} \rightarrow \tau^{KP}\{t \mid G\} =
\frac{\langle 0 \mid e^H \ G \ \mid 0 \rangle}{\langle 0 \mid \ G \ \mid 0
\rangle} =   \\
= \frac{\Delta(\lambda,\tilde\lambda)}
{\Delta^2(\lambda) \Delta^2(\tilde\lambda)}
{\rm det}_{a  b} \frac{\langle 0 \mid   \psi(\lambda_a )
\tilde\psi(\tilde\lambda_ b)\ G\ \mid 0 \rangle}{\langle 0 \mid \ G \ \mid
0 \rangle} .
\end{array}
\label{kpdetrep}
\ee
This function no longer depends on $\bar t$-times and  is
just a KP $\tau$-function.

Matrix element
\be
\varphi(\lambda,\tilde\lambda) =
\frac{\langle 0 \mid \psi(\lambda)\tilde\psi(
\tilde\lambda)\  G\ \mid 0 \rangle}{\langle 0 \mid G \mid 0 \rangle}
\ee
is singular, when $\lambda \rightarrow \tilde\lambda$:
$\varphi(\lambda,\tilde\lambda) \rightarrow
\frac{1}{\lambda - \tilde\lambda}$.
If now in (\ref{kpdetrep}) all $\tilde\lambda \rightarrow \infty$,
\be
\tau^{KP}\{t \mid G\} =
\frac{{\rm det}_{a  b}\varphi_b(\lambda_a )}{\Delta(\lambda)},
\label{KPdetmain}
\ee
where
\be
\varphi_b(\lambda) \equiv \langle 0 \mid \psi(\lambda)
\left(\partial^{ b-1}\tilde\psi\right)(\infty)\
G \ \mid 0 \rangle \sim \lambda^{ b-1} \left(1 +
{\cal O}\left(\frac{1}{\lambda}\right)\right).
\ee
This is the main determinant representation of KP $\tau$-function in Miwa
parametrization.

Starting from representation (\ref{KPdetmain}) one can restore the
corresponding matrix ${\cal H}^{KP}_{ij}$ in eq.(\ref{addkptau1}) \cite{Toda}:
\be
{\cal H}^{KP}_{ij}\{t\} = \oint z^i\varphi_{-j}(z) e^{\sum_k t_kz^k} dz,
\label{addkptau3}
\ee
i.e.
\be
T_{lj}^{KP} = \oint z^l\varphi_{-j}(z).
\ee
Then obviously $\displaystyle{\frac{\partial {\cal H}^{KP}_{ij}}{\partial t_k}
= {\cal H}^{KP}_{i+k,j}}$. Now we need to prove that the $\tau$-function is
given at once by $\frac{{\rm det}\
\varphi_a (\lambda_\delta)}{\Delta(\lambda)}$ and ${\rm Det} {\cal
H}^{KP}_{ij}\{t\}$. In order to compare these two expressions one should take
$t_k = \frac{1}{k}\sum_a ^n \lambda_a ^{-k}$, so that
\be
\exp\left({\sum_{k>0} t_kz^k}\right) = \prod_{a  =1}^n
\frac{\lambda_a }{\lambda_a  - z} =
\left(\prod_a ^n \lambda_a \right)
\sum_a  \frac{(-)^a }{z-\lambda_a }
\frac{\Delta_a (\lambda)}{\Delta(\lambda)},
\ee
where
\be
\Delta_a (\lambda) = \prod_{\stackrel{\alpha>\beta}{\alpha,\beta\neq
a }}(\lambda_\alpha - \lambda_\beta) =
\frac{\Delta(\lambda)}{\prod_{\alpha\neq a }(\lambda_\alpha -
\lambda_a )},
\ee
and
\be
\left.{\cal H}^{KP}_{ij}\right|_
{t_k = \frac{1}{k}\sum_a ^n \lambda_a ^{-k}} =
\left(\prod_a ^n \lambda_a \right)
\sum_a  \frac{(-)^{a +1}\Delta_a (\lambda)}{\Delta(\lambda)}
\lambda_a ^i\varphi_{-j}(\lambda_a ).
\label{derivkpvertod1}
\ee
As far as $n$ is kept finite, determinant of the infinite-size matrix
(\ref{derivkpvertod1}),
$\displaystyle{\left.{\rm Det}_{i,j<0}{\cal H}^{KP}_{ij}\right|_{t_k =
\frac{1}{k}\sum_a ^n \lambda_a ^{-k}} = 0}$
since it is obvious from (\ref{derivkpvertod1}) that
the rank of the matrix is equal to $n$. Therefore let us consider the maximal
non-vanishing determinant,
\be
\new
\begin{array}{c}
\left.{\rm Det}_{-n\leq i,j<0}{\cal H}^{KP}_{ij}\right|_{t_k =
\frac{1}{k}\sum_a ^n \lambda_a ^{-k}} = \\ =
\left(\prod_a ^n \lambda_a \right)^n
{\rm det}_{ia } \left(
\frac{(-)^{a +1}\Delta_a (\lambda)}{\lambda_a ^i\Delta(\lambda)}
\right)
\cdot {\rm det}_{a  j}\varphi_{j}(\lambda_a )
=  \\ =
\frac{{\rm det}_{a  j}\varphi_{j}(\lambda_a )}{\Delta(\lambda)}.
\end{array}
\label{derivkpvertod2}
\ee
We used here the fact that determinant of a matrix is a product of
determinants and reversed the signs of $i$ and $j$. Also used were some
simple relations:

\be
\new
\begin{array}{lc}
 \prod_{a =1}^n \frac{\Delta_a (\lambda)}{\Delta(\lambda)} =
\frac{1}{\Delta^2(\lambda)}, \nn \\
 {\rm det}_{ia }\frac{1}{\lambda_a ^i} = \left(\prod_a ^n
\lambda_a \right)^{-1}\Delta(1/\lambda), \\
 \Delta(1/\lambda)  = \prod_{\alpha>\beta}\left(\frac{1}{\lambda_\alpha} -
\frac{1}{\lambda_\beta}\right) = (-)^{n(n-1)/2} \Delta(\lambda)
\left(\prod_a ^n\lambda_a \right)^{-(n-1)},  \\
{\rm thus}  \\
 \left(\prod_a ^n\lambda_a \right)(-)^{n(n-1)/2}
\prod_{a =1}^n \frac{\Delta_a (\lambda)}{\Delta(\lambda)}
{\rm det}_{ia }\frac{1}{\lambda_a ^i} = \frac{1}{\Delta(\lambda)}.
\end{array}
\ee
Since (\ref{derivkpvertod2}) is true for any $n$, one can claim that
in the limit $n \rightarrow \infty$ we recover the statement, that
$\tau^{KP}\{t\} = {\rm Det}_{i,j<0} {\cal H}_{ij}^{KP}$ with ${\cal
H}_{ij}^{KP}$ given by eq.(\ref{derivkpvertod1}) (that formula does
not refer directly to Miwa parametrization and is defined for any $t$
and any $j<0$ and $i$). This relation between $\varphi_a $'s and
${\cal H}_{ij}^{KP}$ can now be used to introduce negative times
$\bar t_k$ according to the rule (\ref{addkptau2}). Especially simple
is the prescription for zero-time: ${\cal H}_{ij} \rightarrow {\cal
H}_{i+N,j+N}$, when expressed in terms of $\varphi$ just implies that

\be
\frac{{\rm det}\ \varphi_a (\lambda_b )}{\Delta(\lambda)} \rightarrow
\frac{{\rm det}\ \varphi_{a  +N}(\lambda_b )}
{({\rm det}\Lambda)^N\Delta(\lambda)}.
\ee
Generalizations of  (\ref{addkptau3}), like

\be
{\cal H}_{ij}\{t,\bar t\} = \oint \oint z^i\bar z^j
\langle 0 \mid \psi(z) \ G\ \tilde\psi(\bar z)\mid 0 \rangle
e^{\sum_{k} (t_kz^k + \bar t_k\bar z^k)} dzd\bar z,
\ee
can be also considered.

\subsection{1-Matrix model versus Toda-chain hierarchy}

At the end of this section we use an explicit example of dicrete 1-matrix
model \cite{GMMMO} to demonstrate how a more familiar Lax description of
integrable hierarchies arises from determinant formulas.
Lax representation appears usually after some coordinate system is chosen
in the Grassmannian. In the example which we are now considering this
system is introduced by the use of orthogonal polinomials.

Formalism of orthogonal polinomials was intensively used at the early
days of the theory of matrix models. It is applicable to
scalar-product eigenvalue models (see \cite{UFNmm} for details about
this notion) and allows to further transform (diagonalize) the
remaining determinants into products. In variance with both reduction
from original $N^2$-fold matrix integrals to the eigenvalue problem,
which (when possible) reflects a physical phenomenon - decoupling of
angular (unitary-matrix) degrees of freedom (associated with
$d$-dimensional gauge bosons), - and with occurence of determinant
formulas which reflects integrability of the model,  orthogonal
polinomials appear more as a technical device. Essentially orthogonal
polinomials are necessary if wants to explicitly separate dependence
on the the size $N$ of the matrix in the matrix integral
("zero-time") from dependencies on all other time-variables and to
explicitly construct variables, which satisfy Toda-like equations.
However, modern description of integrable hierarchies in terms of
$\tau$-functions does not require explicit separation of the
zero-time and treats it more or less on the equal fooring with all
other variables, thus making the use of orthogonal polinomials
unnecessary. Still this technique remains in the arsenal of the
matrix model theory\footnote{ Of course, one can also use this link
just with the aim to put the rich and beautifull mathematical theory
of orthogonal polinomials into the general context of string theory.
Among interesting problems here is the matrix-model description of
$q$-orthogonal polinomials.  } and we now briefly explain what it is
about.

In the context of the theory of scalar-product matrix models orthogonal
polinomials naturally arise when one notes that after
partition functions appears in a simple determinantal form,

\be
Z_N = \frac{1}{N!} \prod_{k=1}^N \int d\mu_{h_k,\bar h_k}
 {\rm Det}_{ik} h_k^{i-1} {\rm Det}_{jk} \bar h_k^{j-1} =   \nn \\
= {\rm Det}_{ij} \int d\mu_{h,\bar h} h^{i-1}\bar h^{j-1} =
      {\rm Det}_{ij} \langle h^{i-1} \mid  \bar h^{j-1} \rangle,
\label{dimamodet}
\ee
(of which eq.(\ref{1mamoev}) is a simple example),
any linear change of basises $h^i \rightarrow Q_i(h) = \sum_kA_{ik}h_k,
\ \bar h^j \rightarrow \bar Q_j(\bar h) = \sum_l B_{jl}\bar h^l$ can
be easily performed and $Z \longrightarrow Z\cdot {\rm det} A\cdot
{\rm det} B$.  In particular, if $A$ and $B$ are triangular with
units at diagonals, their determinants are just unities and $Z$ does
not change at all. This freedom is, however, enough, to diagonalize
the scalar product and choose polinomials $Q_i$ and $\bar Q_j$ so
that

\be
\langle Q_i(h) \mid \bar Q_j(\bar h) \rangle = e^{\phi_i}\delta_{ij}.
\label{orthopol}\label{orthocond}
\ee
$Q_i$ and $\bar Q_j$ defined in this way up to normalization are called
orthogonal polinomials. (Note that $\bar Q$ does not need to be a {\it
complex} conjugate of $Q$: "bar" does not mean complex conjugation.)
Because of above restriction on the form of matrices
$A$ and $B$ these polinomials are normalized so that

\be
Q_i(h) = h^i + \ldots;\ \ \ \bar Q_j(\bar h) = \bar h^j + \ldots
\nn
\ee
i.e. the leading power enters with the $unit$ coefficient. From
(\ref{dimamodet}) and (\ref{orthopol}) it follows that

\be
Z_N = {\rm Det}_{o<i,j\leq N} \langle h^i|\bar h^j\rangle =
\prod_{i=1}^N e^{\phi_{i-1}}.
\label{ZNprod}
\ee
This formula is essentially the main outcome of orthogonal
polinomials theory fro matrix models: it provides complete separation
of the $N$-dependence of $Z$ (on the size of the matrix) from that on
all other parameters (which specify the shape of potential, i.e. the
measure $d\mu_{h,\bar h}$), this information is encoded in a rather
complicated fashion in $\phi_i$.  As was already mentioned, any
feature of matrix model can be examined already at the level of
eq.(\ref{dimamodet}), which does not refer to orthogonal polinomials
and thus they are not really relevant for the subject.

Consider now the case of the {\it local} measure, $d\mu_{h,\bar h} =
d\mu_h\delta(h,\bar h)$, when $\bar Q_i = Q_i$.  The local measure is
distinguished by the property that multiplication by (any function
of) $h$ is Hermitean operator:

\be
\langle hf(h)\mid g(\bar h)\rangle =
\langle f(h) \mid \bar h g(\bar h)\rangle, \ \ {\rm if} \
d\mu_{h,\bar h} \sim \delta(h-\bar h).
\ee
This implies further that the coefficients $c_{ij}$ in the recurrent
relation
\be
hQ_i(h) = Q_{i+1}(h) + \sum_{j=0}^i c_{ij}Q_j(h)
\ee
are almost all vanishing. Indeed: for $j<i$
\be
\new
\begin{array}{c}
c_{ij} = \frac{\langle hQ_i(h) \mid Q_j(\bar h)\rangle}{\langle Q_j(h) \mid
Q_j(\bar h)\rangle} =
\frac{\langle Q_i(h) \mid \bar hQ_j(\bar h)\rangle}{\langle Q_j(h) \mid
Q_j(\bar h)\rangle} = \\ =
\delta_{i,j+1}\frac{\langle Q_i(h) \mid Q_i(\bar h)\rangle}{\langle Q_j(h)
\mid Q_j(\bar h)\rangle} =
\delta_{j,i-1} e^{\phi_i -\phi_{i-1}}.
\end{array}
\ee
In other words, polinomials, orthogonal w.r.to a local measure are
obliged to satisfy the "3-term recurrent relation":

\be
hQ_n(h) = Q_{n+1}(h) + c_nQ_n(h) + R_nQ_{n-1}(h)
\label{3termrel}
\ee
(the coefficient in front of $Q_{n+1}$ can be of course changed by
the change of normalization). Parameter $c_n$ vanishes if the measure
is even (symmetric under the change $h \rightarrow -h$), then
polinomials are split into two orthogonal subsets: even and odd in
$h$. Partition function (\ref{ZNprod}) of the $one$-component model
can be expressed through parameters $R_i =  e^{\phi_i -\phi_{i-1}}$
of the 3-term relation:

\be
Z_N = Z_1 \prod_{i=1}^{N-1} R_i^{N-i},
\ee
thus defining a one-component matrix model (i.e. particular shape of
potential), associated with any system of orthogonal polinomials.

Coming back to the 1-matrix model (\ref{1mamoev}), one can say that
all the information is contained in the determinant formula
(\ref{todatau}) together
with the rule (\ref{todaeqforH}), which defines time-dependence of
${\cal H}_{ij}^f = \langle h^i\mid h^j\rangle  = \hat{\cal H}_{i+j}^f$:
\be
\new
\begin{array}{c}
\frac{\partial {\cal H}_{ij}^f}{\partial t_k} =
 {\cal H}_{i+k,j}^f = {\cal H}_{i,j+k}^f ,\ \ {\rm or} \\
\frac{\partial \hat{\cal H}_i^f}{\partial t_k} =
 \hat{\cal H}_{i+k}^f.
\end{array}
\ee
The possibility to express everything in terms of ${\cal H}_i^f$ with a
single matrix index $i$ is the feature of Toda-chain reduction of generic
Toda-lattice hierarchy.

However, in order to reveal the standard Lax representation we need to go
into somewhat more involved considerations. Namely, we consider
representation of two operators in the basis of orthogonal polinomials.
First,
\be
h^k Q_n(h) = \sum_{m=0}^{n+k}
\frac{\langle n \mid h^k \mid m \rangle}{\langle m \mid m \rangle}
Q_m(h) =
\sum_{m=0}^{n+k} \gamma_{nm}^{(k)}Q_m(h)
\ee
(here the simplified notation is introduced for
$\langle n \mid f(h) \mid m \rangle \equiv
\langle Q_n \mid f(h) \mid Q_m \rangle$ and
$\displaystyle{\gamma_{nm}^{(k)} \equiv
\frac{\langle n \mid h^k \mid m \rangle}{\langle m \mid m \rangle}}$.)
Second,
\be
\new
\begin{array}{c}
\frac{\partial Q_n(h)}{\partial t_k} =
-\sum_{m=0}^{n-1}
\frac{\langle n \mid h^k \mid m \rangle}{\langle m \mid m \rangle}
Q_m(h) =
-\sum_{m=0}^{n-1}\gamma_{nm}^{(k)}Q_m(h), \\
\frac{\partial \phi_n}{\partial t_k} =
\frac{\langle n \mid h^k \mid n \rangle}{\langle n \mid n \rangle}
= \gamma_{nn}^{(k)}.
\end{array}
\ee
(These last relations arise from differentiation of orthogonality
condition (\ref{orthocond}):
\be
\new
\begin{array}{c}
e^{\phi_n}\frac{\partial \phi_n}{\partial t_k}\delta_{nm} =
\frac{\partial \langle Q_n \mid Q_m \rangle}{\partial t_k} =
\nn \\
= \langle \frac{\partial Q_n}{\partial t_k} \mid Q_m \rangle +
\langle Q_n \mid \frac{\partial Q_m}{\partial t_k} \rangle +
\langle Q_n\mid h^k \mid Q_m \rangle
\end{array}
\ee
by looking at the cases of $m<n$ and $m=n$ respectively.)

{}From these relations one immediately derives the Lax-like formula:
\be
\frac{\partial \gamma_{nm}^{(k)}}{\partial t_q} =
- \sum_{l=m-k}^{n-1} \gamma_{nl}^{(q)}\gamma_{lm}^{(k)} +
\sum_{l=m+1}^{n+k} \gamma_{nl}^{(k)}\gamma_{lm}^{(q)}
\label{laxrep1mamo}
\ee
or, in a matrix form,
\be
\frac{\partial \gamma^{(k)}}{\partial t_q}
= [ R\gamma^{(q)}, \gamma^{(k)}],
\ee
where

\be
R\gamma_{mn}^{(k)} \equiv \left\{
\begin{array}{c}
-\gamma_{mn}^{(k)} \ \ {\rm if} \ m>n, \\
\gamma_{mn}^{(k)} \ \ {\rm if} \ m<n
\end{array}
\right.
\ee
(We remind that usually $R$-matrix acts on a function
$f(h) = \sum_{n = -\infty}^{+\infty}f_nh^n$ according to the rule:
$Rf(h) = \sum_{n\geq l}f_nh^n - \sum_{n<l}f_nh^n$ with some "level" $l$.)
These $\gamma^{(k)}$ are not symmetric matrices, but one can also
rewrite all the formulas above in terms of symmetric ones:

\be
{\cal L}_{mn}^{(k)} \equiv e^{\frac{1}{2}(\phi_n - \phi_m)}
\gamma_{mn}^{(k)}
= \frac{\langle m \mid h^k \mid n \rangle}{\sqrt{\langle m \mid m \rangle
\langle n \mid n \rangle}}
\ee

{}From eqs.(\ref{laxrep1mamo}) one can easily deduce Toda-equations for
$\phi_n$:
\be
\new
\begin{array}{c}
\frac{\partial^2\phi_n}
{\partial t_k\partial t_l} = \frac{\partial}{\partial t_k}
\frac{\langle n \mid h^l \mid n \rangle}{\langle n \mid n \rangle}
=  \\
= \left( \sum_{m>n} - \sum_{m<n}\right)
\frac{\langle n \mid h^k \mid m \rangle \langle m \mid h^l \mid n \rangle}
{\langle m \mid m \rangle\langle n \mid n \rangle},
\end{array}
\ee
where the r.h.s. can be expressed in terms of $R_m = e^{\phi_m - \phi_{m-1}}$.
In particular,
\be
\frac{\partial^2\phi_n}{\partial t_1\partial t_1} =
R_{n+1} - R_n = e^{\phi_{n+1} - \phi_n} - e^{\phi_n - \phi_{n-1}}.
\ee

Let  us also mention that in this formalism the Ward identities (Virasoro
constraints) follow essentially from the relation
\be
\left( \frac{\partial}{\partial h}\right)^\dagger =
- \frac{\partial}{\partial h} -
\sum_{k>0} kt_k h^{k-1},
\ee
where Hermitean conjugation is w.r.to the scalar product $\langle\ \mid\
\rangle$. For example, this relation implies, that
\be
\langle Q_n \mid \frac{\partial Q_n}{\partial h} \rangle =
- \langle \frac{\partial Q_n}{\partial h} \mid Q_n \rangle
- \sum_{k>0} kt_k \langle Q_n \mid h^{k-1} \mid Q_n \rangle.
\ee
Now we note that $\frac{\partial Q_n}{\partial h}$ is a polinomial of
degree $n-1$, thus
${\langle Q_n \mid \frac{\partial Q_n}{\partial
h} \rangle = 0}$. (In fact $$\displaystyle{
\frac{\partial Q_n}{\partial h} = -\sum_{k>0} kt_k
\left(\sum_{m=0}^{n-1} \gamma_{nm}^{(k-1)}Q_m \right) =
-\sum_{k>0} kt_k \frac{\partial Q_n}{\partial t_{k-1}}}.)$$
Also we recall that $\langle Q_n \mid h^{k-1} \mid Q_n \rangle =
\langle Q_n \mid Q_n \rangle \frac{\partial \phi_n}{\partial t_{k-1}}$,
and obtain:
\be
\sum_{k>0} kt_k\frac{\partial \phi_n}{\partial t_{k-1}} = 0
\ee
for any $n$. This should be supplemented by relation $\frac{\partial
\phi_n}{\partial t_0} = \phi_n$. In order to get the lowest Virasoro
constraint (string equation), $L_{-1}Z_N = 0$ or $L_{-1}\log Z_N = 0$
it is enough just to sum over $n$ from $0$ to $N-1$.

For more details about 1-matrix model, Toda-chain hierarchy and
application of the formalism of orthogonal polinomials in this
context see \cite{GMMMO}.

\section{$\tau$-function as a group-theoretical quantity \label{se5}}
\setcounter{equation}0

This section contains some remarks about the general notion of
$\tau$-funcion on the lines suggested in ref.\cite{UFNmm}.
Examples below are taken from \cite{GKLMM} and \cite{KMM}.

As mentioned in the beginning of the previous section we define
the (generalized) $\tau$-function as
the generating functional of all the matrix elements of a given
group element $g \in G$ in a given representation $R$:

\be
\tau_R (t,\bar t|g) \equiv \sum_{\{m,\bar m\}\in R}
s^{R}_{m,\bar m}(t,\bar t) <m |g| \bar m>
\label{taugen}
\ee
The choice of functions $s^{R}_{m,\bar m}(t,\bar t)$
is the main ambiguity in the definition of $\tau$-function
and needs to be fixed in some clever way, not yet known
in full generality. The only {\it a priori} requirement is that
it is indeed a generating functional, i.e. there should be some
($g$-independent) operators ${\cal M}$, acting on $t,\bar t$-variables,
which  allow to extract all particular matrix elements
once $\tau_R(t,\bar t)$ is known:
$$
<m |g| \bar m> = \ll {\cal M}_{R,m,\bar m}(t,\bar t)|
\tau_R (t,\bar t|g) \gg
$$

The ambiguity in the choice of $s^{R}_{m,\bar m}(t,\bar t)$
can be partly fixed (at least in the case of the highest weight
representations $R$) by the requirement that
\be
\tau_R (t,\bar t|g)  =
<{\rm vac}_R| U(t) g \bar U(\bar t)| {\rm vac}_R >
\label{tau}  \label{taugen2}
\ee
where operators $U$ and $\bar U$ do not depend on $R$.
In order to be even more specific one can further request that
evolution operators are group elements, i.e.
\be
\Delta U(t) = U(t) \otimes U(t) =
\left(U(t)\otimes I\right) \left(I\otimes U(t)\right)
\label{clacoU}
\ee
where $\Delta$ denotes group comultiplication law. In the
case of Lie algebras $\Delta(T_{{\vec\alpha}}) = T_{{\vec\alpha}}\otimes I
+ I\otimes T_{{\vec\alpha}}$, and (\ref{clacoU}) is true at least
for the evolution operators in KP/Toda systems. Later we shall
see that in the case of quantum groups  it can be natural to slightly
modify the condition (\ref{clacoU}).

Remarkably, the $\tau$-function defined in (\ref{taugen}),
always satisfies a family of non-linear equations \cite{GKLMM}, relating
$\tau_R$ with different $R$'s, which reflect just the fact
that matrix elements of the {\it same} group element in
different representations are not independent. Conventional
bilinear Hirota equation for KP/Toda $\tau$-functions is
nothing but particular case of this generic construction\footnote{
Note, that it is somewhat different from approach, advocated
by V.Kac \cite{Kac} (see also \cite{UFNmm}), which makes
use of Casimir operators and is less universal than the one
to be described below (using intertwining operators).
},
which has two (essentially identical) interpretations: in terms
of fundamental representations of $GL(\infty)$ and in
terms of the level $k=1$ Kac-Moody $\hat{U(1)}$ algebra.

\subsection{From intertwining operators to bilinear equations\label{bieq}}

The following construction \cite{GKLMM} in terms of intertwining
operators is the general source of bilinear equations for the
$\tau$-function (\ref{taugen}). One can easily recognize the standard
free-fermion derivation of Hirota equations for KP/Toda
$\tau$-functions as a particular example (with $G$ being the level
$k=1$ Kac-Moody algebras $\hat G_{k=1}$, $V$ a fundamental
representation, and $W$ - the simplest fundamental representation
corresponding to the very left root of the Dynkin diagram).
Construction below involves a lot of arbitrariness. In order to make
the consideration more transparent, we formulate our construction
explicitly for finite-dimensional Lie algebras and their
$q$-counterparts.

Bilinear equations which we are going to derive are relating
$\tau$-functions (\ref{taugen}) for four different Verma modules $V,\
\hat V,\ V', \hat V'$. Given $V,\ V'$, every allowed choice of $\hat
V,\ \hat V'$ provides a separate set of bilinear identities. Of
course, not all of these sets are actually independent and can be
parametrized by source modules $V$ and $V'$ and by a weight of
finite-dimensional representation.  Also different choices of
positive root systems and their ordering in (\ref{tau}) provides
equations in somewhat different forms.  A more invariant description
of the minimal set of bilinear equations for given $G$ would be
clearly interesting to find.

1. Our starting point is embedding of Verma module $\hat V$ into the
tensor product $V\otimes W$, where $W$ is some irreducible
finite-dimensional representation of $G$ (in the case of Kac-Moody
algebra evaluation representation should be used).  Once $V$ and $W$
are specified, there is only finite number of choices for $\hat V$.

Now we define right vertex operator of the $W$-type
as homomorphism of $G$-modules:

\be\label{inttw}
E_R: \ \ \hat V \longrightarrow V\otimes W.
\ee

This intertwining operator can be explicitly continued to the whole
representation once it is constructed for the vacuum
(highest-weight) state:

\be\label{vacuum}
\hat V = \left\{ | {\bf n_{\bf\alpha}} \rangle_{\hat V} =
 \prod_{{\bf\alpha}>0}
 \left(\Delta (T_{-{\bf\alpha}}\right)^{n_{\bf\alpha}}
 | {\bf 0} \rangle_{\hat V} \right\},
\ee
where comultiplication $\Delta$ provides the action of $G$ on
the tensor product of representations, and

\be\label{vacuum2}
|{\bf 0} \rangle_{\hat V} =
\left( \sum_{\{p_{\bf\alpha},i_{\bf\alpha}\}}
 A\{p_{\bf\alpha},i_{\bf\alpha}\}
 \left(\prod_{{\bf\alpha}>0}
 \left. (T_{-{\bf\alpha}}\right)^{p_{\bf\alpha}}\otimes
 \left. (T_{-{\bf\alpha}}\right)^{i_{\bf\alpha}}\right) \right)
| {\bf 0} \rangle_V \otimes | {\bf 0} \rangle_W.
\ee
For finite-dimensional $W$'s, this gives every
$| {\bf n_{\bf\alpha}} \rangle_{\hat V}$ in a form of {\it finite}
sums of states $| {\bf m_{\bf\alpha}} \rangle_{V}$ with
coefficients, taking values in elements of $W$.

2. The next step is to take another triple, defining a left vertex
operator,

\be\label{inttw2}
\bar E'_L: \ \ \hat V' \longrightarrow W^* \otimes V',
\ee
Note the change of ordering at the r.h.s., this is different from
$V'\otimes W^*$ in the case of quantum groups.  The product $W\otimes
W^*$ of the module $W$ and its conjugate contains {\it unit}
representation of ${\cal G}$.  The projection to this unit
representation

\be\label{proj}
\pi: \ \ W\otimes W^* \longrightarrow I
\ee
is explicitly provided by multiplication of any element
of $W\otimes W^*$ by
\be\label{explproj}
\pi = \left._W \langle {\bf 0} | \otimes \left._{W^*} \langle {\bf 0} |
\left( \sum_{\{i_{\bf\alpha},i'_{\bf\alpha}\}}
\pi\{i_{\bf\alpha},i'_{\bf\alpha}\}
 \left(\prod_{{\bf\alpha}>0}
 \left. (T_{+{\bf\alpha}}\right)^{i_{\bf\alpha}}\otimes
 \left. (T_{+{\bf\alpha}}\right)^{i'_{\bf\alpha}}\right) \right)
\right.\right.
\ee
Using this projection, if it is not occasionally orthogonal to the
image of $E\otimes E'$, one can build a new intertwining
operator
\be\label{Gamma}
\Gamma: \ \
\hat V \otimes \hat V' \stackrel{E\otimes E'}{\longrightarrow}
V \otimes W \otimes W^* \otimes V'
\stackrel{I\otimes \pi \otimes I}{\longrightarrow} V \otimes V',
\ee
which possesses the property
\be\label{CRGamma}
\Gamma (g\otimes g) = (g\otimes g) \Gamma
\label{Ggg=ggG}
\ee
for any group element $g$ such that
\be
\Delta(g)=g\otimes g.
\ee

3. It now remains to take a matrix element of (\ref{Ggg=ggG})
between four states,
\be\label{CRGamma2}
\left._{V'} \langle k' | \left._V \langle k |
(g\otimes g) \Gamma | n \rangle_{\hat V}
 |n' \rangle_{\hat V'} \right.\right. =
\left._{V'} \langle k' | \left._V \langle k |
\Gamma (g\otimes g) | n \rangle_{\hat V}
|n' \rangle_{\hat V'} \right.\right.
\ee
and rewrite this identity in terms of generating functions
(\ref{taugen}).

\subsection{The case of KP/Toda $\tau$-functions  \label{funrept}}

We do not present here the standard derivation of Hirota equations in
the free-fermion formalism, because it is both well known and easily
recognizable in the general picture from the previous subsection.
Instead we describe here its slight variation -- starting from the
fundamental representations of $SL(\infty)$.  The reason why this
case is the closest one to the standard integrable hierarchies is
that, in variance with generic Verma modules for group $G \neq
SL(2)$, the fundamental representations are generated by subset of
the {\it mutually commuting} operators, not by entire set of
generators from maximal nilpotent subalgebra. We describe the basic
construction for $G = SL(n)$, since in this case the (finite)
Grassmannian construction is the most similar to the conventional
infinite-dimensional ($G = \hat{U(1)}$) situation.

The Lie algebra $SL(n)$ is generated by operators $T_{\pm{\vec\alpha}}$
and Cartan operators $H_{\vec\beta}$, such that
$[H_{\vec\beta}, T_{\pm{\vec\alpha}}] =
\pm{1\over 2}({{\vec\alpha}{\vec\beta}})
T_{\pm{\vec\alpha}}$. All elements of all representations are
eigenfunctions of $H_{\vec\beta}$, $H_{\vec\beta}|{\vec\lambda}\rangle
= {1\over 2}({\vec\beta{\vec\lambda}})|{\vec\lambda}\rangle$.
The highest weight of representation $F^{(k)}$ is ${{\vec\mu}}_k$.
Vectors ${{\vec\mu}}_k$'s are ``dual'' to the {\it simple} roots
${{\vec\alpha}}_i$, $i = 1,\ldots,r$: $({{\vec\mu}}_i{{\vec\alpha}}_j) =
\delta_{ij}$, and ${{\vec\rho}} = \frac{1}{2}\sum_{\vec\alpha >0}
{\vec\alpha} = \sum_i {{\vec\mu}}_i$.

There are as many as $r \equiv rank\ G = n-1$ fundamental
representations of $SL(n)$.
Let us begin with the simplest fundamental representation
$F \equiv F^{(1)}$ -- the $n$-plet, which consists of the states
\be
\psi_i = T_{-(i-1)}\ldots T_{-2}T_{-1}\psi_1, \ \ \ i = 1,\ldots,n.
\label{psii}
\ee
Moreover
\be
T_{-i}\psi_j = \delta_{ij}\psi_{i+1},
\label{psiii}
\ee
and the weights are given by

\be
{{\vec\lambda}}(\psi_i) = {{\vec\mu}}_1 - {{\vec\alpha}_1} - \ldots -
{{\vec\alpha}}_{i-1},
\ee
where ${{\vec\mu}}_1$ is the highest weight of $F^{(1)}$.  Here
$T_{\pm i} \equiv T_{\pm{\vec\alpha}_i}$ are generators, associated
with the simple roots. Let us denote the corresponding basis in
Cartan algebra $H_i = H_{{\vec\alpha}_i}$, and
${H_i}|{{\vec\lambda}}\rangle = {1\over
2}({{{\vec\alpha}}_i{{\vec\lambda}}}) |{{\vec\lambda}}\rangle =
\lambda_i|{{\vec\lambda}}\rangle$.  Then

\be\label{lambda}
\lambda_i^{(j)} \equiv \lambda_i(\psi_j) = \frac{1}{2}(\delta_{ij} -
\delta_{i,j-1}).
\ee
This formula, together with (\ref{psii}) and defining commutation
relations between the positive and negative simple-root generators
implies that $||\psi_i||^2 = 1$, and, with the help of the classical
comultiplication formula, $\Delta(T) = T\otimes I + I\otimes T$, one
immediately obeserves that the antisymmetric combinations
$\psi_{[1}\ldots \psi_{k]}$ are all the highest weight vectors (i.e
are annihilated by all $\Delta_k(T_{+i})$ and, thus by all the
$\Delta_k(T_{+{\vec\alpha}})$). These combinations are the highest
vectors of all the other fundamental representations $F^{(k)}$, which
are thus skew powers of $F = F^{(1)}$:

\be\label{frep}
F^{(k)} = \left\{ \Psi^{(k)}_{i_1\ldots i_k} \sim
 \psi_{[i_1}\ldots \psi_{i_k]} \right\}
\ee
{}From this description it is clear that $0 \leq k \leq n$,
moreover $F^{(0)}$ and $F^{(n)}$ are respectively the singlet and
dual singlet representations.

According to (\ref{psiii}) one can also describe all the states
of $F^{(1)}$ in terms of a single generator $T_-$, which is a sum of
those for all the $r$ {\it simple} roots of $G$,
$T_- = \sum_{i=1}^r T_{-{\vec\alpha}_i}$:
\be\label{simplfrep}
\psi_i = T_-^{i-1}| 0 \rangle_F, \ \ i = 1,\ldots,n.
\ee
Looking at explicit form (\ref{frep}) of the states in $F^{(k)}$ it is
easy to see that they can be all generated from the highest weight
one, $\Psi_{12\ldots k}$ by the operators
\be\label{coprodFR}
R_k(T_-^i) \equiv T_-^i\otimes I \otimes \ldots \otimes I +
I\otimes T_-^i \otimes \ldots \otimes I +
I\otimes I \otimes \ldots \otimes T_-^i.
\ee
These operators obviously commute with each other. For
given $k$ exactly $k$ of them (with $i = 1,\ldots,k$) are
independent. However, they are neither (linear combinations of)
the generators of Lie algebra acting in $F^{(k)}$, nor even their
algebraic functions (note that
$R_k (T_-^i) \neq (\left(R_k(T_-)\right)^i$). If one wants to
make clear that $F^{(k)}$ is indeed a representation of $G$, it is
better to say that it is generated by another set of operators,

\be
\Delta^{k-1}({\cal T}_i), \ \ \ \
{\cal T}_i \equiv \sum_{{{\vec\alpha}}:
{h({{\vec\alpha}}) = i}} T_{-{\vec\alpha}}
\label{calTi}
\ee
The ``height'' $h({{\vec\alpha}})$ is the number of items in linear
decomposition of the root ${{\vec\alpha}}$ in simple roors
${{\vec\alpha}}_i$ (in $F^{(1)}$ where all the generators
$T_{{\vec\alpha}}$ are represented by $n\times n$ matrices ${\cal
T}_i$ are matrices with all zero entries except for units at the
$i$-th subdiagonal). In particular, $T_- = {\cal T}_1$. Operators
(\ref{calTi}) are obviously generators of $G$, instead their mutual
commutativity is somewhat less transparent.  Since both sets
(\ref{calTi}) and (\ref{coprodFR}) generate $F^{(k)}$ it is a matter
of convenience which of them is used in particular considerations. In
dealing with KP/Toda hierarchies the explicitly commuting set
(\ref{coprodFR}) is more convenient.  It is exactly the lack of such
equivalence of two sets which makes consideration of KP/Toda
hierarchies more subtle in the quantum ($q\neq 1$) case, see
s.\ref{quaKP} below.

The intertwining operators which are of interest for us are

\be\label{inttwfrep}
I_{(k)}:\ \ F^{(k+1)} \longrightarrow F^{(k)} \otimes F, \nn \\
I^*_{(k)}:\ \ F^{(k-1)} \longrightarrow F^*\otimes F^{(k)},
\ \ {\rm and} \nn \\
\Gamma_{k|k'}:\ \ F^{(k+1)}\otimes F^{(k'-1)} \longrightarrow
F^{(k)} \otimes F^{(k')}.
\ee
Here

\be\label{frep2}
F^* = F^{(r)} = \left\{\psi^i \sim \epsilon^{ii_1\ldots i_r}
  \psi_{[i_1}\ldots \psi_{i_r]} \right\}, \nn \\
I_{(k)}:\ \ \Psi^{(k+1)}_{i_1\ldots i_{k+1}} =
            \Psi^{(k)}_{[i_1\ldots i_k}\psi^{\phantom{fgh}}_{i_{k+1}]}, \nn \\
I^*_{(k)}:\ \ \Psi^{(k-1)}_{i_1\ldots i_{k-1}} =
            \Psi^{(k)}_{i_1\ldots i_{k-1}i}\psi^i,
\ee
and $\Gamma_{k|k'}$ is constructed with the help of embedding
$I \longrightarrow F \otimes F^*$, induced by the pairing
$\psi_i \psi^i$: the basis in linear space $F^{(k+1)}\otimes
F^{(k'-1)}$, induced by $\Gamma_{k|k'}$ from that in
$F^{(k)}\otimes F^{(k')}$ is:

\be\label{prodfrep}
\Psi^{(k)}_{[i_1\ldots i_k}\Psi^{(k')}_{i_{k+1}]i'_1\ldots i'_{k'-1}}.
\ee
Operation $\Gamma$ can be now rewritten in terms of matrix elements

\be\label{gkdet}
g^{(k)}\left({{i_1\ldots i_k}\atop{j_1\ldots j_k}}\right) \equiv
\langle \Psi_{i_1\ldots i_k} | g | \Psi_{j_1\ldots j_k} \rangle=
\det_{1\leq a,b\leq k} g^{i_a}_{j_b}
\ee
as follows:

\be\label{gkgk}
g^{(k)}\left({{i_1\ldots i_k}\atop{[j_1\ldots j_k}}\right)
g^{(k')}\left({{i'_1\ldots i'_k}\atop
{j_{k+1}]j'_1\ldots j'_{k'-1}}}\right) = \nn \\ =
g^{(k+1)}\left({{i_1\ldots i_k[i'_{k'}}
\atop{j_1\ldots j_{k+1}}}\right)
g^{(k'-1)}\left({{i'_1\ldots i'_{k'-1}]}\atop
{j'_1\ldots j'_{k'-1}}}\right)
\ee
This is the explicit expression for eq.(\ref{CRGamma}) in the case of
fundamental representations, and it is certainly identically true for any
$g^{(k)}$ of the form (\ref{gkdet}).\footnote{
To see this directly it is enough to rewrite the l.h.s. of
(\ref{gkgk}) as
$$
g^{(k)}\left({{i_1\ldots i_k}\atop{[j_1\ldots j_k}}\right)
g^{[i_k'}_{j_{k+1}]}
g^{(k'-1)}\left({{i'_1\ldots i'_{k-1}]}\atop{j'_1\ldots j'_{k-1}}}\right)
$$
(expansion of the determinant $g^{(k')}$ in the first column) and now
 the first two factors can be composed
into $g^{(k+1)}$ (expansion of the determinant $g^{(k+1)}$ in the first
row), thus giving the r.h.s. of (\ref{gkgk}).}

Let us note that one can use the minors (\ref{gkdet})  to construct
local coordinates in the Grassmannian. Bilinear Plucker
relations satisfied by these coordinates are nothing but defining
equations of the Grassmannian consisting of all the $k$-dimensional
vector subspaces of $n$-dimensional vector space. Parametrizing
determinants (\ref{gkdet}) by time variables (see (\ref{H})), one gets
a set of bilinear differential equations on the generating function
of these Plucker coordinates, which is just a $\tau$-function \cite{Ohta}.

Now let us introduce time-variables and rewrite (\ref{gkgk}) in terms
of $\tau$-functions. We shall denote time variables through
$s_i, \bar s_i$, $i = 1,\ldots,r$ in order to emphasize their
difference from generic $t_{\vec\alpha}, \bar t_{\vec\alpha}$
labeled by all the positive roots ${\vec\alpha}$ of $G$. Note that in
order to have a closed system of equations we need to introduce all the
$r$ times $s_i$ for all $F^{(k)}$ (though $\tau^{(k)}$ actually depends
only on $k$ independent combinations of these).

Since the highest weight of representation $F^{(k)}$ is
identified as
\be
| 0 \rangle_{F^{(k)}} = |\Psi^{(k)}_{1\ldots k} \rangle,
\ee
we have:

\be\label{tau-k}
\tau^{(k)}(t,\bar t\ |\ g) =
\langle \Psi^{(k)}_{1\ldots k} |
\exp \left(\sum_i t_i R_k(T_+^i)\right)\ g \
\exp\left(\sum_i \bar t_iR_k(T_-^i)\right) |
\Psi^{(k)}_{1\ldots k} \rangle .
\ee

Now,

\be\label{70}
\exp\left(\sum_i t_i R_k(T^i)\right) =
\exp\left(R_k\left(\sum_i t_iT^i\right)\right) = \nn \\
= \left( \exp\left(\sum_i t_i T^i\right)\right)^{\otimes k} =
\left( \sum_j P_j(t)T^j\right)^{\otimes k},
\ee
where we used the definition of Schur polynomials

\be\label{Schur}
\exp\left(\sum_i t_iz^i\right) = \sum_j P_j(t)z^j.
\ee
Essential property of Shur polynomials is that

\be\label{SS}
\frac{\partial}{\partial t_i} P_j(t) =
(\frac{\partial}{\partial t_1})^i P_j(t) = P_{j-i}(t).
\ee
Because of (\ref{70}), we can rewrite the r.h.s. of (\ref{tau-k}) as

\be\label{detrep}
\tau^{(k)}(t,\bar t\ |\ g) =
\nn \\
= \sum_{{i_1,\ldots,i_k}\atop{j_1,\ldots,j_k}}
P_{i_1}(t)\ldots P_{i_k}(t) \langle \Psi^{(k)}_{1+i_1,2+i_2,\ldots,k+i_k}
|\ g\ | \Psi^{(k)}_{1+j_1,2+j_2,\ldots,k+j_k} \rangle
P_{j_1}(\bar t)\ldots P_{j_k}(\bar t) = \nn \\ =
\det_{1\leq \alpha,\beta \leq k} H^\alpha_\beta(t,\bar t),
\ee
where

\be\label{H}
H^\alpha_\beta(t,\bar t) = \sum_{i,j} P_{i-\alpha}(t)g^i_j
P_{j-\beta}(\bar s).
\ee
This formula can be considered as including infinitely many
times $s_i$ and $\bar s_i$, and it is only due to the finiteness
of matrix $g^i_j \in SL(n)$ that $H$-matrix is additionally constrained
\be\label{excon}
\left(\frac{\partial}{\partial t_1}\right)^n H^\alpha_\beta = 0, \nn \\
\ldots \nn \\
\frac{\partial}{\partial t_i}H^\alpha_\beta = 0, \ \ {\rm for}\ \
i \geq n.
\ee
The characteristic property of $H^\alpha_\beta$ is that it satisfies
the following ``shift'' relations (see (\ref{SS})):
\be\label{der}
\frac{\partial}{\partial t_i}H^\alpha_\beta  = H^{\alpha +i}_\beta, \ \ \
\frac{\partial}{\partial \bar t_i} H^\alpha_\beta =
H^\alpha_{\beta + i}.
\ee

Coming back to bilinear relation (\ref{gkgk}), it can be easily
rewritten in terms of $H$-matrix: it is enough to convolute them
with Schur polynomials. For the sake of convenience let us denote
$H\left({\alpha_1\ldots\alpha_k}\atop{\beta_1\ldots \beta_k}\right)
= \det_{1\leq a,b \leq k} H^{\alpha_a}_{\beta_b}$. In accordance with
this notation
$\tau^{(k)} = H\left({1\ldots k}\atop{1\ldots k}\right)$, while bilinear
equation turns into:

\be
H\left({\alpha_1\ldots \alpha_k}\atop{[\beta_1\ldots \beta_k}
\right)
H\left({\alpha'_k\alpha'_1\ldots \alpha'_{k-1}}\atop
{\beta_{k+1}]\beta'_1\ldots\beta_{k-1}'}\right) =
H\left({\alpha_1\ldots \alpha_k[\alpha'_k}\atop
{\beta_1\ldots\beta_k\beta_{k+1}}
\right)
H\left({\alpha'_1\ldots\alpha_{k-1}]'}\atop{\beta'_1
\ldots\beta'_{k-1}}\right).
\ee
Just like original (\ref{gkgk}) these are just matrix identities,
valid for any $H^\alpha_\beta$. However, after the switch from $g$ to
$H$ we, first, essentially represented the equations in the
$n$-independent form and, second, opened the possibility to rewrite
them in terms of time-derivatives.

For example, in the simplest case of
\be
\alpha_i = i, \ \ i = 1,\ldots, k'; \nn \\
\beta_i = i, \ \ i = 1,\ldots,k+1; \nn \\
\alpha'_i = i, \ \ i = 1,\ldots,k-1,\ \ \alpha'_k = k+1; \nn \\
\beta'_i = i, \ \ i = 1,\ldots,k-1
\ee
we
get:
\be
H\left({1\ldots k}\atop{1\ldots k}\right)
H\left({k+1,1\ldots k-1}\atop{k+1,1\ldots k-1}\right) -\\-
H\left({1\ldots k-1, k}\atop{1\ldots k-1, k+1}\right)
H\left({k+1,1\ldots k-1}\atop{k,1\ldots,k-1}\right) =\\=
H\left({1\ldots k+1}\atop{1\ldots k+1}\right)
H\left({1\ldots k-1}\atop{1\ldots k-1}\right)
\ee
(all other terms arising in the process of symmetrization vanish).
This in turn can be represented through $\tau$-functions:
\be\label{hirotafr}
\partial_1\bar\partial_1\tau^{(k)} \cdot \tau^{(k)} -
\bar\partial_1\tau^{(k)} \partial \tau^{(k)} =
\tau^{(k+1)}\tau^{(k-1)}.
\ee
This is the usual lowest Toda-lattice equation. For finite $n$
the set of solutions is labeled by $g \in SL(n)$ (rather than
$SL(\infty)$) as a result of additional constraints (\ref{excon}).

We can now use the chance to illustrate the ambiguity of definition
of $\tau$-function, or, to put it differently, that in the choice of
time-variables. Eq.(\ref{hirotafr}) is actually a corollary of {\it
two} statements: the basic identity (\ref{gkgk}) and the particular
choice of evolution operators in eq.(\ref{tau}), which in the case of
(\ref{tau-k}) implies (\ref{H}) with $P$'s being ordinary Schur
polynomials (\ref{Schur}). At least, in this simple situation (of
fundamental representations of $SL(n)$) one could define
$\tau$-function not by eq.(\ref{tau}), but just by eq.(\ref{detrep}),
with

\be
H^\alpha_\beta(t,\bar t) \longrightarrow
{\cal H}^\alpha_\beta(t,\bar t) = \sum_{i,j} {\cal P}_{i-\alpha}(t)\
g^i_j\ {\cal P}_{j-\beta}(\bar t)
\ee
with {\it any} set of independent functions (not even polynomials)
${\cal P}_\alpha$. Such

\be
\tau^{(k)}_{{\cal P}} = \det_{1\leq \alpha,\beta \leq k}
{\cal H}^\alpha_\beta
\label{detB}
\ee
still remains a generating function for all matrix elements of $G=SL(n)$
in representation $F^{(k)}$. This freedom should be kept in
mind when dealing with ``generalized $\tau$-functions''. As a simple
example, one can take ${\cal P}_\alpha(t)$ to be $q$-Schur polynomials,

\be
\prod_i e_q(t_iz^i) = \sum_j P^{(q)}_j(t)z^j,  \ \ {\rm or} \nn \\
\prod_i e_{q^i}(t_iz^i) = \sum_j \hat P^{(q)}_j(t)z^j,
\ee
which satisfy

\be
D_{t_i} P^{(q)}_j(t) = (D_{t_1})^i P^{(q)}_j(t) = P^{(q)}_{j-i}(t).
\ee
where $D$ are finite-difference operators.
Then instead of (\ref{der}) we would have:

\be
D_{t_i}{\cal H}^\alpha_\beta = {\cal H}^{\alpha +i}_\beta, \ \
D_{\bar t_i}{\cal H}^\alpha_\beta = {\cal H}^\alpha_{\beta+i}
\ee
and

\be
\tau_{P^{(q)}}^{(k)} (t,\bar t | g) =
\det_{1\leq \alpha,\beta \leq k} D_{t_1}^{\alpha -1}
D_{\bar t_1}^{\beta -1} {\cal H}^1_1(t,\bar t).
\ee
So defined $\tau$-function satisfies {\it difference} rather than
differential equations \cite{Sat,MMV}:

\be
\tau^{(k)}\cdot D_{t_1}D_{\bar t_1}\tau^{(k)}- D_{t_1}\tau^{(k)}\cdot
D_{\bar t_1}\tau^{(k)}=\tau^{(k-1)}\cdot M^+_{t_1}M^+_{\bar t_1}
\tau^{(k+1)},\\ \ldots\ \ \ .
\ee
We emphasize, however, that this is just another description
of the $SL(n)$, not $SL_q(n)$ $\tau$-function, if it is interpreted
as a generating function of matrix elements. In particular,
this $\tau$-function takes $c$- rather than $q$-number values.
Still, as concerns its {\it times}-, not $g$-dependence,
it has something to do with the $SL_q(n)$ group, in the
spirit of relation between $q$-hypergeometric functions and quantum
groups (see, for example, \cite{Vinet}).

\subsection{Example of $SL(2)_q$}

Construction from subsection \ref{bieq} is immediately applicable to
the case of quantum groups, the only thing one should keep in mind is
that our definition (\ref{taugen}) gives $\tau$ as an element of
``coordinate ring'' ${\cal A}(G_q)$, not just a $c$-number. If one
wants to obtain a $c$-number $\tau$-function for $q\neq 1$ it is
necessary to restrict the construction further to particular
representation of coordinate ring (this last step will not be
discussed in this paper).  We present here in full detail the
simplest possible example of $SL(2)_q$ \cite{GKLMM}.

\subsubsection{Bilinear identities}

To begin with, fix the notations. We consider generators $T_+$,
$T_-$ and $T_0$ of $U_q(SL(2))$ with commutation relations

\be
q^{T_0} T_\pm q^{-T_0} = q^{\pm 1}T_\pm, \nn \\
\phantom. [T_+,T_-] = \frac{q^{2T_0}-q^{-2T_0}}{q - q^{-1}},
\ee
and comultiplication

\be\label{coprod}
\Delta(T_\pm) = q^{T_0} \otimes T_\pm + T_\pm \otimes q^{-T_0}, \nn \\
\Delta(q^{T_0}) = q^{T_0}\otimes q^{T_0}.
\ee
Verma module $V_\lambda$ with highest weight $\lambda$ (not obligatory
half-integer), consists of the elements
\be\label{notation1}
| n \rangle_\lambda \equiv T_-^n|0 \rangle_\lambda, \ \ n\geq0,
\ee
such that
\be\label{notation}
T_- | n \rangle_\lambda = | n+1 \rangle_\lambda, \nn \\
T_0 | n \rangle_\lambda =
 (\lambda - n) | n \rangle_\lambda , \nn \\
T_+ | n \rangle_\lambda \equiv
b_n(\lambda) | n-1 \rangle_\lambda, \nn \\
b_n(\lambda) = [n][2\lambda + 1 - n], \ \ \
 [x] \equiv \frac{q^x - q^{-x}}{q - q^{-1}}, \nn \\
|| n ||^2_\lambda \equiv \left._\lambda\langle n | n \rangle\right._\lambda
= \frac{[n]!\ \Gamma_q(2\lambda +1)}{\Gamma_q(2\lambda +1-n)}
\stackrel{\lambda \in {\bf Z}/2}{=}
\frac{[2\lambda]![n]!}{[2\lambda -n]!}.
\ee
Now,
\be\label{coprod2}
\left(\Delta(T_-)\right)^n =
q^{nT_0}\otimes T_-^n + [n]T_-q^{(n-1)T_0}\otimes T_-^{n-1}q^{-T_0}
+ \ldots + \nn \\
+ [n] T_-^{n-1}q^{T_0}\otimes T_-q^{-(n-1)T_0} + T_-^n\otimes q^{-nT_0}.
\ee

Let us manifestly derive equations (\ref{CRGamma2})
taking for $W$ an irreducible
spin-${1\over 2}$ representation of $U_q(SL(2))$. Then $\hat
V=V_{\lambda\pm \frac{1}{2}}$, $V=V_\lambda$ and the highest weights of
$\hat V$ in $W\otimes V$ or $V\otimes W$ are\footnote{Hereafter we omit
the symbol of tensor product from the notations of the states $|+\rangle
\otimes|0\rangle_\lambda$ etc.}:

\be\label{hiweight1}
|0\rangle_{\lambda +\frac{1}{2}} =
| + \rangle | 0 \rangle_\lambda, \ \ |+\rangle\equiv|0\rangle_{{1\over 2}},\\
{\rm or}\ \ \ | 0 \rangle_\lambda | + \rangle ;
\ee

\be\label{hiweight2}
|0\rangle_{\lambda - \frac{1}{2}} =
| + \rangle | 1 \rangle_\lambda -
q^{(\lambda + \frac{1}{2})}[2\lambda]| -
\rangle | 0 \rangle_\lambda, \ \ |-\rangle\equiv|1\rangle_{{1\over 2}}, \\
{\rm or} \ \left(q \rightarrow q^{-1}\right) \ \ \
 | 1 \rangle_\lambda | + \rangle -
q^{-(\lambda + \frac{1}{2})}[2\lambda] | 0 \rangle_\lambda
| - \rangle.
\ee
Entire Verma module is generated by the action of $\Delta(T_-)$:

\be\label{state1}
| n \rangle_{\lambda + \frac{1}{2}} =
\left(\Delta(T_-)\right)^n |0\rangle_{\lambda + \frac{1}{2}}
\longrightarrow \nn \\
q^{n/2} \left( | + \rangle| n \rangle_\lambda +
   q^{-(\lambda +\frac{1}{2})}[n]
| - \rangle | n-1 \rangle_\lambda \right), \nn \\
{\rm or}\ \ \
q^{-n/2} \left( | n \rangle_\lambda | + \rangle +
   q^{(\lambda +\frac{1}{2})}[n]
| n-1 \rangle_\lambda | - \rangle \right);
\ee

\be\label{state2}
| n \rangle_{\lambda - \frac{1}{2}} =
\left(\Delta(T_-)\right)^n |0\rangle_{\lambda - \frac{1}{2}}
\longrightarrow \nn \\
q^{n/2} \left( | + \rangle| n+1 \rangle_\lambda +
   q^{(\lambda +\frac{1}{2})}[n-2\lambda]
| - \rangle | n \rangle_\lambda \right), \nn \\
{\rm or}\ \ \
q^{-n/2} \left( | n+1 \rangle_\lambda | + \rangle +
   q^{-(\lambda +\frac{1}{2})}[n-2\lambda]
| n \rangle_\lambda | - \rangle \right);
\ee

Step 2 to be made in accordance with our general procedure
is to project the tensor product of two different $W$'s
onto singlet state
$S = |+\rangle |-\rangle - q|-\rangle|+\rangle$:\footnote{This
state is a singlet of $U_q(SL(2))$. In the case of $U_q(GL(2))$ one
should account for the $U(1)$ non-invariance of $S$. This is the origin
of the factor  ${\rm det}_q g$ at the r.h.s. of the final equation
(\ref{qeqA}).}

\be\label{qdet}
(A| + \rangle + B| - \rangle)\otimes
(| + \rangle C + | - \rangle D) \longrightarrow
AD-qBC.
\ee

With our choice of $W$ we can now consider
two different cases:

($A$) both $\hat
V=V_{\lambda-\frac{1}{2}}$ and $\hat
V'=V_{\lambda'-\frac{1}{2}}$, or

($B$) $\hat
V=V_{\lambda-\frac{1}{2}}$ and $\hat
V'=V_{\lambda'+\frac{1}{2}}$:

\underline{Case A}:

\be\label{projA}
| n \rangle_{\lambda - \frac{1}{2}}
| n' \rangle_{\lambda ' - \frac{1}{2}} \longrightarrow \nn \\
\longrightarrow q^{\frac{n'-n-1}{2}} \left(
[n'-2\lambda']q^{\lambda '}
| n+1 \rangle_\lambda | n' \rangle_{\lambda '} -
[n-2\lambda]q^{-\lambda}
| n \rangle_\lambda | n'+1 \rangle_{\lambda '}\right).
\ee

\underline{Case B}:

\be\label{projB}
| n \rangle_{\lambda + \frac{1}{2}}
| n' \rangle_{\lambda ' - \frac{1}{2}} \longrightarrow \nn \\
\longrightarrow q^{\frac{n'-n-1}{2}} \left(
[n'-2\lambda']q^{\lambda '}
| n \rangle_\lambda | n' \rangle_{\lambda '} -
[n]q^{+\lambda +1}
| n-1 \rangle_\lambda | n'+1 \rangle_{\lambda '}\right).
\ee

Now we proceed to the step 3. Consider any ``group element'', i.e. an
element $g$ from some extension of $U_q(G)$, which possesses the
property:

\be\label{grel}
\Delta (g)=g\otimes g,
\ee
and take matrix elements of the formula (\ref{CRGamma}):

\be\label{me1}
\phantom._{\lambda '}\langle k' |\phantom._{\lambda}\langle k |
\left(g\otimes g\ \Gamma = \Gamma\ g\otimes g \right)
| n \rangle_{\hat\lambda} | n' \rangle_{\hat\lambda '}.
\ee
The action of operator $\Gamma$ can be represented as:

\be\label{me2}
\Gamma | n \rangle_{\hat\lambda} | n' \rangle_{\hat\lambda'}
= \sum_{l,l'} | l \rangle_\lambda | l' \rangle_{\lambda '}
\Gamma(l,l'| n,n'),
\ee
and in these terms (\ref{me1}) turns into:

\be\label{me3}
\sum_{m,m'} \Gamma (k,k'| m,m')
\frac{|| k ||^2_\lambda
 || k' ||^2_{\lambda '}}
 {|| m ||^2_{\hat\lambda}
 || m' ||^2_{\hat\lambda '}}
\langle m | g | n \rangle_{\hat\lambda}
\langle m' | g | n' \rangle_{\hat\lambda '} = \nn \\
= \sum_{l,l'}
\langle k | g | l \rangle_\lambda
 \langle k' | g | l' \rangle_{\lambda '}
\Gamma(l,l'| n,n').
\label{bilimat}
\ee

In order to rewrite this as a difference equation, we
use our definition of $\tau$-function:

\be\label{tau2}
\tau_\lambda(t,\bar t| g) \equiv
\langle \lambda | e_q(tT^+) g e_q(\bar tT^-) | \lambda \rangle =
\sum_{m,n} \langle m | g | n \rangle_\lambda
\frac{t^m}{[m]!}\frac{\bar t^n}{[n]!}.
\ee
Then, one can write down the generating formula for the equation
(\ref{bilimat}), using the manifest form (\ref{projA})-(\ref{projB})
of matrix elements $\Gamma(l,l'| n,n')$:

\underline{Case A}:

\be\label{qeqA}
\sqrt{M_{\bar t}^- M_{\bar t'}^+}
\left( q^{\lambda '} D_{\bar t}^{(0)} \bar t'D_{\bar t'}^{(2\lambda ')} -
  q^{-\lambda } \bar t D_{\bar t}^{(2\lambda)} D_{\bar t'}^{(0)}\right)
\tau_\lambda (t,\bar t | g)\tau_{\lambda '}(t',\bar t'| g) = \nn \\
= [2\lambda][2\lambda']({\rm det}_q g) \sqrt{M_{t}^- M_{t'}^+}
\left( q^{-(\lambda+\frac{1}{2})}t' - q^{(\lambda' + \frac{1}{2})}t\right)
\tau_{\lambda - \frac{1}{2}}(t,\bar t | g)\tau_{\lambda ' - \frac{1}{2}}
(t',\bar t'| g).
\ee
Here
$
D_t^{(\alpha)} \equiv
\frac{q^{-\alpha}M^+_t - q^\alpha M^-_t}{(q - q^{-1})t}
$
and $M^{\pm}$ are multiplicative shift operators,
$M^{\pm}_tf(t)=f(q^{\pm 1}t)$.

\underline{Case B}:

\be\label{qeqB}
\sqrt{M_{\bar t}^- M_{\bar t'}^+}
\left( q^{\lambda '} \bar t' D_{\bar t'}^{(2\lambda ')} -
  q^{(\lambda +1)} \bar t D_{\bar t'}^{(0)}\right)
\tau_\lambda (t,\bar t | g)\tau_{\lambda '}(t',\bar t'| g) = \nn \\
= \frac{[2\lambda ']}{[2\lambda +1]} \sqrt{M_{t}^- M_{t'}^+}
\left( q^{\lambda '} tD_{t}^{(2\lambda +1)} -
  q^{\lambda} t' D_{t}^{(0)}\right)
\tau_{\lambda + \frac{1}{2}}(t,\bar t | g)\tau_{\lambda ' -
\frac{1}{2}}(t',\bar t'|g).
\ee

Let us note that the derivation of these equations can be presented
in the form which looks even closer to conventional free-fermion
formalism.  It is possible to represent operator $\Gamma$ in
component form as $E_1^R\otimes E_2^L-qE_2^R\otimes E_1^L$, where
$E_{i}$'s are components of the vertex operator (given by fixing
different vectors from $W$). Then the equation (\ref{Ggg=ggG}) can be
rewritten

\be\label{J}
\left._{\hat V}
\langle 0|e_q(tT_+)\ E_1^Rg\ e_q(\bar t T_-)|0\rangle_V
\cdot
_{\hat V'}\langle 0|e_q(tT_+)\ E_2^Lg\ e_q(\bar t
T_-)|0\rangle_{V'}\right.-\\-
q\left._{\hat V}\langle 0|e_q(tT_+)\ E_2^Rg\
e_q(\bar t T_-)|0\rangle_V
\cdot
_{\hat V'}\langle 0|e_q(tT_+)\
E_1^Lg\ e_q(\bar t T_-)|0\rangle_{V'}\right.
=\\=
\left._{\hat V}
\langle 0|e_q(tT_+)\ gE_1^R\ e_q(\bar t T_-)|0\rangle_V
\cdot
_{\hat V'}\langle 0|e_q(tT_+)\ gE_2^L\ e_q(\bar t
T_-)|0\rangle_{V'}\right.-\\-
q\left._{\hat V}\langle 0|e_q(tT_+)\ gE_2^R\
e_q(\bar t T_-)|0\rangle_V
\cdot
_{\hat V'}\langle 0|e_q(tT_+)\ gE_1^L\
e_q(\bar t T_-)|0\rangle_{V'}\right.
\ee
We can easily obtain commutation relations of $E_i$'s with generators
of algebra as well as their action on vacuum states. Then, it is
straightforward to commute $E_i$'s with $q$-exponentials in the
expression (\ref{J}) and represent the result of the commutation by the
action of difference operators. Of course, the results (\ref{qeqA}) and
(\ref{qeqB}) are reproduced in this way.

\subsubsection{Solution to bilinear identities}

In this particular case (of $SL(2)_q$) one can easily
evaluate the $\tau$-function explicitly, and let us use this
possibility to show how bilinear equations are satisfied. The
fact that our $\tau$-function is operator-valued will be of
course of principal importance.

Let us begin from the case of $\lambda = \frac{1}{2}$. Then
\be
\tau_{{1\over 2}}(t,\bar t|g) =
\langle + | g | + \rangle  + \bar t \langle + | g | - \rangle
+ t \langle - | g | + \rangle + t\bar t \langle - | g | - \rangle =\nn \\
= a + b\bar t + ct + dt\bar t,
\ee
where  $a,b,c,d$ are elements of the matrix
\be
{\cal T} = \left(
\begin{array}{cc}
 a& b \\ c& d
\end{array}
\right)
\ee
with the commutation relations dictated by ${\cal T}{\cal T}{\cal R}=
{\cal R}{\cal T}{\cal T}$ equation \cite{FTR}

\be
ab = qba, \nn \\
ac = qca, \nn \\
bd = qdb, \nn \\
cd = qdc, \nn \\
bc = cb, \nn \\
ad - da = (q - q^{-1})bc.
\label{core}
\ee
If $b$ or $c$ or both are non-vanishing, $\tau_{{1\over 2}}(t,\bar
t|g)$ with different values of time-variables $t,\bar t$ do not
commute.  Still such $\tau_{{1\over 2}}(t,\bar t| g)$ does satisfy
the same bilinear identity (\ref{qeqA}), moreover, for this to be
true it is essential that commutation relations (\ref{core}) are
exactly what they are.  Indeed, the l.h.s. of the equation
(\ref{qeqA}) is equal to

\be\label{check}
-q^{{1\over 2}}\sqrt{M^-_{\bar t}}(b+dt)\sqrt{M^+_{\bar t'}}(a+ct')+
q^{-{1\over 2}}\sqrt{M^-_{\bar t}}(a+ct)\sqrt{M^+_{\bar t'}}(b+dt')=\\=
(q^{-{1\over 2}}ab-q^{{1\over 2}}ba)+
(q^{-{1\over 2}}cd-q^{{1\over 2}}dc)tt'+
(q^{-{1\over 2}}cb-q^{{1\over 2}}da)t+
(q^{-{1\over 2}}ad-q^{-{1\over 2}}bc)t'=\\
=(q^{-{1\over 2}}t'-q^{{1\over 2}}t){\rm det}_qg,
\ee
which coincides with the r.h.s. of the equation (\ref{qeqA}).

To perform the similar check for any half-integer-spin representation,
let us note that the corresponding
$\tau$-function can be
easily written in terms of $\tau_{{1\over 2}}$. Indeed,
\be
| n \rangle_{\lambda} = \left(q^{T_0}\otimes T_- +
T_- \otimes q^{-T_0}\right)^n| 0 \rangle_{\lambda -\frac{1}{2}}\otimes
|0 \rangle_\frac{1}{2} = \nn \\
= q^{-n/2}\left( |n \rangle_{\lambda-\frac{1}{2}} \otimes
|0\rangle_\frac{1}{2} +
[n] q^{\lambda } |n-1\rangle_{\lambda -\frac{1}{2}}\otimes
| 1 \rangle_\frac{1}{2}\right); \\
\phantom._{\lambda}\langle n | =
\phantom._{\lambda-\frac{1}{2}}\langle 0 | \otimes
\phantom._\frac{1}{2} \langle 0 | =
\left(q^{T_0}\otimes T_+ + T_+\otimes q^{-T_0}\right)^n =
\nn \\ =
q^{-n/2} \left( \phantom._{\lambda-\frac{1}{2}} \langle n | \otimes
\phantom._\frac{1}{2}\langle 0 | + [n]q^{\lambda}
\phantom._{\lambda-\frac{1}{2}} \langle n-1 | \otimes
\phantom._\frac{1}{2}\langle 1 | \right).
\ee
Thus
\be
\phantom._{\lambda }\langle k | g | n \rangle_{\lambda} =
q^{-\frac{k+n}{2}}\left[
\phantom._{\lambda-\frac{1}{2}} \langle k | g | n
\rangle_{\lambda-\frac{1}{2}}
\phantom. \langle + | g | + \rangle +
q^{\lambda }[n]\phantom._{\lambda-\frac{1}{2}}\langle k | g | n-1
\rangle_{\lambda-\frac{1}{2}} \langle + | g | - \rangle +\right.\\+ \left.
q^{\lambda }[k]\phantom._{\lambda-\frac{1}{2}}\langle k-1 | g | n
\rangle_{\lambda-\frac{1}{2}} \langle - | g | + \rangle +
q^{2\lambda}[k][n]\phantom._{\lambda-\frac{1}{2}}\langle k-1 | g | n-1
\rangle_{\lambda-\frac{1}{2}} \langle - | g | - \rangle \right]
\ee
or, in terms of generating ($\tau$-)functions:
\be
\tau_{\lambda }(t,\bar t|g) =
\sqrt{M^-_t M^-_{\bar t}}\left(\tau_{\lambda-{1\over 2}}(t,\bar t|g)
\left(a+ q^{\lambda}\bar t b +
q^{\lambda}t c + q^{2\lambda } t\bar t d\right)\right).
\ee
Applying this procedure recursively we get:
\be
\tau_\lambda(t,\bar t|g) =
\tau_{\lambda - {1\over 2}}(q^{-{1\over 2}}t,q^{-{1\over 2}}\bar t | g)
\tau_{{1\over 2}}(q^{\lambda - {1\over 2}}t,
q^{\lambda - {1\over 2}}\bar t | g) = \nn \\
\stackrel{{\rm if}\ \lambda \in \hbox{\bf Z}/2}{=}
\tau_{{1\over 2}}(q^{{1\over 2} - \lambda}t,
q^{{1\over 2} - \lambda}\bar t|g)
\tau_{{1\over 2}}(q^{{3\over 2}-\lambda}t,
q^{{3\over 2} - \lambda}\bar t | g) \ldots
\tau_{{1\over 2}}(q^{\lambda - {1\over 2}}t, q^{\lambda -
{1\over 2}}\bar t | g),
\ee
i.e. for half-integer $\lambda$ $\tau_\lambda$
is a polynomial of degree $2\lambda$ in $a,b,c,d$.

For example,
\be
\tau_1 (t,\bar t |g)
= \tau_{{1\over 2}}(q^{-{1\over 2}}t,q^{-{1\over 2}}\bar t | g)
\tau_{{1\over 2}} (q^{{1\over 2}}t, q^{{1\over 2}}\bar t | g) = \nn \\
= (a + q^{-{1\over 2}}\bar t b + q^{-{1\over 2}} t c + q^{-1}t\bar t d)
(a + q^{{1\over 2}}\bar t b + q^{{1\over 2}}\bar t c + q t\bar t d) =
\nn \\ =
a^2 + (q^{{1\over 2}}ab + q^{-{1\over 2}}ba)\bar t +
(q^{{1\over 2}}ac + q^{-{1\over 2}}ca) t + b^2 \bar t^2 +\\+
(qad + bc + cb + q^{-1}da)t\bar t + c^2 t^2 +
(q^{{1\over 2}}bd + q^{-{1\over 2}}db)t\bar t^2 +\\+
 (q^{{1\over 2}}cd + q^{-{1\over 2}}dc)t^2\bar t+ d^2 t^2\bar t^2
\ee
Using the relations like
\be
q^{{1\over 2}} ab + q^{-{1\over 2}}ba = [2]q^{{1\over 2}}ba =
[2]q^{-{1\over 2}}ab \ \ {\rm etc.}
\ee
one gets for this case
\be\label{tau=1}
\tau_1 (t,\bar t |g) =
a^2+[2]q^{-{1\over 2}}ab\bar t+[2]q^{-{1\over 2}}act+b^2\bar
t^2+([2]qbc+[2]da)t\bar t+c^2t^2+ \nn \\ +
[2]q^{{1\over 2}}dbt\bar
t^2+[2]q^{{1\over 2}}dct^2\bar t+d^2t^2\bar t^2.
\ee
With this explicit expression, one can trivially make the calculations
similar to (\ref{check})
in order to check manifestly equation (\ref{qeqA}) for $\lambda=1$,
$\lambda'={{1\over 2}},\ 1$ and equation (\ref{qeqB}) for
$\lambda= \lambda'= {{1\over 2}}$.

Thus, we showed explicitly (for the case of $SL_q(2)$) that the
quantum bilinear identities have as many solutions as the classical
ones, provided the $\tau$-function is allowed to take values in
non-commutative ring $A(G)$.

\subsection{Comments on the quantum deformation of KP/Toda
$\tau$-functions \label{quaKP}}

As we saw in the previous subsection, the generic construction is
easily applicable to quantum groups, still the problem of quantum
KP/Toda hierarchies deserves separate consideration and is not yet
fully resolved. The problem is, that the evolution operator $U(t)$ in
(\ref{taugen2}) is usually constructed from all the operators of the
algebra, not just from a commuting set - as it happens in particular
case of fundamental representations of $GL(N)$. As result generic
evolution of $\tau$ with variation of $t$'s is not described as a set
of {\it commuting} flows, rather they form a closed, but non-trivial,
algebra. This manifests itself also in the fact that naturally the
number of independent time-variables is rather close to dimension
than to the rank of the group.  The problem of quantum deformation of
KP/Toda hierarchy is to find a deformation which, while dealing with
$\tau$-function for quantum group, is still describable in terms of
few time-variables.  If at all resolvable this is the problem of a
clever choice of the weight functions $s^{R}_{m,\bar m}(t,\bar t)$ in
(\ref{taugen}).  Following \cite{KMM} we shall now demonstrate that
the problem {\it is} resolvable in principle, though at the moment it
is a quantum deformation of somewhat non-conventional description of
KP/Toda system (with evolution, introduced differently from that in
s.\ref{funrept}, and it is not just a change of time-variables:  the
transformation is representation $R$-dependent).

According to \cite{MV} parametrization of group elements which allows
the most straightforward quantum deformation involves only
{\it simple} roots $\pm \vec\alpha_i$, $i = 1,\ldots,r_G$:
\be
g = g_Ug_Dg_L, \\
g_U = \prod_s^<\ e^{\theta_sT_{i(s)}},   \ \ \
g_L = \prod_s^>\ e^{\chi_sT_{-i(s)}}, \ \ \
g_D = \prod_{i=1}^{r_G} e^{\vec\phi\vec H}
\label{genpar}
\ee
Every particular simple root $\vec\alpha_i$ can appear several
times in the product, and there are different parametrizations of
the group elements of such type, depending on the choice of the
set $\{s\}$ and the mapping $i(s)$. Quantum deformation of
{\it such} formula is especialy simple because comultiplication
rule is especialy simple for generators, associated with
{\it simple} roots:
\be
\Delta(T_i) = T_i\otimes q^{-2H_i} + I\otimes T_i, \\
\Delta(T_{-i}) = T_{-i}\otimes I + q^{2H_i}\otimes T_{-i}
\label{quco}
\ee
For $q\neq 1$ any expression of the form (\ref{genpar})
remains just the same, provided exponents in $g_U$ and $g_L$
are understood as $q$-exponents (in the simply-laced case,
$q^{||\vec\alpha_i||^2/2}$-exponents in general), and parameters
$\psi, \chi, \vec\phi$ become non-commuting generators of
``coordinate ring'' ${\cal A}(G_q)$. Actualy they form a kind of a
very simple Heisenberg-like algebra:
\be
\theta_s\theta_{s'} = q^{-\vec\alpha_{i(s)}\vec\alpha_{i(s')}}
\theta_{s'}\theta_s,\ \ \ s<s', \\
\chi_s\chi_{s'} = q^{-\vec\alpha_{i(s)}\vec\alpha_{i(s')}}
\chi_{s'}\chi_s,\ \ \ s<s', \\
q^{\vec\beta\vec\phi}\theta_s = \theta_s q^{\vec\beta\vec\phi}
q^{\vec\beta\vec\alpha_{i(s)}}, \\
q^{\vec\beta\vec\phi}\chi_s =
\chi_s q^{\vec\beta\vec\phi}q^{\vec\beta\vec\alpha_{i(s)}}
\label{comrel}
\ee
These relations imply that $\Delta(g) = g\otimes g$.

The simplest possible assumption about evolution operators
would be to say that, just as it was in the case of the
standard KP/Toda theory (see s.\ref{funrept}),
$U(t)$ is always an object of the type $g_U$,
while $\bar U(\bar t)$ - of the type $g_L$. However, these are no
longer group elements:
$$
\Delta(g_U) \neq g_U \otimes g_U, \ \ \
\Delta(g_L) \neq g_L \otimes g_L,
$$
because of the lack of factors $g_D$. Still the simplest possibility
is to insist on identification of
$U$ and $\bar U$ as objects of the type $g_U$ and $g_L$ respectively,
and explicitly investigate implications of the failure of
(\ref{clacoU}). As result one obtains instead of (\ref{clacoU})
\be
\Delta(U(\xi)) = U^{(2)}_L(\xi) \cdot U^{(2)}_R(\xi),
\ee
where
\be
U(\xi) = \left.\prod_s\right.^<\
{\cal E}_q\left(\xi_sT_{i(s)}\right), \\
U^{(2)}_L = \left.\prod_s\right.^<\
{\cal E}_q\left(\xi_sT_{i(s)}\otimes q^{-2H_{i(s)}}\right)
\neq I\otimes U(\xi), \\
U^{(2)}_R = \left.\prod_s\right.^<\
{\cal E}_q\left(\xi_s I\otimes T_{i(s)}\right) = I\otimes U(\xi)
\ee
and this has some simply accountable implications for determinant
formulas for quantum $\tau$-functions.

In s.\ref{funrept} we essentially used an evolution operator of the type
\be
U(\xi) = \prod_{1\leq i \leq N}^r\prod_{1\leq j <i}\exp\left(
\xi_{ij} T_{i-j}\right)
\label{i(s)}
\ee
where $\xi_{ij}$ are certain functions of only $N$ idependent
variables $t$. While (\ref{i(s)}) is trivial to deform in the
direction of $q\neq 1$, it is a separate (yet unresolved) problem
to find such reduction to only $N$-variables, consistent with the
commutation relations between $\xi_{ij}$,
$$
\xi_{ij}\xi_{i'j'} = q^{-\vec\alpha_{i-j}\vec\alpha_{i'-j'}}
\xi_{i'j'}\xi_{ij}, \ \ \ \{i,j\}<\{i',j'\}.
$$

One can instead use a much simpler evolution,
\be
\hat U(\xi) = \left.\prod_{i=1}^{r_G}\right.^<\
\exp\left(\xi_iT_i\right)
\label{UA}
\ee
This is enough to generate all the states of any fundamental
representation from the corresponding vacuum (highest vector)
state, but
$<vac_{F_n}|\ \hat U(\xi) $ has nothing to do with the usual
$<vac_{F_n}|\ U(t)$, where $U(t)$ is given by (\ref{i(s)}).
It can be better to say, that identification
$ <vac_{F_n}|\ U^{(A)}(\xi) = \ <vac_{F_n}|\ U(t)$
defines a relation $\xi_i(t)$, which explicitly depends on $n$.

One can of course built the theory of KP/Toda hierarchies in terms of
$\xi$-variables instead of conventional $t$-variables, but it can
{\it not} be obtained by just change of time-variables: the whole
construction will look different. Instead this new construction is
immediately deformed to the case of $q\neq 1$: instead of (\ref{UA})
we just write

\be
\hat U(\xi) = \left.\prod_{i=1}^{r_G}\right.^<\
{\cal E}_q\left(\xi_iT_i\right)
\label{UAq}
\ee
where $\xi$'s are non-commuting variables,
\be
\xi_i\xi_j = q^{-\vec\alpha_i\vec\alpha_j}\xi_j\xi_i,
\ \ \ i<j,
\ee
and it is easy to derive quantum counterpart of any statement
of the classical ($q=1$) theory once it is formulated for
$\xi$-parametrization.

In what follows we first briefly describe the con\-ven\-tional KP/Toda
hierarchy in this non-standard para\-metri\-zation, then consider
the cor\-responding quantum deformation and derive the substitute of
deter\-minant formulas for $\tau_n \equiv \tau_{F_n}$ in the case of
$q \neq 1$.

\subsubsection{On the modified KP/Toda hierarchy}

Our first purpose is to demonstrate that all the main ingredients
of description of the classical KP/Toda hierarchy, as described in
s.\ref{funrept}, are preserved if evolution (\ref{UAq}) is used instead
of (\ref{i(s)}), in particular,  there are determinant formulas and a
hierarchy of differential equations.

{}From now on we denote the $\tau$-function associated with the
evolution (\ref{UAq}) through $\hat\tau(\xi,\bar\xi|g)$.  This
$\tau$-function is linear in each time-variable $\xi_i$, hence, it
satisfies simpler determinant formulas and simpler hierarchy of
equations.  Indeed, now we have

\be
\hat\tau_1(\xi,\bar\xi|g) \equiv\ <0_{F_1}|\hat U(\xi) g \hat{\bar U}(\bar
\xi)|0_{F_1}>\ = \sum_{k,\bar k \geq 0} s_k\bar s_{\bar k} <k|g|\bar k>
\label{clatau1s}
\ee
where $s_k = \xi_1\xi_2\ldots \xi_{k}$, $s_0=1$, and

\be
\hat\tau_1^{m\bar m}(\xi,\bar\xi|g)
\equiv\ <m_{F_1}|\hat U(\xi) g \hat{\bar U}(\bar \xi)|\bar m_{F_1}>\ =
\frac{1}{s_m\bar s_{\bar m}} \sum_{{k \geq m}\atop{\bar k \geq \bar m}}
s_k\bar s_{\bar k} <k|g|\bar k>\   = \\
=\frac{1}{s_{m}\bar s_{\bar m}}\sum_{{k \geq
m}\atop{\bar k \geq \bar m}} \frac{\partial}{\partial\log s_k}
\frac{\partial}{\partial\log \bar s_{\bar k}}\tau_1(\xi,\bar\xi|g)=
\frac{1}{s_{m-1}\bar s_{\bar
m-1}}\frac{\partial}{\partial \xi_m} \frac{\partial}{\partial \bar
\xi_{\bar m}}\tau_1(\xi,\bar\xi|g).
\label{clataumm1s}
\ee

Thus,\footnote{
One can compare determinant representations (\ref{detB}) and
(\ref{detA}) to find the connection between different coordinates $t$
and $\xi$.  For every given $n$ the variables $s_k$ are some
functions of $P_j(t)$.  For example, in the simplest case of the
first fundamental representation $F^{(1)}$ we have $\tau_1(t|g)
=\hat\tau_1(\xi|g)$ and $s_k = P_k(t),\ \ \ {\partial\over\partial
t_k}=\sum_i s_{i-k}{\partial\over\partial s_i}$.  However,
identification of $\tau_n(t)$ and $\hat\tau_n(\xi)$ with $n\neq 1$
will lead to different relations between $\xi$ and $t$. Thus the two
different evolutions are {\it not} related just by a change of
time-variables, relation is representation-dependent, and can not be
lifted to the actual KP/Toda case (when $n=\infty$). Two evolutions
provide two equally nice, but not just equivalent descriptions of the
same hierarchy.}

\be
\hat\tau_{n+1} = \det_{0\leq m,\bar m \leq n}\hat\tau_1^{m\bar m}
= \left(\prod_{m=1}^n s_m\bar s_{\bar m}\right)^{-1}
{\rm det}_{m\bar m} \left(
\sum_{{k \geq m}\atop{\bar k \geq \bar m}} s_k\bar s_{\bar k}
<k|g|\bar k>\right) = \\
=  \frac{1}{s_n\bar s_n}
\sum_{k,\bar k \geq n} s_k\bar s_{\bar k}
\det_{0 \leq m,\bar m \leq n-1}\left(
\begin{array}{cc}
g_{m\bar m} & g_{m\bar k} \\
g_{k\bar m} & g_{m\bar m}
\end{array} \right)
\equiv \frac{1}{s_n\bar s_n}
\sum_{k,\bar k \geq n} s_k\bar s_{\bar k}
{\cal D}_{k\bar k}^{(n)}.
\label{detA}
\ee

\subsubsection{$q$-Determinant-like representation}

In this section we demonstrate how the technique developed in the
previous sections is deformed to the quantum case and, in particular,
obtain $q$-determinant-like deformation of (\ref{detA}).  Our
evolution operator (\ref{UAq}) satisties the following
comultiplication rule:

\be
\Delta^{n-1}\left(U\{T_i\}\right) = \prod_{m=1}^n U^{(m)}
\ee
where
\be
U^{(m)} = U\left\{\ I\otimes\ldots
\ldots I\otimes \xi_iT_i \otimes q^{-2H_i}\otimes \ldots\otimes
q^{-2H_i}\right\}
\ee
and $T_i$ appears at the $m$-th place in the tensor product.  Similarly
\be
\bar U^{(m)} = \bar U
\left\{\ q^{2H_i}\otimes\ldots \ldots q^{2H_i}\otimes T_{-i}
\otimes I\otimes \ldots\otimes I\right\}.
\ee

Now let us transform the operator-valued $q$-factors into
$c$-number ones. Let
$$
H_i|\bar j_{F_1}> = h_{i,\bar j}|\bar
j_{F_1}>, \ \ \ <j_{F_1}|H_i = h_{i,j}<j_{F_1}|
$$
(in fact for $SL(N)$ $\
2h_{i,i-1} = +1,\ 2h_{i,i} = -1$, all the rest are vanishing).  Then
\be
\hat\tau_n^{j_1\ldots j_n\bar j_1\ldots\bar j_n}(\xi_i,\bar \xi_i|g)
\equiv \\ \equiv
\left(\otimes_{m=1}^n<j_m|\right)
\Delta^{n-1}(U) g^{\otimes n} \Delta^{n-1}(\bar U)
\left(\otimes_{m=1}^n|\bar j_m>\right)=
\\= \prod_{m=1}^n \ <j_m|
U\left\{\xi_iT_i q^{-2\sum_{l = m+1}^nh_{i,j_l}}\right\}\ g\
\bar U\left\{\bar\xi_iT_{-i}
q^{2\sum_{l=1}^{m-1}h_{i,\bar j_l}}\right\}\ |\bar j_m>\ =\\
= \prod_{m=1}^n \hat\tau_1^{j_m\bar j_m}\left(\xi_i
q^{-2\sum_{l = m+1}^nh_{i(s),j_l}},\bar \xi_i
q^{2\sum_{l=1}^{m-1}h_{i(s),\bar j_l}}\right).
\label{4.42}
\ee

In order to get the analogue of (\ref{detB}),
one should replace antisymmetrization
by $q$-antisymmetrization, since, in quantum case,
fundamental representations are described by $q$-antisymmetrized vectors.
We define $q$-antisymmetrization as a sum over all permutations,
\be
\left( [1,\ldots,k]_q\right) = \sum_P (-q)^{{\rm deg}\ P}
\left(P(1),\ldots,P(k)\right),
\ee
where
\be
{\rm deg}\ P = \#\ {\rm of\ inversions\ in}\ P.
\ee
Then, $q$-antisymmetrizing (\ref{4.42}) with $j_k=k-1,\ \bar j_{\bar
k}=\bar k-1$, one finally gets

\be
\tau_n(\xi,\bar \xi|g) =
\sum_{P,P'} (-q)^{{\rm deg}\ P + {\rm deg}\ P'}\times\\
\times\prod_{m=0}^{n-1}
\tau_1^{P(m)P'(\bar m)}\left(\xi_s
q^{-2\sum_{l = m+1}^{n-1}h_{i(s),P(l)}},\bar \xi_s
q^{2\sum_{\bar l=0}^{m-1}h_{i(s),P'(\bar l)}}\right).
\label{detq}
\ee
This would be just  a $q$-determinant, be there no
$q$-factors which twist the time-variables.\footnote{
Let us remind that the $q$-determinant is defined as
$$
{\rm det}_q A \sim  A^{[1}_{[1}\ldots A^{n]_q}_{n]_q}
= \sum_{P,P'} (-q)^{{\rm deg}\ P + {\rm deg}\ P'}
\prod_{a} A^{P(a)}_{P'(a)}
$$
Note that this is not obligatory the same as
$A^1_{[1}\ldots A^n_{n]_q}$. It is the same only for peculiar
commutation relations of the matrix elements $A_i^j$.}

To make this expression more transparent let us consider the simplest
example of the second fundamental representation:
\be
\tau_2  = \tau_1^{00}(q\xi_1,q^{-1}\xi_2,\xi_i;\
\bar \xi_1,\bar \xi_2,\bar \xi_i) \tau_1^{11}(\xi_1,\xi_2,\xi_i;\
  q\bar \xi_1,\bar \xi_2,\bar \xi_i) - \\
- q\tau_1^{01}(q\xi_1,q^{-1}\xi_2,\xi_i;\
  \bar \xi_1,\bar \xi_2,\bar \xi_i)
\tau_1^{10}(\xi_1,\xi_2,\xi_i;\
  q^{-1}\bar \xi_1,q\bar \xi_2,\bar \xi_i) - \\
- q\tau_1^{10}(q^{-1}\xi_1,\xi_2,\xi_i;\
  \bar \xi_1,\bar \xi_2,\bar \xi_i)
\tau_1^{01}(\xi_1,\xi_2,\xi_i;\
  q\bar \xi_1,\bar \xi_2,\bar \xi_i) + \\
+ q^2\tau_1^{11}(q^{-1}\xi_1,\xi_2,\xi_i;\
  \bar \xi_1,\bar \xi_2,\bar \xi_i)
\tau_1^{00}(\xi_1,\xi_2,\xi_i;\
  q^{-1}\bar \xi_1,q\bar \xi_2,\bar \xi_i).
\ee
This can be written in a more compact form with the help of operators
\be
\hbox{{\hr D}}^L_i\equiv D_i\otimes I,\ \ \
\hbox{{\hr D}}^R_i\equiv \prod_j M_j^{-\vec\alpha_i\vec\alpha_j}\otimes D_i,
\\
\bar {\hbox{{\hr D}}}^L_i\equiv \bar D_i\otimes\prod_j \bar
M_j^{-\vec\alpha_i\vec\alpha_j},\ \ \
\bar {\hbox{{\hr D}}}^R_i\equiv I\otimes \bar D_i.
\ee
These operators have simple commutation relations:
\be
\hbox{{\hr D}}_i^{L}    \hbox{{\hr D}}_j^{R}
= q^{\vec\alpha_i\vec\alpha_j}
\hbox{{\hr D}}_j^{R}\hbox{{\hr D}}_i^{L}, \\
\bar {\hbox{{\hr D}}}_i^{L} \bar {\hbox{{\hr D}}}_j^{R}
= q^{\vec\alpha_i\vec\alpha_j}\bar{\hbox{{\hr D}}}_j^{R}\bar{\hbox{{\hr
D}}}_i^{L}.
\ee
Then,
\be
\tau_2 = \left(M^-_1\otimes\bar M_1^+\right)\left( \hbox{{\hr D}}_1^R-q
\hbox{{\hr D}}_1^L\right)\cdot\left(
\bar{\hbox{{\hr D}}}_1^R-q\bar{\hbox{{\hr D}}}_1^L\right)
\tau_1\otimes \tau_1.
\ee

\section{Conclusion}

These notes combine presentation of some well established
facts with that of more recent and sometime disputable speculations.
There are all reasons to believe that further developements will prove that
the theory of generalized $\tau$-functions and non-perturbative
partition functions can become a flourishing branch of mathematical
physics with applications well beyond the present modest scope
of topological theories and $c<1$ string models and with profound
relations to other fields of the string theory. It is also important that
there are plenty of ``small problems'' at all the levels of this theory,
which are enjoyable to think about.

\section{Acknowledgements}

I am indebted to my coauthors and friends for cooperation
during our work on the subject of matrix models and integrable
systems.

It is a pleasure to thank Gordon Semenoff and Luc Vinet for
the invitation to present this material at the Banff school,
as well as for their hospitality and support.

The text was partly prepared during my stay at CRM, Montreal.

\end{document}